# Smartphone-Based Undergraduate Physics Labs: A Comprehensive Review of Innovation, Accessibility, and Pedagogical Impact

Yiping Zhao
Department of Physics and Astronomy, The University of Georgia, Athens, GA 30602, USA

## Abstract

Smartphone-integrated physics laboratories (SmartIPLs) have emerged as a cost-effective and scalable approach that offers new opportunities for enhancing introductory physics education. By leveraging the sensing, imaging, and computing capabilities of smartphones, SmartIPLs enable hands-on, inquiry-based experiments that can be conducted flexibly, in classrooms, at home, or in remote environments. This review presents a comprehensive synthesis of over 200 documented SmartIPL activities across key physics domains, including mechanics, optics, thermodynamics, electromagnetism, and modern physics. Activities are systematically categorized by topic, sensor modality (sensor-based *versus* camera-based), and instructional design. The pedagogical value of SmartIPLs is examined through the lens of context-based learning, open-ended inquiry, cognitive scaffolding, and student motivation, with particular attention to issues of accessibility and equity. We also analyze implementation challenges such as device heterogeneity, sensor limitations, and integration into traditional curricula. Looking forward, we identify emerging directions for the field, including AI-driven feedback systems, open-source curricular platforms, and expanded applications in modern physics. This review positions SmartIPLs not merely as technological supplements, but as transformative tools for fostering scientific thinking, modeling skills, and epistemic agency in 21st-century physics education.

## 1. Introduction

The rapid evolution of modern society calls for transformative approaches in STEM education that prioritize interdisciplinary problem-solving, scientific modeling, and real-world application.[1] Introductory Physics Laboratories (IPLs), which are intended as foundational experiences for undergraduates entering STEM disciplines, have traditionally struggled to meet these demands. Despite the American Association of Physics Teachers (AAPT) identifying core goals for IPLs, including the cultivation of experimentation skills, analytical reasoning, conceptual understanding, model-data integration, and collaborative learning,[2] many IPLs remain dominated by rigid, procedural formats that emphasize rote compliance over authentic scientific inquiry. Holmes and Wieman demonstrated that such "cookbook" labs often show no measurable impact on student learning outcomes, highlighting the urgent need for pedagogical reform.[3]

Contemporary learning theories, including constructivism and situated cognition, emphasize that meaningful learning emerges from context-rich, interactive environments where students actively construct knowledge through experimentation and collaboration.[4-8] In response, the past three decades have witnessed a surge in physics education reforms centered on active learning, including models such as Studio Physics, Workshop Physics, Studio Physics, and SCALE-UP (Student-Centered Active Learning Environment for Undergraduate Programs).[9-17] These environments integrate lecture, lab, and discussion into unified spaces where students engage in peer instruction, hands-on experimentation, and real-time data analysis. For example, Studio Physics, pioneered by Jack Wilson at Rensselaer Polytechnic Institute, eliminates the traditional separation between lectures and labs by having students work in small teams at shared tables equipped with computers and lab instruments, engaging in hands-on activities and peer instruction.[11] SCALE-UP, developed by Robert Beichner at North Carolina State University, adapts this studio-style format for large-enrollment courses by emphasizing collaborative learning, group problem-solving, and real-world applications in flexible classroom layouts.[18] Numerous studies confirm that these formats lead to improved conceptual understanding, higher retention rates, and more positive attitudes toward physics compared to traditional formats. [9-17, 19-21] Despite their proven effectiveness, these reformed environments face logistical and financial barriers. Implementing SCALE-UP or Studio Physics typically requires specialized classrooms, increased staffing, and significant curricular redesign, constraints that limit scalability, particularly in large-enrollment or resource-constrained institutions.

Against this backdrop, smartphone-integrated physics labs (SmartIPLs) have emerged as a cost-effective, scalable, and pedagogically robust alternative that aligns with the active learning paradigm. Drawing on the ubiquitous availability and multifunctionality of smartphones, SmartIPLs have been shown in prior studies to support authentic, inquiry-driven experiments using built-in sensors or camera-based analysis. They can also support context-based learning, inquiry-based experimentation, and tool-enabled cognitive scaffolding, facilitating deep engagement and epistemic agency.[22-24] SmartIPLs can be deployed in a variety of settings, including conventional classrooms,[24] dorm rooms,[25] or fully remote environments[26], without the need for specialized infrastructure. Applications such as *Phyphox*, *Tracker*, and *Physics Toolbox* provide platforms for real-time data collection, visualization, and analysis, supporting iterative modeling and reflection.[27, 28] The COVID-19 pandemic further accelerated the adoption of SmartIPLs as institutions pivoted to remote instruction. Students used smartphone apps and video capture tools to conduct physics experiments from home, analyzing acceleration, sound, and light data using accessible software.[23, 29, 30] These adaptive practices illustrated the usefulness of SmartIPLs during the COVID-19 pandemic and suggested their applicability for supporting self-directed, hands-on learning. Importantly, post-pandemic studies affirm the enduring educational value of SmartIPLs. Lahme et al. found that smartphone-based labs significantly improved student understanding and engagement, with over 85% of students reporting positive experiences.[31] Tufino and Organtini showed that integrating Arduino and smartphones into mechanics labs enhanced students' ownership of experimental design and improved modeling and data interpretation skills.[32] Rather than replacing traditional labs, SmartIPLs may serve as a complementary modality that extends learning beyond fixed lab schedules and potentially increases accessibility. Their flexible implementation supports open-ended inquiry, project-based learning, and interdisciplinary integration, aligning with broader STEM reform goals and contemporary educational theory.

This review synthesizes over a decade of research and practice surrounding SmartIPLs. We analyze over 200 documented SmartIPL activities spanning mechanics, optics, thermodynamics, electromagnetism, and modern physics. These activities are categorized by topic area, data acquisition method (sensor-based vs. camera-based), and instructional format. In addition to map SmartIPLs against active learning goals, we evaluate empirical evidence on their effectiveness in promoting science process skills, conceptual understanding, student motivation, and equity. We also examine persistent challenges, including sensor variability, device compatibility, and classroom integration, and propose future directions such as artificial intelligence (AI) integration, equity-driven instructional design, and open-source curricular platforms. Ultimately, we examine how smartphones, given their ubiquity and portability, can contribute to ongoing efforts to reimagine experimental physics instruction.

## 2. The Role of Smartphones in Education

### 2.1 Global reach and ubiquity of smartphones

Smartphones have become one of the most widely adopted technologies globally, with approximately 4.88 billion users in 2024, accounting for about 60% of the world's population.[33] This number is projected to rise to over 5.3 billion by 2025. In the United States, smartphone ownership is particularly high among young adults, with 97% of individuals aged 18 to 29 owning a smartphone. This unprecedented level of penetration makes smartphones the most accessible and cost-effective computational and sensing devices available, far outpacing traditional lab equipment in terms of affordability and reach. While 97% of young adults in the United States own smartphones, the statistic presented here only highlight the potential reach and accessibility of these devices, not as a blanket justification for their educational use. The pedagogical integration of smartphones must be intentional and evidence-based, with careful attention to avoiding excessive screen time and ensuring that smartphone use supports active, inquiry-driven learning rather than passive digital engagement[34, 35].

### 2.2 Smartphone hardware and software capabilities

Modern smartphones are engineering marvels that incorporate a wide range of physics concepts across classical and modern domains. Understanding the foundational physics behind smartphone functions not only enriches students' appreciation of everyday technologies but also creates meaningful opportunities for physics education. These include mechanical motion, electromagnetic interactions, optical effects, quantum mechanics, and thermodynamics, as shown in **Table 1**. These embedded physical principles make smartphones highly versatile experimental tools for teaching physics, as they embody real-world applications of abstract theories.

**Table 1** Key physics principles underpinning smartphone components.

| Physics Domain | Principles/Concepts | Smartphone Components |
|---|---|---|
| **Mechanics** | Kinematics and dynamics | Accelerometer, gyroscope (MEMS devices), haptics |
| **Electromagnetism** | EM wave propagation, antennas, signal transmission | WiFi, Bluetooth, cellular antennas, inductive charging |
| **Electronics** | Diodes, transistors, resistors, capacitors, inductors, circuit theory | Processors, sensors, battery charging, touch screens |
| **Optics/ Photonics** | Refraction, diffraction, polarization, interference, image formation | Camera, flashlight, display, ambient light sensors |
| **Magnetism** | Magnetic fields, Hall effect, induction | Magnetometer, compass, Hall sensors, data storage |
| **Quantum Mechanics** | Band theory, tunneling, photoelectric effect | Semiconductor devices, imaging sensors, flash memory |
| **Thermodynamics** | Heat transfer, energy dissipation | Device cooling, thermal sensors |

**Table 2** Nobel Prize-winning contributions to smartphone technology.

| Year | Laureate(s) | Discovery/Contribution | Smartphone Application |
|---|---|---|---|
| **1921** | A. Einstein | Photoelectric effect | CMOS/CCD camera sensors |
| **1956** | Shockley, Bardeen, Brattain | Transistor effect | Integrated circuits, processors |
| **1964** | C. Townes, N. Basov | Laser development | Optical components, facial recognition, LiDAR |
| **1973** | L. Esaki, I. Giaever | Quantum tunneling in semiconductors | Flash memory, solid-state storage |
| **1989** | N. F. Ramsey, H. G. Dehmelt, W. Paul | Ion trapping and atomic time standards | Precision timing in GPS and wireless communication |
| **2000** | J. S. Kilby | Invention of the integrated circuit | System-on-a-chip design in all smartphones |
| **2007** | A. Fert, P. Grünberg | Giant magnetoresistance (GMR) | High-density magnetic storage (e.g., sensors, memory) |
| **2009** | C. Kao | Fiber optic transmission | High-speed internet access and smartphone data transfer |
| **2012** | S. Haroche, D. J. Wineland | Quantum systems control and measurement | Foundation for quantum sensing; underpins future secure communications and sensing in mobile platforms |
| **2014** | I. Akasaki, H. Amano, S. Nakamura | Blue LED technology | Energy-efficient backlighting and OLED/LED smartphone displays |
| **2024** | J. Hopfield, G. Hinton | Foundational work in neural networks and deep learning | AI-based smartphone functions: facial recognition, language translation, voice assistants, and image enhancement |

Several components within smartphones owe their existence directly to fundamental scientific discoveries that were recognized with Nobel Prizes (**Table 2**). These include the physics of semiconductors, electromagnetic communication, optical sensors, and quantum phenomena. These discoveries collectively support almost every functional element of the smartphone—from processing and imaging to sensing and communication. Highlighting these connections may help students appreciate how fundamental physics research underpins technologies with broad societal impact.

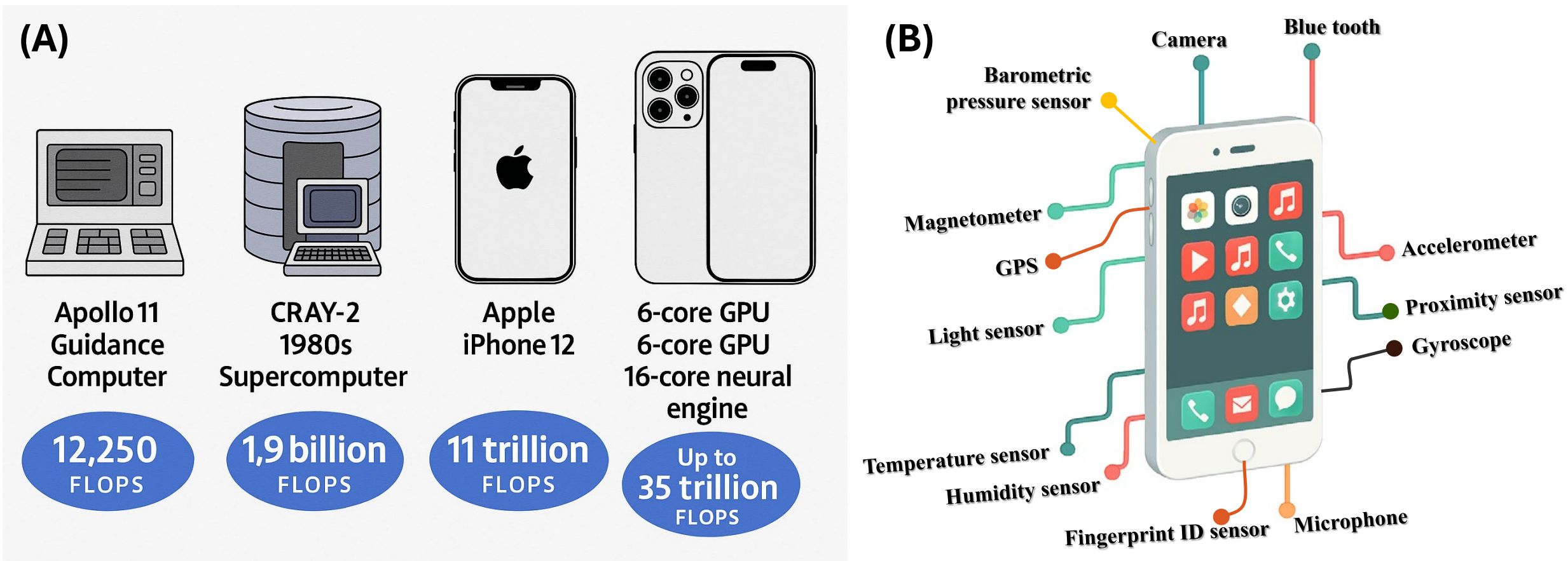

**Figure 1** (A) Comparison of computational power across four iconic systems: the Apollo 11 Guidance Computer, the CRAY-2 supercomputer, the Apple iPhone 12, and iPhone 16. (B) Sensors and other related functions of a smartphone.

Over the past two decades, smartphones have undergone a dramatic transformation, from basic communication devices into powerful, pocket-sized computing platforms with extensive access to digital content. Today's smartphones are equipped with multi-core CPUs, high-performance GPUs, and advanced neural processing engines, enabling real-time AI computation, graphics rendering, and data processing tasks once limited to desktop systems or specialized supercomputers. These devices now rival, and in some cases exceed, the capabilities of computers used in landmark scientific achievements. A striking example of this evolution is illustrated in the cartoon comparison (**Figure 1A**), which contrasts the computing power of several iconic machines across history. The Apollo 11 Guidance Computer, which helped land humans on the Moon in 1969, performed only about 12,250 floating-point operations per second (FLOPS).[36] The CRAY-2 supercomputer, one of the most powerful systems of the 1980s, achieved 1.9 billion FLOPS—roughly 155,000 times more powerful than the Apollo computer.[37] Fast forward to today, and the iPhone 16 Pro, powered by Apple's A17 Pro chip, is capable of delivering up to 35 trillion FLOPS.[38] It features a 6-core CPU, 6-core GPU, and a 16-core neural engine, enabling computational performance that is approximately 2,900 times faster than the CRAY-2 and over 2.8 billion times faster than Apollo's system. Samsung's Galaxy S25, powered by the Snapdragon 8 Elite chip, similarly pushes the boundaries of mobile computing with a custom 8-core CPU, enhanced GPU, and dedicated Hexagon NPU for high-efficiency on-device AI processing. These capabilities enable not just seamless multitasking and immersive media, but also real-time scientific data analysis, machine learning inference, and augmented reality—all within a single handheld device. As smartphones increase in performance, they are becoming increasingly useful platforms for scientific exploration, including portable, sensor-rich laboratories for education.

In parallel, smartphones provide unparalleled access to digital learning resources via cellular and Wi-Fi connectivity. Students can instantly retrieve lab manuals, view instructional videos, collaborate on cloud platforms, and access open-source textbooks, all within a single, portable device. This fusion of processing power and connectivity makes smartphones ideally suited for integrated, inquiry-based learning in both classroom and remote settings.

**Table 3** Key sensors in smartphones and their physics applications.

| Sensor | Physical Quantity Measured | Common Uses in Physics Labs |
|---|---|---|
| **Accelerometer** | Linear acceleration ($m/s^2$) | Kinematics, free fall, harmonic motion, inclined planes |
| **Gyroscope** | Angular velocity (rad/s) | Rotational motion, circular motion, pendulums |
| **Magnetometer** | Magnetic field strength (μT) | Mapping magnetic fields, studying magnetic dipoles |
| **Microphone** | Sound pressure waves (Hz, dB) | Measuring beats, resonance frequencies, Doppler shift |
| **Ambient Light Sensor** | Light intensity (lux) | Transmission through filters, reflection/refraction |
| **Proximity Sensor** | Distance (typically <5 cm) | Triggering near-object detection, optical response timing |
| **Camera** | Visual data (images, video) | Video-based motion tracking, wave interference, diffraction |
| **Barometer** | Atmospheric pressure (Pa or hPa) | Altitude estimation, convection experiments |
| **GPS** | Geolocation, velocity, time | Motion tracking, velocity estimation, large-scale position experiments |
| **Temperature Sensor** | Ambient/device temperature | Newton's law of cooling, thermal gradients |
| **Humidity Sensor** (less common) | Relative humidity (%) | Environmental monitoring in thermodynamics experiments |
| **Touchscreen** | Contact location and force | Interactive simulations, electric field visualization |
| **Infrared Sensor** (on select devices) | Thermal radiation | Thermal imaging and IR-based heat transfer experiments |

Beyond computation and content access, smartphones are also powerful experimental tools, equipped with a suite of built-in sensors (**Figure 1B**) capable of measuring a wide range of physical quantities, as summarized in **Table 3**. These sensors, originally designed for enhancing user interaction and device functionality, have been successfully repurposed to support hands-on physics investigations. Coupled with a growing ecosystem of data acquisition and analysis apps, smartphones offer a low-cost, portable, and highly accessible alternative to traditional laboratory instrumentation.

To harness these sensors for educational use, a growing ecosystem of free or low-cost apps (**Table 4**) has emerged, offering real-time graphing, signal processing, multimodal data collection, and export functionalities. Apps like *Phyphox*, *SPARKvue*, and *Physics Toolbox Suite* convert smartphones into flexible scientific instruments. These tools enable students to design integrated experiments; for example, recording acceleration data using the accelerometer[39-43] while simultaneously capturing video for trajectory analysis, or combining sound and motion data in a Doppler[44, 45] or harmonic oscillator study[46, 47]. Such versatility allows smartphones to function as low-cost, scalable, and accessible alternatives to traditional laboratory instrumentation, reshaping how experimental physics can be taught and experienced in the 21st century.

Throughout this review, we distinguish between smartphone-native apps, designed specifically for data collection, sensor integration, or image analysis on mobile devices, and general-purpose software tools that are often desktop-based or commercial in nature. While tools like MATLAB or GeoGebra have been used in smartphone-related workflows, their inclusion here is primarily to illustrate hybrid models or analytical extensions. Their limitations in accessibility, licensing, and device compatibility are acknowledged and discussed in **Section 5**.

**Table 4** A summary of commonly used Apps for SmartIPLs.

| App Name | Functionality | Physics Topics |
|---|---|---|
| **Phyphox** | Access multiple sensors, real-time graphing, CSV export | Mechanics, optics, magnetism, sound |
| **Physics Toolbox Sensor Suite** | Multifunctional sensor interface with logging capabilities | All domains (mechanics, EM, optics, etc.) |
| **SPARKvue (Vernier)** | Sensor integration, graphing, student lab templates | Mechanics, thermodynamics, pressure |
| **Tracker** | Video analysis software for motion tracking | Projectile motion, oscillations |
| **Audacity / Audio Kit** | Sound recording and FFT-based analysis | Acoustics, resonance, Doppler effect |
| **ImageJ / Avidemux** | Image analysis, spatial measurements | Optics, diffraction, thermal imaging |
| **Keuwlsoft Apps** | Suite including magnetometer, accelerometer, oscilloscope | EM, circuits, waves |
| **SignalScope** | Oscilloscope and waveform generator (external probe compatible) | Electronics and circuits |
| **Flir Tools / ThermoCam Viewer** | Visualizing thermal distributions (paired with FLIR-compatible cameras) | Thermodynamics, convection, blackbody |

## 2.3 From M-learning to SmartIPLs

Smartphones have played a transformative role in the development of mobile learning (m-learning), which is broadly defined as "learning across multiple contexts, through social and content interactions, using personal electronic devices."[48-52] This modality has become a cornerstone of modern online and hybrid education, especially as institutions seek to accommodate increasingly mobile and diverse student populations. Smartphones and tablets, in particular, have emerged as indispensable tools for delivering course content, managing schedules, receiving real-time notifications about assignments and classroom logistics, and facilitating communication between students and instructors. For part-time students or working professionals, m-learning provides the flexibility to engage with academic material without the need to attend traditional, in-person lectures or training sessions. It also enables the formation of collaborative virtual communities that transcend physical classrooms.

Recent trends highlight the growing legitimacy of m-learning. According to one survey, 77% of academic leaders believe that online learning is either equivalent to or superior to face-to-face instruction.[53] Many m-learning platforms have evolved to include interactive features such as multimedia content and virtual reality environments, thereby enhancing engagement and user experience. Nevertheless, these platforms largely remain extensions of traditional distance learning, offering improved interfaces but still lacking a critical element for STEM education: hands-on experimentation. While effective for content delivery and communication, conventional m-learning typically omits the tactile, investigative component that is vital for cultivating scientific reasoning, process skills, and enthusiasm in STEM fields.

Unlike standard m-learning environments, SmartIPLs leverage the full technical potential of smartphones—not just for accessing information but for performing real-world scientific experiments. Equipped with a variety of built-in sensors, smartphones can be transformed into portable laboratory instruments. Students can collect motion data using accelerometers and gyroscopes, record and analyze sound waves, measure temperature changes using thermal cameras, study optics through image and video analysis, and investigate environmental quantities such as magnetic fields, pressure, and light intensity. By supporting sensor-driven, inquiry-based experimentation, SmartIPLs offer opportunities to connect digital learning environments with direct engagement in physical science. They can offer an experiential layer that traditional m-learning lacks, allowing students to engage directly with physical phenomena in a manner that fosters deeper conceptual understanding and support the goals of modern STEM education.

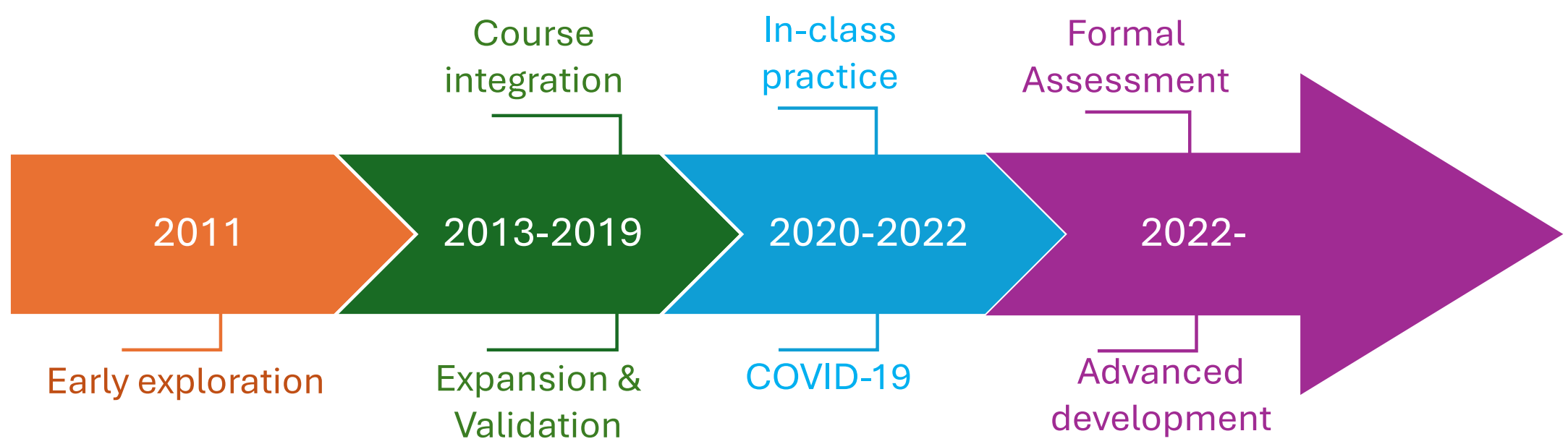

**Figure 2** SmartIPL development timeline.

### 2.5 Timeline of SmartIPL development

The development of SmartIPLs began around 2011 as shown in **Figure 2**, when early adopters demonstrated how the built-in sensors of smartphones could be used to replicate classical physics experiments. One of the earliest examples was a Doppler effect-based measurement of gravitational acceleration using a smartphone's microphone, conducted by Vogt, Kuhn, and Müller, who published their work in *The Physics Teacher*.[45] Shortly thereafter, the same group began systematically exploring the use of accelerometers for analyzing free fall[39], pendulum motion, and oscillatory systems[54]. These pioneering efforts established that smartphones could serve as affordable and accessible tools for physics experiments, particularly in kinematics and basic mechanics.

Between 2013 and 2019, the SmartIPLs grew rapidly, expanding beyond basic kinematics to encompass a wide range of physical phenomena. As smartphones began incorporating more sophisticated sensors, such as gyroscopes, magnetometers, barometers, and light sensors, researchers leveraged these capabilities to explore topics like rotational motion, circular dynamics, magnetic field mapping, and optical effects. For instance, Kuhn and Vogt demonstrated how to use smartphones to investigate radial acceleration,[55] and Sans et al. explored spring-mass oscillations using light sensors.[56] During this period, researchers increasingly validated smartphone-generated data by comparing it to results from conventional laboratory equipment, confirming that smartphones could produce reliable and accurate measurements. Notably, Arribas et al. showed that smartphone magnetometers could effectively verify the inverse cube law of magnetic field decay.[57] Meanwhile, a surge in low-cost and DIY (Do It Yourself) innovations highlighted the real-world applicability of smartphones: Monteiro et al. utilized smartphone barometers to measure vertical velocity in elevators and drones, bridging the gap between classroom theory and practical environments.[58] In the meantime, smartphone-based labs began transitioning from experimental novelties to educational tools. Many studies focused on aligning smartphone-based experiments with standard learning objectives in introductory physics courses. Concurrently, smartphone-based labs began transitioning from experimental tools to structured educational resources. Many studies focused on aligning these experiments with standard learning goals in introductory physics curricula. For instance, at North Carolina State University, Countryman developed the MyTech project, where students used built-in smartphone accelerometers and gyroscopes to collect data on momentum, impulse, and Newton's laws.[59] At the University of Georgia, Zhao integrated smartphone-based optical experiments into an undergraduate Modern Optics course, enabling students to explore interference, diffraction, polarization, and spectroscopy.[23] As evidence of their educational value grew, smartphone labs began to appear not only in university settings, but also in high schools, online courses, and remote learning environments, indicating their development into accessible and scalable tools for supporting hands-on physics education.

The COVID-19 pandemic in 2020 created an unprecedented disruption to laboratory-based physics instruction and accelerated the adoption of smartphones as practical tools for experimental learning. When traditional labs became inaccessible, educators turned to smartphone-based activities delivered through home kits, live demonstrations, and asynchronous modules. Students and teachers adapted by using widely available apps such as *Phyphox*, *Tracker*, and *Physics Toolbox* to continue data collection and analysis outside of conventional settings. Onorato et al. demonstrated distance-learning labs using smartphones to study blackbody radiation[60] and diffraction[61], while at the University of Georgia, a low-cost hybrid Modern Optics Lab was implemented using < $5 kits and student smartphones.[23] Five core labs, on data analysis, reflection, polarization, interference, and diffraction, were conducted remotely, with students designing and executing experiments at home using everyday materials and receiving guidance via Zoom. Despite diverse setups, results were consistently reliable, especially for polarization and diffraction.

By 2022, the field of SmartIPLs had reached a new level of maturity, marked by the publication of two comprehensive books, one cataloging approximately 75 SmartIPLs across various physics domains,[62] and the other focused specifically on optics.[23] Since then, the scope of applications has expanded into more advanced and interdisciplinary territory. Smartphones are now used to explore phenomena such as Planck's law, [60] Raman spectroscopy,[63] and even $\beta-$ decay detection[64] via camera sensor. Emerging technologies like augmented reality (AR) and real-time digital signal processing (DSP) have further enhanced interactivity and analytical depth—for example, AR tools are now used to visualize electric and magnetic fields[65], while Pirinen et al. demonstrated the integration of smartphone accelerometer data with Jupyter Notebooks to teach DSP concepts, bridging physics and computer science.[66] The current trajectory emphasizes hybrid learning environments that combine smartphone-based experimentation with cloud platforms and collaborative technologies, fostering a more connected and interdisciplinary approach to physics education.

**Table 5** Overview of SmartIPLs across major physics domains

| Domain | Key Topics | Number of Publications | Apps & Tools |
|---|---|---|---|
| Classical Mechanics | Kinematics, Projectile Motion, Harmonic Motion, Friction | 76 | Phyphox, SPARKvue |
| Waves & Acoustics | Doppler, Beats, Resonance, Fourier Analysis | 29 | Oscilloscope, Audio Kit |
| Thermal Physics | Newton's Cooling, Convection, Thermal Imaging | 25 | FLIR Tools, VidAnalysis |
| Fluids | Pressure, Surface Tension, Viscosity | 12 | Tracker*, AndroSensor |
| Electromagnetism | Magnetic Fields, Induction | 16 | Magnetometer Apps |
| Electronics | Signal Gen, RC Circuits | 9 | SignalScope, Oscium |
| Optics | Light, Refraction, Diffraction, Polarization | 32 | ImageJ, RGB Color Assist |
| Modern Physics | Beta Decay, Blackbody Radiation | 3 | RadioactivityCounter |
| Astronomy/Geo | ISS Tracking, Planetary Transits | 6 | MATLAB Mobile*, GeoGebra* |

* These platforms are not smartphone-native and may require desktop access and/or licenses.

## 3. Taxonomy of Smartphone-Based Labs

Between 2011 and 2024, approximately 208 studies (excluding books) have reported on the development and implementation of SmartIPLs. **Table 5** summarizes these labs by physics domain, highlighting key topics, the number of publications, and commonly used smartphone apps. The wide distribution of topics illustrates the extensive integration of smartphones across nearly all areas of introductory physics, utilizing both sensor-based and image/video-based experimental approaches. **Table 5** shows that the most developed area by far is Classical Mechanics, with 76 documented studies. Optics, Waves & Acoustics, and thermal physics follow, with 32, 29, and 25 studies respectively. In contrast, several underrepresented domains reveal clear opportunities for further SmartIPL development. Modern Physics, with just 3, is particularly sparse. Fluids, Electromagnetism, and Electronics also show relatively low coverage (around 9–16 studies each), suggesting a need for more diverse experimental designs and app-based tools in these areas. Meanwhile, Astronomy and Geophysics remain niche but promising, with 6 studies leveraging GPS-based apps and satellite tracking tools.

The following subsections summarize detailed SmartIPLs by physics topic.

*Classical mechanics.* SmartIPLs in classical mechanics span 76 experiments as detailed in **Table 6**, covering kinematics, projectile motion, friction, circular motion, and oscillations. Students use smartphone sensors—accelerometers, gyroscopes, magnetometers, and light sensors—to measure parameters like acceleration, velocity, and angular velocity. Video analysis is also common for free fall, projectile motion, and pendulums. Widely used apps include *Phyphox*, *SPARKvue*, *Tracker*, and *AndroSensor*. While these labs reinforce core concepts, additional experiments on work, energy, momentum, and gravitation are recommended for comprehensive coverage.

*Waves and acoustics.* **Table 7** summarizes 29 SmartIPLs focused on acoustics, covering topics such as sound wave analysis, resonance, beats, standing waves, and Fourier analysis. Students use smartphone microphones to measure frequency, amplitude, and interference patterns. Some experiments involve two smartphones—one as a sound source, the other as a detector. Apps like *Audio Kit*, *Oscilloscope*, *Phyphox*, and *Tracker* support signal generation and spectral analysis. Expanding the curriculum to include reflection, diffraction, and absorption would further enrich learning outcomes.

*Thermal physics.* Over 20 smartphone-integrated thermal physics labs have been reported (**Table 8**) demonstrating wide-ranging applications of smartphone-based thermal cameras and sensors. Key topics include Newton's law of cooling, radiative heat loss, emissivity, phase changes, heat conduction, convection phenomena (e.g., Bénard–Marangoni instability), and real-world thermal analysis such as building insulation and energy dissipation in braking systems. Smartphone-connected thermal cameras (e.g., *FLIR ONE*, *Seek Thermal*) and apps like *Phyphox*, *VidAnalysis*, *Framelapse*, and *FLIR Tools* enable students to visualize and quantify temperature distributions and cooling curves through images, videos, and time-series data. Some studies integrate Arduino boards or custom apps for enhanced control and analysis.[67] Despite this diversity, additional SmartIPLs are still needed to cover underexplored areas such as thermal expansion, specific heat capacity, and systematic investigations of heat transfer modes (conduction, convection, radiation) in everyday and industrial contexts.

*Fluidics.* Twelve SmartIPLs (**Table 9**) explore atmospheric pressure, fluid oscillations, drainage, and surface tension. Smartphone barometers, gyroscopes, and cameras are used alongside apps like *Phyphox*, *AndroSensor*, and *Tracker*. Image analysis with *ImageJ* or MATLAB supports studies of droplet dynamics and viscosity. Further development could target laminar/turbulent flow, buoyancy, and capillarity.

*Electromagnetism.* **Table 10** lists around 16 SmartIPLs focused on magnetism and a few on electricity. Labs investigate Biot–Savart law, magnetic field decay, Faraday's law, and RC circuits. Magnetometers and microphones are used with apps like *Phyphox* and *Physics Toolbox Suite*. AR and sensor fusion enhance understanding of vector fields. More experiments are needed on electric fields, Gauss's law, and circuit analysis.

*Electronics.* Nine SmartIPLs are outlined in **Table 11**. Smartphones are used as signal generators and oscilloscopes for studying RC, RLC, and digital circuits. Apps like *SignalScope*, *Signal Generator*, and *Oscium* support analysis. Some labs explore DSP and AR for logic circuits. More coverage is needed on basic electronics, circuits, semiconductor devices, and amplifiers.

*Optics.* Thirty-two labs (**Table 12**) explore intensity, polarization, diffraction, and lens optics. Smartphones function as light meters, cameras, and spectrometers. Apps include *Physics Toolbox Suite*, *Tracker*, *ImageJ*, and *RGB ColorAssist*. Labs demonstrate Malus' law, Brewster's angle, Beer's law, and interference. The coverage spans most of an undergraduate optics course. Other optic labs and applications can be found in Ref. [23].

*Modern physics.* Only three labs are reported to date (**Table 13**). Topics include β− radiation detection via smartphone cameras, blackbody radiation using heated filaments, and SPR imaging. Apps like *RadioactivityCounter* and *Phyphox* are used. This area is still emerging but shows potential for integration with quantum and nuclear physics topics. In parallel, interactive simulations such as PhET,[68] Visual Quantum Mechanics,[69] and Falstad applets,[70] now accessible on mobile browsers, offer intuitive, visual models of quantum phenomena like tunneling, wavefunctions, and energy quantization. Though not smartphone-native, these tools complement hands-on SmartIPLs, especially in remote or low-resource settings. One potential direction for SmartIPLs in modern physics is to integrate real-time sensor data with interactive simulations, providing links between theoretical concepts and both experimental and virtual activities.

*Astronomy and geophysics.* Six SmartIPLs (**Table 14**) address ISS tracking, planetary transits, sunlight variations, and magnetic field surveys. Cameras, light sensors, and magnetometers are used with apps like *Phyphox*, *PocketLabs*, and MATLAB Mobile. *GeoGebra* helps analyze lunar surface features from images.

**Table 6** A list of SmartIPLs for classic mechanics

| Physics Topics | Role of Smartphone | Brief description | Data Analysis | App Name | Ref. |
|---|---|---|---|---|---|
| Kinematics 1D constant motion | Timer Accelerometer | Record the time taking for a student to walk at a constant speed between chairs with known positions | Position *vs*. time Acceleration *vs*. time | iOS: Best Stopwatch, SPARKvue Android: Chronometer | 71 |
| | Magnetometer | Measure and analyze the constant speed of a dynamic car by detecting the magnetic field of magnets placed along a linear track | Magnetic field peak *vs*. time | Physics Toolbox Suite | 72 |
| | Sound generator Microphone | Determine vehicle speed by analyzing the frequency of sound generated by a smartphone via Doppler effect | Velocity vs. time | Audacity | 44 |
| | Display | Use a smartphone paired with an Arduino-based setup via Bluetooth to measure the uniform linear motion of a vehicle | Position *vs*. time Velocity *vs*. time | MIT AppInventor | 73 |
| Kinematics 1D accelerated motion | Video | Analyze the free fall position versus time in a recorded video | Position *vs*. time | Fast Burst Camera Lite | 74 |
| | Sound generator Microphone | Drop a smartphone emitting a constant frequency sound to measure the Doppler shift during free fall, to calculate the acceleration of gravity | Doppler shift *vs* fall time | Audacity, Test Tone Generator, SPEAR | 45 |
| | Accelerometer | Record the acceleration over time as a smartphone falls toward a soft object | Acceleration *vs*. time | iOS: SPARKvue Android: Accelogger | 39-43 |
| | Microphone Timer | Record the height and impact time of a bolt nut falls on a floor | Falling height *vs*. impact time | Android: Smart Voice Recorder | 40, 75 |
| | | Record the time of a free fall of a metal ball from a fixed height | Falling height *vs*. impact time | Voice memo Audio Time | 76 |
| | | Record the relationship between the falling height versus falling time of a free fall of a metal ball use the sound sensor of the smartphone as a stopwatch | Falling height *vs*. falling time | Phyphox | 77 |
| | | Use photoresistor modified headphone to measure the time taken for an object to pass through a specified distance. | Distance *vs*. time | AudioTime | 78 |
| | Light sensor Video | Determine the acceleration and speed of a light-emitting object on an inclined plane using a smartphone's light sensor; Video analysis of the motion. | Illuminance *vs*. distance Illuminance *vs*. time Distance *vs*. time | AndroSensor | 79 |
| | Barometer | Measure the vertical velocities and accelerations of elevators, pedestrians climbing stairs, and drones by means of smartphone barometer | Pressure *vs*. time Distance *vs*. time | Physics Toolbox AndroSensor | 58 |
| | Magnetometer | Record the magnetic field of a bar magnet during its free fall from a fixed height | Magnetic field *vs*. time Distance *vs*. time | Phyphox | 80 |
| | | Record the time of peak magnetic fields during the free fall of a magnetic ruler | Magnetic field *vs*. time Distance *vs*. time | Phyphox | 81 |
| | Sound generator | Record the sound produced by a smartphone during its free fall and analyzed the frequency shift versus time. | Frequency vs. time<br>Velocity vs. time | Audacity<br>SPEAR | 45, 82 |
| | GPS<br>Accelerometer | Analyze position and velocity data using A-GPS and an accelerometer, and calculated and compare normal and tangential acceleration components | 3D Position *vs*. time | Custom-built | 83 |
| Projectile | Video | Analyze the two components of positions versus time of projectile object via video | Positions *vs*. time | MeasureDynamics | 84 |
| | | | | Fast Burst Camera Lite | 74 |
| | Sound sensor | Measure the time taken for a ballistic ball took off from a rail on a table | Height *vs*. time | Phyphox | 85 |
| | | Record the time between two consecutive bounces of a ball using a sound sensor, with the initial height of the ball predetermined | Coefficient of restitution | iOS: Oscilloscope<br>Audacity | 40, 44, 86 |
| Friction | Accelerometer | Measure the accelerations of smartphone moving in an inclined plane with different tilting angle and determined *g*. | Acceleration | iOS: SPARKvue<br>Android: Accelogger | 43 |
| | | Measure *g* and determined the coefficient of kinetic friction as the iPhone slid down an inclined plane. | | iOS: SPARKvue<br>Clinometer | 87, 88 |
| | | Measure the acceleration of an object subjected to a constant force on various flat surfaces and on an inclined surface to determine the coefficient of kinetic friction. | Acceleration *vs*. time | Physics Toolbox Sensor Suite | 89 |
| | | Determine the drag resistance coefficients of different vehicles | Acceleration *vs*. time<br>Acceleration *vs*. speed | Accelerometer Data Pro | 90 |
| | Sound generator<br>Microphone | Use the Doppler effect to study one phone (emitting a fixed frequency) sliding down an inclined plane towards another phone at the bottom, and to determine the coefficient of kinetic friction. | Frequency *vs*. time<br>Speed *vs*. time<br>Friction coefficient *vs*. inclined angle | Phyphox<br>Science Journal | 91 |
| | Angle meter | Measure the critical angle to start sliding a smartphone on an inclined surface and determined the maximum coefficient of static friction | Critical angle | Physics Toolbox Sensor Suite | 92, 93 |
| Radial acceleration | Accelerometer | Measure radial acceleration in a controlled lab environment and on a playground merry-go-round | Radial acceleration *vs* speed & radius | iOS: SPARKvue<br>Android: Accelogger | 43, 55 |
| | | Measure the radial acceleration and the damping process due to friction on a spinning disc | Radial acceleration *vs* tine & radius | iOS: SPARKvue | 94 |
| | Accelerometer<br>Gyroscope | Explore the stable and unstable rotational dynamics of a smartphone when tossed about its principal axes, and analyze the rotational stability of different axes | Angular velocities and acceleration components | SPARKvue | 95 |
| | | Measure Coriolis acceleration using a smartphone sliding on a rotating track, verifying the dependence of Coriolis acceleration on the track's angular velocity and the smartphone's sliding speed | Coriolis acceleration, angular velocity | Data Collection | 96 |
| Angular Velocity and Acceleration | Accelerometer Gyroscope | Measure the acceleration and angular velocity of a physical pendulum by attaching a smartphone to a rotating bicycle wheel, and analyze its motion and phase space trajectories | Radial & tangential acceleration, angular velocity & displacement | AndroSensor Tracker | 97, 98 |
| | | Measure the angular velocity of a slamming door experiment to investigate different friction models | Angular velocity *vs* time | Phyphox | 99 |
| | | Measure and verify the relationship between angular velocity and centripetal acceleration during the rotation of a merry-go-round at various distances from the center | Centripetal acceleration vs angular velocity | | 100 |
| | Gyroscope | Study the motion of a rolling cylinder on a slope | Angular velocity *vs* time | Physics Toolbox Gyroscope | 101 |
| | | Study the rotational motion of a low friction wheel and three rotation chairs, and model the effects of frictional torque | Angular velocity *vs* time, frictional torque | AndroSensor | 102 |
| | | Toss a smartphone in the air to investigate the stability of its rotational dynamics about different axes and the conservation of angular momentum and rotational kinetic energy | Angular velocity *vs* time | Phyphox | 103 |
| | Accelerometer | Investigate the mechanics of preventing spills by measuring the acceleration components experienced by a liquid container on a SpillNot device during oscillatory motion | Radial & tangential acceleration | AndroSensor | 104 |
| | Magnetic field sensor | Measure the average angular velocity of a slow-spinning grill by using a smartphone's magnetic field sensor | Angular velocity *vs* time | Physics Toolbox Sensor | 105 |
| Torque | Accelerometer | Measure the oscillation period of a balance pan with varying masses to determine the inertial mass | Oscillation period *vs* mass | Physics Toolbox Suite Tracker | 106 |
| | | Investigate the rotational dynamics and frictional forces in the slamming of a door using a smartphone | Angular velocity, friction | SPARKvue | 107 |
| | | Investigate the dynamics of a variable mass Atwood's machine | Acceleration & velocity *vs* time | Vernier Graphical Analysis | 108 |
| | | Measure the period of oscillations of a rod with an attached mass at different distances to demonstrate the parallel axis theorem | Oscillation period *vs* distance | Physics Toolbox Suite | 109 |
| | Video | Investigate the friction affecting the motion of a rod and a bicycle wheel by analyzing the amplitude damping of their oscillatory movements | Oscillation amplitude *vs* time | Tracker | 46 |
| Rolling | Gyroscope | Investigate the rolling motion of a hollow cylinder on an inclined plane to determine the coefficients of static and kinetic friction | Angular speed & acceleration *vs* time | Sensorlog | 110 |
| | | Study the motion of a hallow cylinder on an inclined plane | Angular velocity *vs* time | Physics Toolbox Gyroscope | 101 |
| | | Study the rolling motion of a rolling cylinder on a slope | Angular velocity *vs* time | | 111 |
| | Gravity sensor<br>Gyroscope | Analyze the rolling motion of a cylindrical object down an inclined plane | Angular position *vs* time | Rolling Cylinder (custom-built) | 112 |
| Spring-Mass Oscillation | Accelerometer | Investigate free and damped harmonic oscillations and extract spring constant and damping constant | Acceleration *vs* time | Accelerometer Monitor | 113 |
| | | Analyze the oscillatory motion of a spring pendulum by using a smartphone's accelerometer to measure acceleration, allowing for the determination of key parameters such as the spring constant and the period of oscillation | Acceleration *vs* time | iOS: SPARKvue<br>Android: Accelogger | 54 |
| | Light sensor | Analyze the oscillatory motion of a system with two coupled springs using a smartphone's light sensor | Light intensity *vs* time | Physics Toolbox Light Sensor | 56 |
| | | Measure the period of oscillation of a spring-mass system to calculate the spring constant | Oscillation period | Physics Toolbox Sensor Suite | 114 |
| | Light sensor<br>Accelerometer | Study the damped harmonic motion of a mass-spring system | Light intensity *vs* time | Phyphox | 67 |
| | Accelerometer | Measure the oscillatory motion of a spring system | Acceleration *vs* time | iOS: Vibsensor<br>Android: Accelerometer Monitor | 115 |
| | | Measure an unknown mass by attaching it to a smartphone suspended from a rubber band and use the period of oscillation to determine the mass | Acceleration *vs* time | Phyphox | 116 |
| Pendulum | Accelerometer | Use the accelerometer of the smartphone to determine the oscillation period of a pendulum | Acceleration *vs* time | iOS: SPARKvue<br>Android: Accelogger | 43 |
| | | Analyze acceleration in free and damped harmonic oscillations and pendulum systems | Acceleration *vs* time | iOS: SPARKvue<br>Android: Accelogger | 117 |
| | | Measure and model the motion of a pendulum | Angle *vs* time | InduLab | 118 |
| | | Study the oscillatory motion of a pendulum | Acceleration *vs* time | MATLAB Mobile | 119 |
| | Video | Investigate the relationship between the mass of a projectile and the resulting displacement of a ballistic pendulum | Maximum vertical displacement *vs* mass | | 120 |
| | Accelerometer<br>Gyroscope | Investigate the frequency doubling effect in a pendulum by measuring and analyzing the angular velocity and acceleration | Angular acceleration *vs* time | Phyphox | 121 |
| | Proximity sensor | Measure the period of a pendulum's oscillation and calculate the gravitational acceleration | Oscillation period | Phyphox | 122 |
| | Gyroscope | Measure the rotational oscillation of a torsion pendulum to determine the shear modulus of the wire material | Angular frequency *vs* time | Phyphox | 123 |
| | Gyroscope<br>Magnetometer | Measure the angular velocity and displacement of a gravity pendulum to construct a phase plot that illustrates the pendulum's dynamic behavior | Angular velocity & displacement *vs* time | Phyphox | 124 |
| | Accelerometer | Study the motion of a pendulum to analyze its nonlinear dynamics | Acceleration *vs* time | Phyphox | 125 |
| | | Analyze the forces and accelerations experienced by a person on a playground swing | Acceleration *vs* time | Phyphox | 126 |
| | | Study the oscillatory motion of a pendulum | Acceleration *vs* time | SPARKvue<br>Vernier Graphical Analysis | 42 |
| Other Oscillation | Accelerometer | Measure and analyze the oscillatory motion of an elevator to explore the relationship between oscillation period and cable length | Oscillation period | iOS: SPARKvue<br>Android: Accelogger | 127 |
| | | Study two-dimensional harmonic oscillations to generate and analyze Lissajous curves resulting from the superposition of harmonic motions in perpendicular directions | Acceleration components *vs* time | Accelerometer Toy | 128 |
| | Video | Investigate the oscillatory behavior of fluid inside a straw after the fluid is suddenly released | Fluid level *vs* time | Tracker | 47 |
| Coupled Oscillation | Accelerometer | Analyze coupled oscillations quantitatively in a system of two gliders connected by springs, and explore different modes of oscillation | Acceleration *vs* time | Accelerometer Monitor | 129 |
| | Video | Use an asymmetrical double torsion pendulum to study coupled harmonic motion | Angular displacement *vs* time | Tracker<br>PASCO Capstone | 130 |
| Energy and Work | Gyroscope | Measure and analyze the rotational energy of a physical pendulum | Angular velocity *vs* time | AndroSensor | 131 |
| | Image | Use a smartphone equipped with a thermal camera to visually demonstrate energy transfer during a physical activity | Temperature | | 132 |
| | Accelerometer | Measure the centripetal acceleration of a pendulum to verify the conservation of mechanical energy | Acceleration *vs* time | Physics Toolbox Accelerometer | 133 |
| | Gyroscope<br>Accelerometer | Study the motion of a torsion pendulum to analyze the relationship between kinetic and potential energy during the oscillation | Angular displacement, velocity, and acceleration *vs* time | SensorLog | 134 |

**Table 7** A list of SmartIPLs for waves and acoustics

| Physics Topics | Role of Smartphone | Brief description | Data Analysis | App Name | Ref. |
|---|---|---|---|---|---|
| Sound generation and analysis | Microphone Display and FFT analyzer | Capture and analyze different types of sound waves such as tones, sounds, noise, and bangs to help students understand the physical differences between these acoustic phenomena | Frequency Amplitude | iOS: Audio Kit | 135 |
| | | Capture and analyze the sound of knuckle cracking to understand the bioacoustic phenomenon of oscillating gas bubbles in the synovial fluid | Frequency | iOS: Oscilloscope | 136 |
| | | Measure and analyze the sound frequencies emitted by vibrating rods of different materials, shapes, and lengths, and compare the results with theoretical predictions | Frequency | Android: AudiA | 137 |
| | | Record and analyze the sounds produced by a piano, a quiet house, an airplane cabin, a brass tube, a glass bowl, a metal cup, a metal spoon, an open flame, and a beating heart | Frequency Intensity | iOS: Oxford SpectrumView Android: Advanced Spectrum Analyzer Pro | 138 |
| | | Measure and analyze the acoustic response of different classroom environments, especially reverberation time and speech intelligibility | Reverberation time and speech intelligibility | APM Tool Lite | 139 |
| | | Determine the Shepard scale illusion and explore the auditory illusion and the underlying acoustic properties | Frequency | Android: Shepard Illusion or Phyphox | 140 |
| | | Demonstrate multiple acoustic labs such as calibrating smartphones for sound pressure level measurements, mapping noise levels across a university campus, and measuring vibration levels to estimate engine RPM | Sound pressure level, reverberation time, and vibration frequency | NoiseCapture APM Tool Lite Phyphox | 141 |
| | | Observe and understand sound phenomena, such as sound produced by vibrations, the mechanical nature of sound waves, and the characteristics of sound like loudness and pitch | Sound intensity, frequency, and amplitude | Phypox Physics Toolbox Suite | 142 |
| Sound interaction | Sound generator Display and FFT analyzer | Use earbuds and smartphones to explore fundamental physics phenomena such as beat frequency, interference patterns, resonance conditions, and Doppler shifts | Frequency, beat & resonance frequency, interference pattern, and Doppler shift | iOS: Physics Toolbox Tone Generator, Audacity | 143 |
| | Microphone | Visualize sound directivity patterns by measuring the angular dependence of sound levels around a sound source to study loudspeaker directivity and microphone polar patterns | Sound levels (in dB) *vs* orientation angle | Polar Pattern Plotter (Custom designed) | 144 |
| | | Measure and analyze acoustic beats generated by the superposition of two sound waves with slightly different frequencies to illustrate wave interference and beat | Amplitude *vs* time/frequency | Android: Spectroid, AudiA Advanced Spectrum Analyser | 82, 145 |
| | Sound generator | Understand the working principles of tuning forks and explore concepts like acoustic interference and beat | Frequency and intensity | Frequency<br>Sound Generator | 146 |
| Standing Wave | Sound generator<br>Microphone | Replicate the classical Kundt's tube experiment, where students can observe standing waves in a column of air, measure the distance between nodes and antinodes, and calculate the speed of sound | Wavelength<br>Sound speed | iOS: Audio Kit, Oscilloscope<br>Android: Signal Generator & Soundbeam | 147 |
| | Microphone | Measure the harmonic resonances in metal rods to capture and analyze the sound produced by the rods when they are struck | Resonant frequency | Spectrum View Plus | 148 |
| | Sound generator<br>Microphone | Explore the relationship between the fundamental frequency and wavelength in resonance tubes by generating and measuring sound frequencies | Fundamental frequency | Android: TrueTone<br>Advanced Spectrum Analyzer PRO | 149 |
| | Sound generator<br>Microphone | Measure the sound pressure profiles of standing waves inside a tube closed at one end, and visualize and analyze the formation of nodes and antinodes within the tube | Sound pressure level *vs* position | Frequency<br>Oscope | 150 |
| Helmholtz resonator | Sound generator<br>Microphone | Measure the acoustic response of a Helmholtz resonator using a glass beaker filled with different gases, and determine the speed of sound in different gases | Resonant frequency | iOS: Spektroskop<br>Android: Advanced Spectrum Analyzer | 151 |
| | | Study the Helmholtz resonance phenomenon in a kitchen setting, specifically the change in pitch of the sizzling sound produced in a frying pan as the lid is manipulated | Resonant frequency | iOS: SpectrumView<br>NoiseGenerator | 152 |
| | Microphone<br>Display and FFT analyzer | Investigate the phenomenon of side window buffeting in a car, where the airflow over an open window generates low-frequency, high-amplitude sound waves from a Helmholtz resonator model | Buffeting frequency<br>Car speed | iOS: Spectrum View<br>Android: AndroSensor | 153 |
| | | Measure the resonant frequencies of a telescopic vacuum cleaner pipe at various lengths to calculate the speed of sound in air | Resonant frequency vs length of the pipe | Physics Toolbox, Phyphox, Spectroid or Advanced Spectrum | 154 |
| | | Measure the resonance frequencies of a tea bottle acting as a Helmholtz resonator, demonstrating the relationship between the volume of air in the bottle and the frequency of sound produced | Resonant frequency *vs* air volume | Advanced Spectrum Analyzer PRO | 155 |
| Speed of sound | Sound generator<br>Microphone | Measure the speed of sound in air by generating sound waves inside a pipe partially immersed in water | Wavelength *vs* position | Function generator<br>SpectrumView | 156 |
| | | Measure how long it takes for a sound signal to travel between two smartphones placed at a known distance apart | Time delay | Phyphox | 157 |
| | | Measure the speed of sound in air by using a smartphone to generate and record sound waves inside a cardboard tube | Resonant frequency | Function Generator<br>Smart Recorder | 158 |
| | | Generate sound within a closed tube and measure the resulting sound intensity at various points to identify the resonant frequencies | Sound intensity vs resonant position | Frequency Generator<br>iOS: Decibel X<br>Android: Sound Meter | 159 |
| | Microphone | Measure the resonant frequencies of sound waves generated inside a telescopic vacuum cleaner pipe at different pipe lengths | Resonant frequency | Physics Toolbox<br>Phyphox; Spectroid<br>Advanced Spectrum | 154 |
| Doppler effect | Sound generator<br>Microphone | Drop a smartphone emitting a constant frequency sound to measure the Doppler shift during free fall, to calculate the acceleration of gravity | Doppler shift *vs* fall time | Audacity, Test Tone Generator<br>SPEAR | 45 |
| | | Measure the Doppler shift of sound to calculate the speed of sound | Frequency shift | Audio Kit<br>Spektro | 160 |
| | Microphone | Measure the Doppler effect in various types of linear motions | Frequency shift *vs* speed | Frequency Analyzer<br>(Custom designed) | 161 |

**Table 8** A list of SmartIPLs for thermal physics

| **Physics Topics** | **Role of Smartphone** | **Brief description** | **Data Analysis** | **App Name** | **Ref.** |
|---|---|---|---|---|---|
| Temperature | Image<br>Video | Use smartphone-attached thermal cameras (SATC) to visualize fundamental thermal processes (conduction, convection, radiation, and phase changes) through guided and inquiry-based school experiments. | Temperature *vs* time<br>Temperature distribution | FLIR ONE | 162 |
| | Image<br>Video | Use SATC to investigate everyday thermal phenomena, emphasizing how visual thermal feedback shapes interpretation and learning | Temperature *vs* time<br>Temperature distribution | FLIR ONE | 163 |
| | Video | Use SATC to monitor the cooling of metal cylinders with different emissivities and investigate how surface properties affect radiative heat loss | Temperature *vs* time | FLIR ONE | 164 |
| | Video | Use SATC to capture leg temperature changes during a "wall sit" exercise and visualize internal energy dissipation in static systems. | Temperature *vs* time | SeeK Thermal | 132 |
| | Image<br>Video | Use SATC to explore thermal effects such as frictional heating, cooling, convection, radiation, emissivity, and environmental thermal profiles in both educational and real-world contexts | Temperature *vs* time<br>Temperature distribution | Not mentioned | 165 |
| | Video | Use SATC to monitor and analyze the cooling process of a liquid, demonstrating Newton's Law of Cooling | Temperature *vs* time | Framelapse<br>VidAnalysis | 166 |
| | Image<br>Video | Use a SATC to investigate artistic pigments and paintings via infrared imaging, linking physics concepts like thermal emission and emissivity with art analysis in a project-based learning activity. | Temperature *vs* time<br>Temperature distribution | FLIR ONE | 167 |
| | Image<br>Video | Evaluate SATC in terms of measurement principles, image quality, calibration accuracy, and their overall educational suitability as affordable physics lab tools | Temperature *vs* time<br>Temperature distribution | FLIR ONE<br>Seek Thermal | 168 |
| | Image | Use SATC to monitor water cooling in flasks of different volumes and analyze cooling curves using Newton's Law of Cooling | Temperature *vs* time | Framelapse<br>VidAnalysis | 59 |
| | | Use smartphones and Arduino sensors to explore mechanical oscillations, light transmission, beam deformation, and heat loss in everyday materials | Rate of heat loss | Phyphox | [67] |
| | Image | Use SATC to identify thermal inefficiencies in buildings by visualizing insulation performance and heat leakage | Temperature distribution | FLIR ONE | [169] |
| | Image | Use smartphone internal sensors and thermal cameras to examine how temperature varies in smartphones under different workloads and cooling conditions | Temperature *vs* time | CoolTool | [170] |
| | Image<br>Video | Use SATC to perform low-cost experiments visualizing heat transfer in common materials and support conceptual understanding of thermodynamics | Temperature *vs* time<br>Temperature distribution | FLIR ONE | [171] |
| | Display | Use SATC to visualize and compare temperature distributions in cylindrical bodies during cooling, highlighting lumped vs. non-lumped system behavior | Temperature distribution | FLIR Tools | [172] |
| | Video | Use SATC to record temperature evolution in metal rods heated at one end, enabling analysis of heat conduction and estimation of thermal diffusivity | Temperature *vs* time<br>Temperature distribution | FLIR Tools | [173] |
| | | | | | |
| | Image | Use SATC to measure surface temperatures of pigs in groups, applying emissivity compensation through altered-spectrum thermography | Temperature | Custom-developed Android App | [174] |
| | Display | Use interactive thermal videos on the Telelab platform to explore temperature variations and distributions in a variety of everyday and experimental contexts | Temperature *vs* time<br>Temperature distribution | Telelab | [175] |
| Heat transfer | Video | Use smartphone cameras to record thermochromic color changes in PDMS samples in contact with materials of varying thermal properties, analyzing heat transfer and specific heat capacity | Temperature *vs* time | Custom image-processing App | [176] |
| | Video | Use SATC to conduct and analyze heat transfer experiments involving radiation, conduction, and convection in familiar materials and systems | Temperature *vs* time<br>Temperature distribution | FLIR ONE | [177] |
| Convection | Video | Use SATC to analyze the cooling behavior of a heated copper plate under different air stream velocities and quantify convective heat loss | Temperature *vs* time | Vernier Thermal Analysis Plus | [178] |
| | Display | Visualize and measure the formation of convection cells in a heated fluid layer to study Bénard-Marangoni instability | Wavelength *vs* fluid layer's thickness | FLIR Tools | [179] |
| | Image<br>Video | Use SATC to record the temperature of a magnetic brake disc before, during, and after a free-fall amusement ride, visualizing heating and cooling dynamics | Temperature *vs* time<br>Temperature distribution | FLIR ONE | [180] |
| | Image<br>Video | Use SATC to analyze real-world heat transfer scenarios, including radiation and insulation effectiveness in everyday objects | Temperature *vs* time<br>Temperature distribution | FLIR ONE | [181] |
| | Image<br>Video | Use SATC to visualize resistance heating in pencil-drawn resistor circuits on various substrates, illustrating Ohm's Law and Joule's Law | Temperature distribution | Not mentioned | 182 |
| | Image<br>Video | Use SATC to observe temperature evolution and latent heat behavior in phase change materials during melting and solidification | Temperature *vs* time<br>Temperature distribution | FLIR ONE Pro<br>Thermal Image Extractor and ThermoVision Studio | 183 |

**Table 9** A list of SmartIPLs for fluidics

| Physics Topics | Role of Smartphone | Brief description | Data Analysis | App Name | Ref. |
|---|---|---|---|---|---|
| Pressure | Barometer | Measure atmospheric pressure at different heights to determine the density of air | Pressure *vs* height | Phyphox | 26 |
| | | Measure and analyze the pressure changes during the inflation and deflation of a rubber balloon, comparing the results with theoretical predictions | Pressure *vs* time | Barometer Graph<br>Physics Toolbox Sensor Suite<br>SideSync | 184 |
| | | Investigate the pressure changes experienced in a moving train as it travels through a tunnel | Pressure *vs* time | AndroSensor | 185 |
| | | Investigate the relationship between depth and pressure in various fluids, and help students to derive Stevin's law and understand fluid pressure concepts | Pressure *vs* depth | Physics Toolbox Suite | 186 |
| | | Measure the pressure exerted by water at different depths, allowing students to verify Stevin's law and explore the relationship between pressure and depth in a fluid | Pressure *vs* depth | Android: Physics Toolbox Sensor Suite, Sensor Suite, DS Barometer<br>iOS: Barometer & Altimeter, Bar-o-Meter, Barometro Pro | 187 |
| | | Simulate a space mission, allowing students to design and test prototypes that can withstand pressure differences, thereby understanding the concept of air pressure and its effects in a space environment | Air pressure change | Phyphox | |
| | Barometer<br>GPS | Measure pressure and altitude by mounting a smartphone on a quadcopter, allowing students to compare the data with standard atmospheric models | Pressure *vs* altitude | AndroSensor | 188 |
| | Video | Record and analyze the oscillatory motion of fluid in a drinking straw when released from a capped state, modeling the dynamics using Newton's second law | Fluid level *vs* time | Tracker | 47 |
| Flow | Gyroscope<br>Video | Analyze the parabolic shape formed by the free surface of a liquid in a rotating frame and to correlate the surface curvature with the angular velocity of the rotation | Concavity vs angular velocity | AndroSensor<br>Tracker | 189 |
| | Video | Investigate the fluid draining process from a container, focusing on the effects of viscosity and the size of the draining hole on the fluid flow | Fluid height *vs* time | Tracker | 190 |
| Surface Tension | Image | Capture images of droplets suspended from a tube tip | Diameter of pendant droplet | ImageJ | 191 |
| | | Photograph and analyze liquid surface waves, utilizing image recognition techniques to measure wave properties and calculate the liquid's surface tension coefficient | Wavelength and frequency | MATLAB | 192 |
| | Video | Measures the viscosity of air by recording and analyzing the motion of a soap film moving through a funnel | Distance *vs* time | ImageJ | 193 |

**Table 10** A list of SmartIPLs for electromagnetism

| Physics Topics | Role of Smartphone | Brief description | Data Analysis | App Name | Ref. |
|---|---|---|---|---|---|
| Capacitor | Video | Utilizes hydraulic analogues of electrical circuits to visually demonstrate concepts like Ohm's law and the charging of a capacitor | Volume flow rate, hydraulic resistance, and hydraulic capacitance | ImageJ | 194 |
| | Light Meter | Measure the brightness of an LED in an RC circuit, providing an inexpensive method to study the capacitor's charge and discharge processes | LED intensity | Physics Toolbox Suite | 195 |
| | Microphone | Measure the RC time constant by analyzing the sound intensity decay as a capacitor discharge through a piezoelectric buzzer | Sound intensity *vs* time | Physics Toolbox Suite<br>iNVH & Phyphox | 196 |
| Magnetic field | Magnetometer | Verify the Biot–Savart law by measuring the magnetic field produced by a short current-carrying segment and analyzing its dependence on current and distance | Magnetic field *vs* current & distance | Phyphox | 197 |
| | | Measure the magnetic field of small magnets at different distances to explore the inverse cubic relationship between magnetic field strength and distance | Magnetic field *vs* distance | iOS: Magnetometer<br>Android: Magnetometer Metal Detector, Physics Toolbox Magnetometer | 57 |
| | | Measure the Earth's magnetic field components and calculate the magnetic dip angle to visualize the vector nature of the Earth's magnetic field | Different components of magnetic field | iOS: Magnetic Field Detector, Magnetometer, & Teslameter<br>Android: Physics Toolbox Magnetometer & Bubble Level | 198 |
| | | Measure and analyze the magnetic field strength produced by a circular current, allowing for a comparison between experimental data and theoretical predictions | Magnetic field *vs* location | Android: Sensor Box, 3D Compass, & Magnetometer | 199 |
| | | Measure and analyze the magnetic field distribution of a linear quadrupole, providing a hands-on approach to understanding magnetic field variations in complex configurations | Magnetic field *vs* distance | iOS: Magnetometer<br>Android: Physics Toolbox Sensor Suite | 200, 201 |
| | | Measure the magnetic fields produced by small-scale analogs of accelerator magnets and compares these measurements with theoretical predictions and simulations | Magnetic field *vs* distance | Phyphox | 202 |
| | Magnetometer<br>Accelerometer | Measure the magnetic field generated by a current-carrying coil and calculate the field's spatial dependence using data from the sensors | Magnetic field *vs* distance | Androsensor | 203 |
| | Magnetometer<br>Video | Use a smartphone's magnetometer to measure magnetic field strength and its camera, combined with "Tracker", to determine the charge-to-mass ratio (e/m) of the electron by analyzing its curved trajectory in a magnetic field | Magnetic field, radius of curvature of the electron's path | Keuwlsoft Gauss Meter<br>Tracker | 204 |
| | Externally built teslameter | Construct a low-cost teslameter using an Arduino and a smartphone application to measure and analyze the magnetic field produced by a system of Helmholtz coils | Magnetic field *vs* current | Custom Built App. | 205 |
| | Magnetometer<br>Accelerometer<br>Gyroscope<br>Camera | Employ smartphone sensors and augmented reality apps to enhance the visualization and understanding of three-dimensional magnetic fields and motion graphs | Visualize magnetic field vectors | Magna AR, LiDAR Motion, Physics Toolbox Sensor Suite | 65 |
| Faraday's Law | Magnetometer | Study Faraday's law of induction by measuring the time-dependent magnetic field and the corresponding induced electromotive force | Magnetic field *vs* time | Physics Toolbox Sensor Suite | 206 |
| | Video | Explore the effects of eddy currents on the motion of rare-earth magnets rolling down a conductive incline | Position of the magnet *vs* time | Avidemux | 207 |
| | Audio port | Construct an inductive metal detector using a smartphone's audio port to explore electromagnetic principles | Induced electromotive force | Android: Simple Tone Generator, SmartScope<br>iOS: Audio Function Generator, SignalScope X | 208 |

**Table 11** A list of SmartIPLs for electronics

| Physics Topics | Role of Smartphone | Brief description | Data Analysis | App Name | Ref. |
|---|---|---|---|---|---|
| Device | Signal Generator | Design a low-cost signal generator assembly using a smartphone or tablet with an amplifier circuit | Signal frequency & amplitude | Signal Generator | 209 |
| | Display | Convert a smartphone into a portable oscilloscope | Voltage *vs* frequency | SignalScope, AudioScope, & oScope | 210 |
| | | To construct a virtual oscilloscope for measuring time constants in RC and LR circuits | Relaxation time constant | Oscium | 211 |
| Circuit | Signal generator Oscilloscope | Use two smartphones, one as a signal generator and the other as an oscilloscope, to study the frequency response and resonance behavior of an RLC series circuit | Voltage *vs* frequency | Physics Sensors Toolbox Suite Phyphox | 212 |
| | Headphone port | Measure temperature changes by interfacing smartphone with an external thermistor and analyzing signal amplitude variations | Amplitude of the returning sine wave | Custom built | 213 |
| | Optical flash | Use the smartphone's optical flash and photodiode sensors to teach about optical sensing, signal conditioning, and process control, including variable-frequency strobe modulation and Morse code transmission. | Output voltage | | 214 |
| Digital Circuit | | Use a smartphone-based augmented reality system to enhance the learning of digital electronics by automatically identifying logic gate ICs and displaying relevant circuit information | Relationships between various logic gate ICs | Custom built | 215 |
| | Accelerometer | Use a smartphone's accelerometer data and a Jupyter Notebook to introduce students to digital signal processing concepts through the practical application of the Discrete Fourier Transform on real-world vibration data | Frequency | Phyphox | 66 |
| | Microphone | Use a smartphone connected to an external microphone and a stethoscope chestpiece to record and analyze heart sounds | Heart rate | Custom built | 216 |

**Table 12** A list of SmartIPLs for optics

| Physics Topics | Role of Smartphone | Brief description | Data Analysis | App Name | Ref. |
|---|---|---|---|---|---|
| Light Property | Light meter | Investigate the inverse square law by measuring how light intensity decreases with increasing distance from a point source | Light intensity *vs* distance | Physics Toolbox Sensor Suite | 217 |
| | | Evaluate the luminous efficacy and efficiency of different optical sources using a smartphone's ambient light sensor | Light intensity *vs* distance | Sensor Box | 218 |
| | | Measure the illuminance of a linear fluorescent tube, confirming that illuminance follows an inverse-distance law rather than the inverse-square law | Light intensity *vs* distance | Physics Toolbox Suite | 219 |
| | Light meter<br>Gyroscope | Verify Malus' law by measuring the intensity of polarized light passing through a polarizer at various angles | Light intensity *vs* polarization angle | Physics Toolbox | 220 |
| | Video | Utilize video analysis with a smartphone to investigate Malus' law by measuring the intensity of polarized light as a function of the polarizer's angle | Light intensity *vs* polarization angle | Tracker | 221 |
| | Light source | Observe polarized light using a smartphone screen and human naked eye, demonstrating Haidinger's brush and the optical activity of various materials | Observe intensity change *vs* polarization angle | Haidinger's Brush | 222 |
| | Light meter | Measure Brewster's angle for different materials by analyzing the reflected light intensity at varying angles of incidence | Light intensity *vs* incident angle | Lux Meter<br>Light Meter | 223 |
| Ray Optics | Camera | Determine the focal length of a smartphone camera by capturing and analyzing images of an object placed at different distances | Focal length | | 224 |
| | | Investigate the thin lens equation using a smartphone camera, exploring both its successes and limitations | Focal length | ImageJ | 225 |
| | | Demonstrate a simple method for measuring the focal length of a smartphone camera lens using a ruler and image analysis | Object distance & height, image height in pixels | | 226 |
| | | Use water droplets on a tablet screen as convex lenses to verify the lens equation and lens-maker's equation through image analysis of magnified screen pixels | Magnification *vs* radius of water droplet lens | ImageJ | 227 |
| | | Determine the focal length of both converging and diverging lenses through a two-step imaging method | Image height vs distance | ImageJ | 228 |
| | Camera<br>GPS | Use smartphones, GPS, and 3D printing to demonstrate how local noon varies with longitude, thereby explaining the need for time zones | Shadow & time | | 229 |
| Beer's Law | Camera | Measure the color of the liquid to determine the concentration of a solution | Color intensity change | ImageJ | 230 |
| | Light source<br>Light meter | Demonstrate light absorption following the Beer–Lambert law and light scattering | Light intensity *vs* dye concentration | RGB<br>ColorAssist | 231 |
| | Light meter | Create a low-cost, mobile, paper-based colorimeter that uses a smartphone's light sensor to measure absorbance and concentration of liquid dyes | Light transmission intensity | | 232 |
| | | Demonstrate the exponential decay of light intensity (following the Lambert law) as it passes through successive sheets of paper | Light transmission intensity | Physics Toolbox suite or Phyphox, Tracker | 233 |
| | Camera | Use a simple, low-cost setup for colorimetric and fluorometric analysis using smartphones | Color intensity | ColorGrab | 234 |
| Diffraction & Interference | Light source<br>Grating<br>Camera | Investigate the phenomenon of selective light transmittance in a glue stick, which appears white near a light source and gradually changes to orange and then red toward the opposite end | Light intensity *vs* distance of glue stick | Tracker | 61 |
| | Grating | Explore diffraction and interference patterns using smartphone screens as reflective diffraction gratings | Separation of diffraction spots | | 235 |
| | Mirror | Construct a low-cost Mach-Zehnder interferometer using Lego, household items, and Arduino detectors to demonstrate the wave-like nature of light through visible fringe patterns sensitive to temperature-induced refractive index changes | Shift in interference fringes | | 236 |
| | Video | Study a Michelson interferometer influenced by external sound | Interference pattern *vs* frequency | Phyphox and Physics Toolbox | 237 |
| Spectrometer | Camera | Use a 3D-printable spectrophotometer that integrates a smartphone camera to teach the principles of UV-Vis spectroscopy, such as the Beer-Lambert Law, by measuring the absorbance of light in various solutions | Intensity *vs* wavelength | ImageJ | 238 |
| | | Utilize a smartphone-based spectrophotometer to teach principles of visible range spectroscopy | light intensity, absorbance, and transmittance of various solutions | Apache Cordova and Node.js | 239 |
| | | Fabricate a low-cost, high-resolution smartphone spectrometer using paper-based housing, a lab-made narrow slit, and a holographic diffraction grating | Intensity *vs* wavelength | ImageJ | 240 |
| | | Measure the light intensity and spectral characteristics of a heated tungsten filament, demonstrating the Planck radiation law and Wien's displacement law | Spectra of tungsten filament at different temperature | Physics Toolbox<br>Phyphox | 60 |
| | | Use a homemade spectrometer and a smartphone camera to measure the spectra of discrete and continuous light sources | Intensity *vs* wavelength | | 241 |
| | | Make a sinusoidal diffraction grating using a holographic method and then using a smartphone-based optical spectrograph to observe and analyze optical spectra | Intensity *vs* wavelength | | 242 |
| | | Use a low-cost apparatus comprising laser diodes, diffraction gratings, and a smartphone to measure the Raman spectrum of water by detecting the inelastically scattered light | Intensity *vs* Raman shift | ImageJ<br>Tracker | 63 |
| Others | Camera<br>Display | Investigate the principles of additive color mixing and color reproduction using smartphone displays and digital cameras | Spectral composition of colors | | 243 |
| | Camera | Demonstrate how to use a smartphone for fundus photography, allowing students to explore the principles of ophthalmoscopy and understand the optical setup needed to observe the retina | Capture clear images of the eye's fundus | | 244 |
| | | Use a smartphone with a 3D-printed magnification attachment to distinguish between the functionalities of a magnifying glass and a microscope | Photograph | | 245 |

**Table 13** A list of SmartIPLs for modern physics

| Role of Smartphone | Brief description | Data Analysis | App Name | Ref. |
|---|---|---|---|---|
| Camera | Demonstrate how smartphones and tablet PCs can be employed for $\beta^-$ spectroscopy, using their camera sensors to detect $\beta^-$ radiation and analyze the energy and intensity of electrons emitted by a radioactive source | Intensity of $\beta-$ radiation | RadioactivityCounter | 64 |
| | Measure the light intensity and spectral characteristics of a heated tungsten filament, demonstrating the Planck radiation law and Wien's displacement law | Spectra of tungsten filament at different temperature | Physics Toolbox<br>Phyphox | 60 |
| | Use a smartphone-based device to perform surface plasmon resonance imaging for real-time mapping of concentration distributions near solid-liquid interfaces | SPR intensity *vs* target concentration | MATLAB<br>Simulink | 246 |

**Table 14** A list of SmartIPLs for astronomy

| Role of Smartphone | Brief description | Data Analysis | App Name | Ref. |
|---|---|---|---|---|
| Video | Observe and record the transit of the International Space Station to calculate its orbital velocity through video analysis and basic astronomical measurements | Angular velocity & orbital velocity | | 247, 248 |
| Light meter | Utilize a smartphone's light sensor to simulate and analyze planetary transits | Light flux *vs* time | Physics Toolbox Sensor Suite | 249 |
| | Measure changes in luminosity on a globe model, helping students understand how Earth's axial tilt and latitude affect seasonal variations in sunlight | Light flux *vs* latitude | Galactica Luxmeter | 250 |
| Magnetometer | Perform classroom-sized geophysical experiments by mimicking magnetic surveys, enabling students to learn about magnetic field anomalies and data processing | Magnetic field components | MATLAB Mobile | 251 |
| Camera | Use smartphone to take photographs of the Moon through a telescope to teach about non-Euclidean geometry by calculating lunar surface features like crater diameters and mountain ranges | Distances between craters, diameters of craters, & areas of lunar features | GeoGebra | 252 |
| Light meter<br>Accelerometer | Conduct home-based astronomy experiments that model exoplanet detection and analysis | Illuminance *vs* time | Phyphox<br>SPARKvue, PocketLabs | 253 |

*Books on SmartIPLs.* In addition, several books dedicated to smartphone or mobile device-based lab and project design have been published. In 2013, Mike Westerfield released "*Building iPhone and iPad Electronic Projects*"[254], a practical guide introducing the techBASIC programming environment for creating interactive electronic projects using iOS devices. This book explores applications utilizing built-in sensors, bluetooth low energy (BLE), and Wi-Fi, with projects ranging from a metal detector to a BLE-enabled model rocket. In 2015, Jason Kinser authored "*Kinematic Labs with Mobile Devices*" [255], the first book to use smartphones and tablets to conduct physics labs. It features 13 labs, including topics such as acceleration of an elevator, tension, the Atwood machine, inelastic collisions, pendulum, Kepler's third law. Each lab employs smartphones, laptops, and household items, facilitating cost-effective physics experiments outside traditional lab settings. In 2018, Auer et al. edited "*Cyber-Physical Laboratories in Engineering and Science Education*"[256], which explores integrating cyber-physical systems into educational labs. It covers theoretical foundations, implementation strategies, and case studies in areas such as automation engineering and additive manufacturing, aiming to enhance remote and virtual lab experiences. In 2021, Organtini published "*Physics Experiments with Arduino and Smartphones*"[257], a comprehensive undergraduate text that integrates modern programming (Python), open-source microcontrollers (Arduino), and smartphone sensors into physics education, guiding students through hands-on experiments in mechanics, oscillations, wave phenomena, data acquisition, and statistical analysis. In 2022, Kuhn and Vogt edited "*Smartphones as Mobile Minilabs in Physics*"[62], compiling over 70 smartphone-based physics experiments. These experiments can be setup easily with straightforward data analysis, covering concepts in mechanics, optics, astrophysics, and more. Also in 2022, Zhao and Phang compiled "*Use of Smartphones in Optical Experimentation*"[23], demonstrating smartphone-based optical labs with low-cost or 3D-printed components. This book includes the demonstrations of fundamental geometric and physical optical principles such as the law of reflection, the law of refraction, image formation equations, dispersion, Beer's law, polarization, Fresnel's Equations, optical rotation, diffraction, interference, blackbody radiation, etc. as well as many practical applications, such as the design of a monochromator and spectrometers, the uses of the Gaussian beam of a laser, the monitoring of water pollution, and understanding the colors of LED lights, butterflies, peacock feathers, plants, and flowers, as well as estimating the temperature of incandescent lamp or sun, etc.

**4. Pedagogical Design Principles in SmartIPLs**

Over the past decade, the use of smartphones in physics education has shifted from niche pedagogical experiments to a more widely studied instructional approach, as shown in **Figure 2**. In the early 2010s, most studies were focused on feasibility: could a smartphone measure acceleration during free fall or detect frequency changes during oscillation? The answer was yes. Yet, while these early efforts were instrumental in establishing technical viability, they provided little quantitative data on student learning outcomes. Instead, they laid the foundation for a broader reimagining of experimental physics instruction, particularly around accessibility, engagement, and contextual learning. By the latter part of the 2010s, research began shifting from feasibility to pedagogy. Studies such as Astuti et al. explored the use of "pocket mobile learning" to develop critical thinking in physics[258], while Moosvi et al. investigated the effectiveness of remote labs versus traditional face-to-face labs, incorporating smartphone apps into the learning experience[259]. The study by Countryman in North Carilina State University found that students in smartphone-based labs (the MyTech group) did not outperform traditional students in kinematics graphing skills, but did show increased recognition of physics as relevant to real life.[260] These students reported greater engagement and informal use of tools beyond class, highlighting a shift in how learners perceive the function and relevance of physics. This trend toward increased cognitive and emotional engagement was corroborated by Hochberg et al., who reported that students, especially those previously underperforming, experienced greater interest and confidence when given autonomy through smartphone-based activities.[261] Similarly, Nikolopoulou & Kousloglou[262] and Luh Sukariasih et al.[263] found that smartphone-enabled group inquiries promoted collaboration, metacognitive awareness, and learner autonomy. These outcomes suggest that SmartIPLs may contribute to greater engagement with science, although findings vary across contexts. The historical trajectory of SmartIPLs suggests a shift toward more student-centered and tool-enabled pedagogical approaches, with mobile-first designs increasingly integrated into instructional practices. As Darmendrail et al. emphasized, The convergence of embedded sensors, processing power, and real-time feedback in smartphones has supported their use as instructional devices that may extend, rather than merely replicate, traditional lab tools.[264] The COVID-19 pandemic accelerated this transformation. As institutions worldwide moved to remote instruction, SmartIPLs rapidly shifted from optional innovations to essential platforms. This shift prompted a wave of systematic studies assessing their pedagogical effectiveness using validated instruments.

To understand how SmartIPLs foster deeper engagement, modeling skills, and epistemic agency, the following examine their instructional design, implementation strategies, and documented educational outcomes across diverse learning environments.

### 4.1 Foundations and Learning Theories

The pedagogical foundation of SmartIPLs is built on three interwoven pillars: context-based learning, inquiry-based active learning, and tool-enabled cognitive scaffolding. These pillars reflect decades of research in science education, emphasizing relevance, epistemic agency, and equitable access to authentic scientific practices. Together, they transform mobile devices from passive data collectors into catalysts for scientific modeling, reflection, and discovery.

#### 4.1.1 Context-based learning (CBL): connecting physics to everyday life

Context-based learning situates scientific inquiry within students' everyday experiences, rendering abstract physics concepts tangible and personally meaningful. Smartphones with their embedded sensors and ubiquitous presence serve as ideal tools for grounding experimentation in familiar settings: walking upstairs, riding a bike, playing music, or even jumping on a trampoline.

Darmendrail et al. conceptualized smartphones as Mobile Digital Experimental Tools (MDETs) that collect real-world data in situ, enabling authentic, embodied learning in everyday environments.[264] González et al. further emphasized the value of nontraditional learning spaces such as stairwells or skateparks as platforms for physical inquiry[24], while Rediansyah demonstrated successful implementation of sensor-based motion labs using *Phyphox* in Indonesian high schools.[265]

CBL-driven SmartIPLs have also been shown to enhance intrinsic motivation, science identity, and curricular alignment. For example, Luh Sukariasih et al.[263] and Zapana & Martínez [266] found that culturally and locally relevant experiments, tailored to the physical and social context of students, significantly improved engagement and comprehension. Even in special education contexts, such as physics instruction for blind students[267], smartphones facilitated inclusive participation through adapted audio or haptic feedback.

By embedding physics into students' lived environments, SmartIPLs grounded in CBL not only improve conceptual understanding but also bridge the gap between classroom learning and real-world phenomena.

### 4.1.2 Inquiry-based active learning (IBL): students as scientists

SmartIPLs inherently support inquiry-based learning by enabling students to design, conduct, and revise experiments using tools, i.e., smartphones, they already understand. Rather than following step-by-step lab manuals, students pose questions, make predictions, test models with real-time data, and iterate their procedures in response to findings. Wong et al. enabled high school students to rediscover scientific laws with mobile sensors and programmable robots,[268] while Liu et al. documented improved reasoning from iterative cycles of prediction and experimentation.[269] Countryman & Paesler introduced the MyTech curriculum to demystify smartphone sensors, enhancing students' mechanistic understanding and modeling abilities.[270] Stampfer et al. demonstrated the power of student-led exploration prior to instruction,[27] while Orban & Teeling-Smith [271] and Koenig & Bake [28] incorporated coding (Python) and reflective cycles into their remote SmartIPL environments, showing that deep inquiry is possible even outside the physical lab. Zhao integrated classical optics experiments with open-ended design and personal data collection, allowing students to act as investigators.[23]

These inquiry-rich designs echo the values outlined by Holmes et al., who warned that overly procedural labs diminish students' sense of agency.[272] Instead, SmartIPLs designed with IBL principles can replicate authentic scientific practice, encouraging prediction, measurement, reflection, and revision, as endorsed by Motiwalla[273] and González et al.[24]

### 4.1.3 Tool-enabled scaffolding: smartphones as performance supports

Modern smartphones offer a unique convergence of data acquisition, visualization, and feedback tools, making them powerful platforms for cognitive scaffolding. Rather than serving as opaque "black-box" instruments, apps like *Phyphox*, *Physics Toolbox*, etc. (see **Table 4**) allow learners to process noisy data, visualize time-series plots, apply fits, and perform in-situ error analysis, all during the experiment itself. This vision is consistent with Motiwalla's mobile learning model, which emphasizes personalization, interactivity, and just-in-time feedback as key features of effective mobile learning environments.[273] Students receive immediate, visual cues about their measurements, helping them recognize discrepancies and improve experimental design in real time. The benefits extend to both students and educators.

Milner-Bolotin et al.[274] and Kaps et al.[275] found that multimodal representations, visual, textual, numeric, enhanced learning and supported large-enrollment courses. Zapana and Martínez [266] and Kramer[276] noted that teachers benefited from intuitive design tools and curricular scaffolds, enhancing both implementation and innovation. Stampfer et al.[27] and Koenig & Bake[28] showed how apps scaffolded theory-building and error recognition, even when students worked independently from home. Darmendrail & Müller[277] illustrated how students used smartphone data to investigate everyday phenomena like fridge vibrations or vertical jumps and fit models to real-world systems, reinforcing the role of tools in bridging informal exploration and formal modeling.

SmartIPLs thus can not only act as laboratory tools, but also serve as embedded tutors, guiding learners through scientific exploration while reducing cognitive overload. The devices support both novices and advanced learners by adapting to instructional needs and enabling scaffolding at multiple levels: conceptual, procedural, and metacognitive.

### 4.2 Instructional Design Principles

The pedagogical effectiveness of SmartIPLs rests not only on their mobile and technological affordances, but also on how they are intentionally designed to support scientific learning. Drawing on principles of constructivist education[4, 5] and mobile performance support[273], SmartIPLs incorporate three overarching instructional strategies: open-ended experimentation, layered cognitive & procedural scaffolding, and scientific modeling & critical reflection. These principles are mutually reinforcing: open-endedness fosters curiosity, scaffolding ensures learners are supported as they explore, and modeling provides a structure for interpreting and communicating findings.

#### 4.2.1 Open-ended experimentation: shifting from verification to exploration

Traditional physics labs often follow rigid protocols that emphasize confirmation of known results. In contrast, SmartIPLs can promote open-ended experimentation, where students can take ownership of both the questions they explore and the methods they use. By leveraging the mobility, flexibility, and multimodal sensors of smartphones, students are empowered to test hypotheses in diverse environments, from staircases and skate parks to kitchens and playgrounds. This open-ended format is not unstructured. Instead, it could replace fixed lab scripts with guided exploration, offering clear objectives while allowing students to select variables, iterate on techniques, and interpret their own findings. This structure supports a wide range of learners: novices benefit from accessible entry points, while more advanced students can deepen their investigations through iterative design and data analysis.

Multiple studies have affirmed the pedagogical value of this design shift. In smartphone-based optics labs, undergraduates explored blackbody radiation, diffraction, or polarization using smartphone cameras and lenses, refining their setups based on the data they collected.[23] Similarly, González et al. [137], Larnder et al. [278], and Becker et al. [279] demonstrated how undergraduate and high school students were empowered to conduct self-directed investigations using smartphone sensors. González et al. emphasized that open design spaces encouraged students to take on the role of experimenters, strengthening their identity as scientists.[24] Wong et al. confirmed that high school students could independently rediscover physical laws through autonomous sensor-data collection.[268] Darmendrail & Müller showcased creative explorations, including measuring joint cracking or monitoring environmental noise via smartphones, highlighting real-world relevance.[277] Kramer reported that when preservice teachers were given freedom to design experiments, from friction measurements to acoustic analyses, they not only developed instructional creativity but also modeled scientific reasoning processes authentically.[276]

Ultimately, the shift toward open-ended experimentation in SmartIPLs reflects a broader transformation in physics education, from confirmatory exercises to epistemic learning environments where students construct, test, and revise their own understanding. By empowering learners to engage with real data, make design decisions, and iterate their models, SmartIPLs embody the principles of active learning and foster ownership of the scientific process.

#### 4.2.2 Layered scaffolding: supporting learners through tools, curriculum, and pedagogy

While SmartIPLs emphasize learner autonomy and creativity, their effectiveness depends equally on intentional scaffolding that supports students as they develop experimental and conceptual competence. Rather than leaving students to navigate open-ended tasks unaided, SmartIPLs should embed support structures across three interrelated levels: technological, curricular, and instructional. This layered scaffolding model ensures that students receive appropriate guidance while retaining ownership of their learning process:

***Technological Scaffolding***: Smartphone apps like *Phyphox*, *Physics Toolbox* act as embedded instructional tools. These platforms offer features like real-time data visualization, sensor-specific calibration tools, and built-in calculators, which reduce cognitive load and provide immediate feedback. Their functionalities enable learners to troubleshoot their experiments in real time and develop stronger data collection and interpretation strategies.[24, 27] Customizable sensor combinations, QR-code-activated experiments, and remote control features further allow instructors to tailor levels of support based on student needs and learning environments.[27]

***Curricular Scaffolding***: Curriculum design in SmartIPLs needs to incorporate goal-oriented prompts, sequenced investigations, and reflective checkpoints to guide students through increasingly complex inquiries. The MyTech curriculum, for instance, introduces students to the principles of smartphone sensor operation through scaffolded exercises that gradually progress toward open exploration. [270] Similarly, structured lab manuals and learning modules allow students to build foundational understanding before tackling open-ended design challenges. Mazzella demonstrated the effectiveness of a gradual release model, in which students first complete guided physics labs before transitioning to independent investigations.[42]

***Instructional Scaffolding***: Instructors play a pivotal role in sustaining learner momentum. Through just-in-time coaching, Socratic questioning, and facilitated group discussions, teachers help students clarify conceptual difficulties and refine their experimental approaches.[28] In hybrid or asynchronous settings, pre-recorded tutorials, interactive checklists, and discussion forums replicate this support. Kaps et al. showed that layered video tutorials and graduated feedback systems allowed university students to transition from structured replication to exploratory work in large-enrollment physics courses.[275] O'Brien emphasized the importance of aligning digital tools with pedagogical intent, enabling cognitive engagement even in remote, post-pandemic classrooms.[280] Similarly, Zhao required student teams to prepare project proposals and then guided them through detailed discussions on research principles, experimental design, and data analysis,[23] modeling how structured guidance can gradually transition into independent, inquiry-driven practice.

These scaffolded approaches align closely with the Vygotsky's Zone of Proximal Development[5] and Bruner's spiral curriculum model[281], both of which emphasize sequenced support and the gradual transfer of responsibility from teacher to learner. As students demonstrate increased confidence and competence, scaffolds are intentionally faded, fostering the development of scientific autonomy. Moreover, scaffolding in SmartIPLs not only benefits learners but also supports future instructors. Kramer found that that preservice teachers who participated in scaffolded SmartIPL training gained confidence in both conducting and facilitating open-ended experiments.[276] Y. Sirisathitkul and C. Sirisathitkul further highlighted that when mobile-assisted inquiry was embedded within strong pedagogical frameworks, it promoted metacognitive self-regulation, curiosity, and conceptual understanding.[282]

In sum, layered scaffolding ensures that SmartIPLs provide "soft guidance" enough to keep students motivated and on track without diminishing their sense of control. It creates a flexible instructional architecture where support can be dialed up or down depending on learner needs, ultimately promoting a path from supported inquiry to independent scientific investigation.

#### 4.2.3 Modeling and critical thinking: making sense of complex, noisy data

A defining pedagogical strength of SmartIPLs lies in their emphasis on scientific modeling and critical thinking, not simply on data collection. Unlike idealized textbook datasets, real-world measurements from smartphone sensors are often noisy, variable, and context-sensitive. Rather than viewing this as a limitation, SmartIPLs harness these imperfections as a powerful catalyst for developing students' data literacy, sense-making, and scientific reasoning. In well-designed SmartIPLs, students are not passive recipients of procedures; they are active modelers who formulate hypotheses, refine predictions, and interpret discrepancies between theory and measurement. This modeling-centered approach mirrors authentic scientific inquiry and cultivates a deeper understanding of the physical world.

***Modeling Practices in SmartIPLs*:** SmartIPLs routinely engage learners in a sequence of modeling-rich activities

- <u>Prediction and testing</u>: Students are encouraged to generate theoretical expectations, such as using kinematic equations to predict motion, and compare these predictions to experimental outcomes.
- <u>Graphical and visual analysis</u>: Apps like *Phyphox* display real-time plots (e.g., acceleration vs. time), prompting learners to identify trends, detect outliers, and consider sources of variation.
- <u>Error reflection and iteration</u>: Students reflect on environmental factors, sensor limitations, and their own experimental designs, iteratively refining their methods and models.

Rather than treating discrepancies as failures, SmartIPLs promote productive failure, students evaluate unexpected results to sharpen their understanding of both the system and their tools.

***Empirical Evidence of Critical Thinking and Modeling***: Holmes et al. argued that lab instruction should prioritize scientific reasoning over rote proceduralism,[272] and SmartIPLs exemplify this shift. Countryman & Paesler showed how the MyTech app demystifies accelerometer data through hands-on sensor modeling.[270] Similarly, González et al. [24] and Wong et al. [268] reported that smartphone-enabled investigations support hypothesis testing, iterative refinement, and evidence-based reasoning. Darmendrail & Müller showcased creative modeling tasks such as the vertical jump experiment, where students fit acceleration curves to theoretical motion models and evaluate deviations.[277] Luh Sukariasih et al. documented similar outcomes in high school settings, where students collaboratively assessed the reliability and accuracy of sensor-generated data.[263] Orban & Teeling-Smith demonstrated how smartphone-based labs can also serve as gateways to computational modeling.[271] In their study, students analyzed motion data and used Python simulations to test and revise both conceptual models and their code, linking physical and computational thinking in powerful ways. This integrated approach fostered error diagnosis, conceptual abstraction, and algorithmic thinking, highlighting the synergy between programming and physical experimentation. Milner-Bolotin et al. emphasized the unique affordances of modern smartphones: students simulate, visualize, and experiment in real time, transforming passive demonstrations into active, model-based learning experiences.[274] This emphasis on modeling extends beyond mechanics and motion. As detailed by Zhao & Phang, smartphone-based optical labs also foster rich opportunities for model construction, testing, and refinement.[23] These projects require students to propose and set up their own projects, reconcile theory with real data, diagnose discrepancies, and refine models. The authors emphasize that such ownership fosters deeper understanding and mirrors authentic scientific reasoning.

***Multi-Representational Reasoning and Epistemic Agency***: SmartIPLs also encourage multi-representational reasoning, as students translate between numerical data, graphical trends, physical intuition, and verbal explanations. Students interpret sensor-based measurements not as isolated values but as dynamic systems represented across multiple modalities. For example, in vertical jump experiments, students relate acceleration-time plots to embodied motion and theoretical models.[277] In motion tracking labs, learners compare raw data to kinematic predictions and reconcile unexpected deviations.[268, 270] Even in coding-integrated investigations, smartphone-derived data become inputs for Python simulations, prompting translation between physical, graphical, and algorithmic domains.[271] These layered representations help to deepen understanding and enable learners to diagnose error, refine models, and construct meaning from complex phenomena.

Crucially, these representational practices can support the cultivation of epistemic agency, the ability to engage in science not as rote procedure, but as inquiry-driven sense-making. SmartIPLs position students as investigators who define their own questions, test hypotheses, and revise interpretations. Across studies, students are seen taking ownership of the full experimental arc, from setup and prediction to troubleshooting and explanation, mirroring authentic scientific reasoning. Whether exploring optics phenomena in student-designed projects[23], iterating on noisy sensor data, [24, 263] or guiding their peers through open-ended tasks[276], learners are empowered not only to do science, but to think like scientists. By embedding modeling, reasoning, and reflection into everyday experimentation, SmartIPLs create a participatory space where students construct knowledge with autonomy and intentionality. These practices are consistent with long-standing calls for science as inquiry[283, 284] and demonstrate how digital tools can help learners meaningfully engage with the messiness and creativity inherent in authentic scientific practice.

### 4.3 Implementation Models of SmartIPLs: Tools, Curriculum, and Institutional Integration

SmartIPLs offer a highly adaptable framework that can be tailored to a wide range of instructional goals, lab formats, learning environments, and student populations. From basic verification labs to open-ended design projects, SmartIPLs can be implemented using different types of mobile hardware and software, serving as an effective complement or replacement to traditional lab equipment. This section outlines the primary implementation models of SmartIPLs, organized by experimental method, lab type, instructional delivery format, and pedagogical modality.

#### 4.3.1 SmartIPL methods: sensor-based and camera-based approaches

SmartIPLs employ two primary methodological approaches (**Figure 3**): (1) Sensor-based experiments, which utilize the smartphone's built-in sensors for real-time data acquisition, and (2) Camera-based experiments, which leverage the smartphone's camera for video or image capture, with data analyzed post hoc. Each method aligns with distinct pedagogical objectives and offers complementary strengths, enabling SmartIPLs to address a wide range of conceptual, procedural, and analytical goals in physics education.

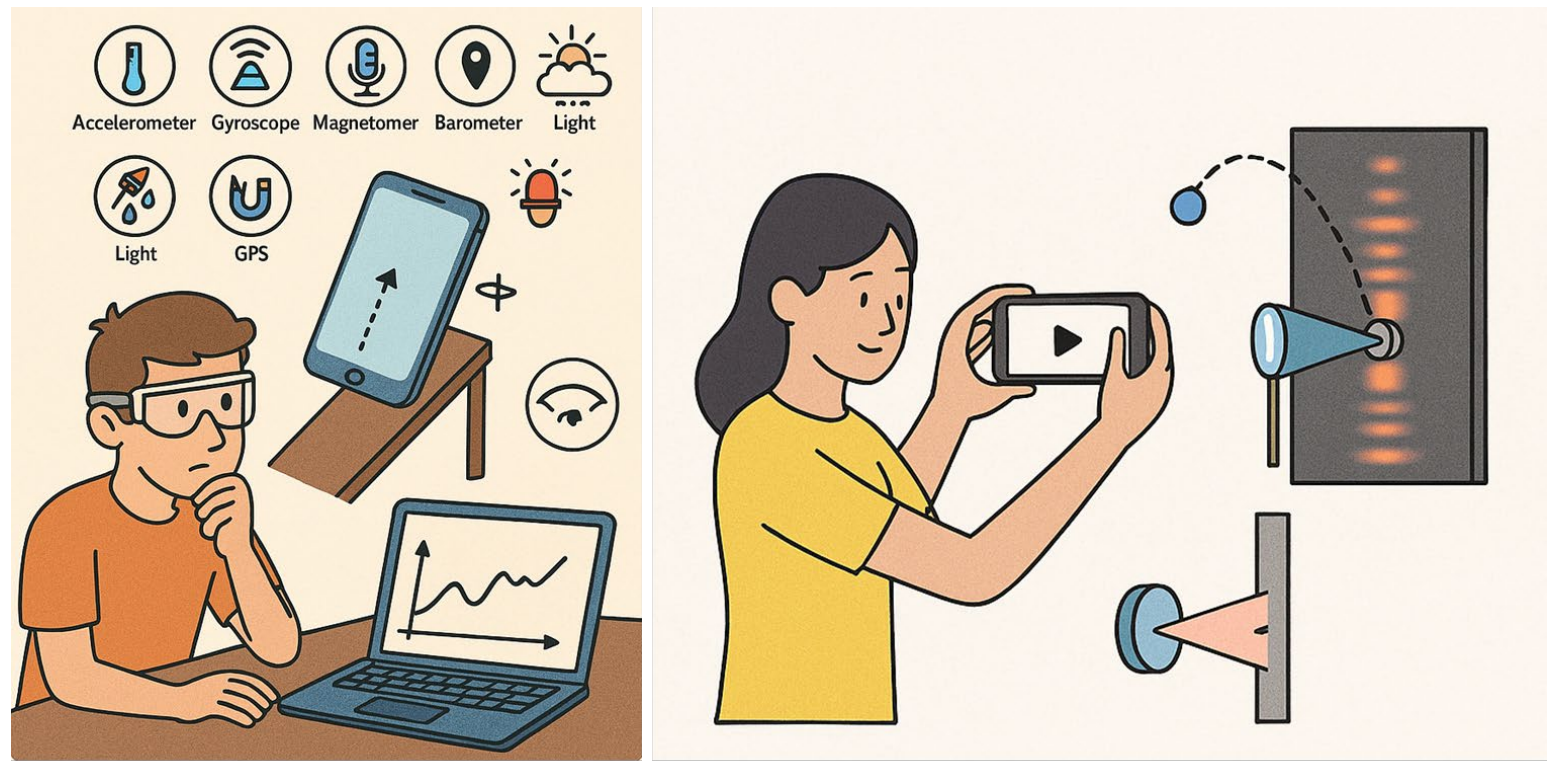

Sensor-based experiments    Camera-based experiments

**Figure 3** Two primary SmartIPL approaches: (A) built-in sensor-based experiments and (B) camera-based experiments.

***Sensor-based experiments.*** Sensor-based labs comprise approximately 78% of published SmartIPLs (**Figure 4A**) and capitalize on the wide array of sensors found in modern smartphones, including accelerometers, gyroscopes, magnetometers, microphones, barometers, light sensors, and GPS modules. For instance, accelerometers (29%) enable measurements of linear and angular motion in setups such as free fall and oscillations.[39-43, 87-90] Microphones (15%) support acoustic timing and sound-based labs,[40, 44, 45, 86] while gyroscopes (11%) assist in rotational dynamics studies. [97-103] Magnetometers (10%), [80, 81, 197-202], light sensors (6%),[195, 217-219] barometers (4%), [58] GPS (2%),[229] and proximity sensors (1%)[122] further expand the experimental repertoire. These sensors are often paired with versatile apps like *Phyphox* (36% usage), *Physics Toolbox Suite* (19%), and *Audacity* (7%) (see **Figure 4B**), which enable students to collect, visualize, and export high-resolution data in real-time. This immediacy fosters technical engagement but may limit opportunities for modeling or critical reflection if experiments become overly automated.

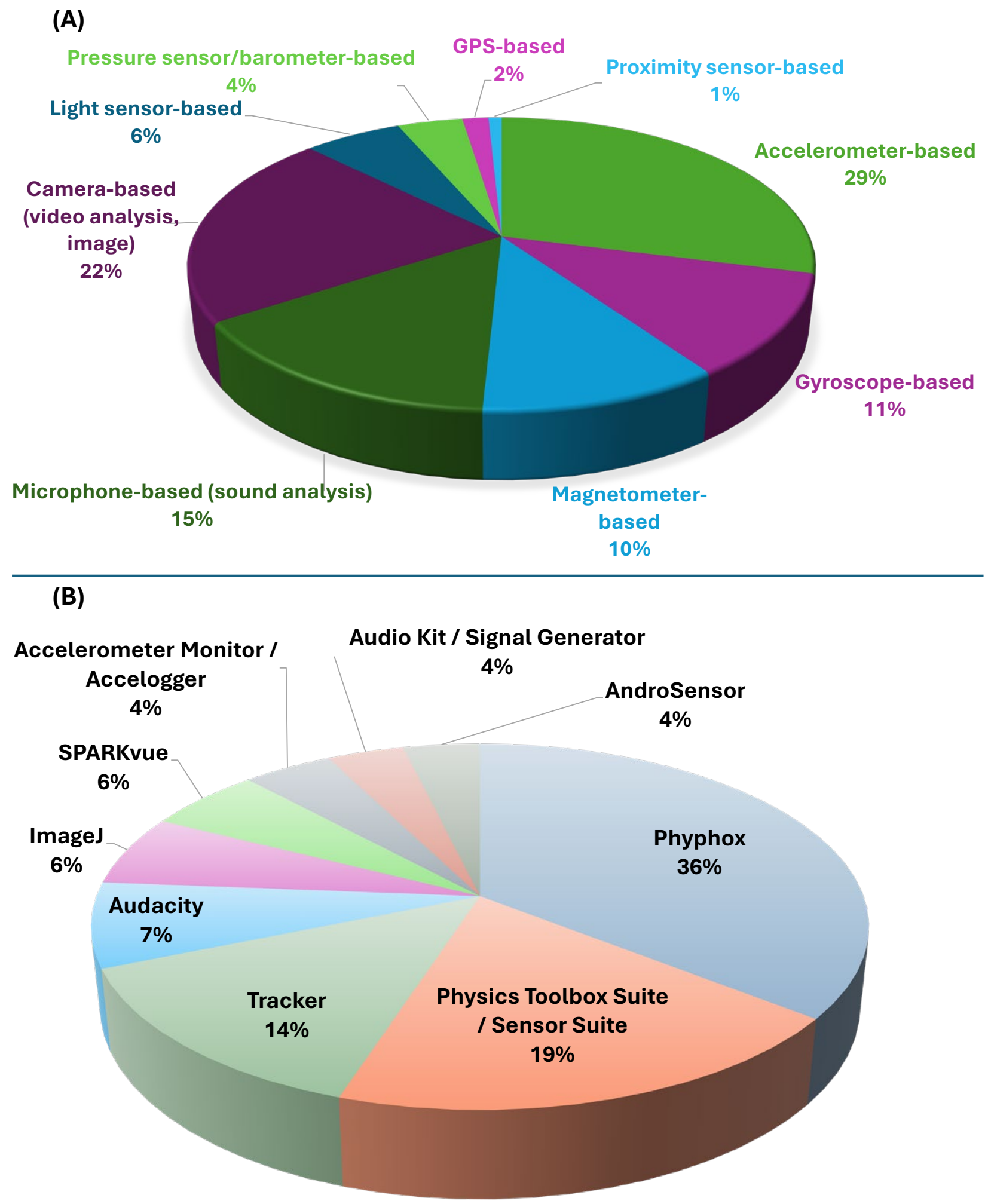

**Figure 4** Distribution of (A) sensor types and (B) common Apps or software tools used in SmartIPL literature.

***Camera-based experiments.*** Accounting for 22% of SmartIPLs (**Figure 4A**), camera-based experiments center on observation, interpretation, and modeling through post-experimental analysis. These labs are particularly effective in mechanics, [74, 84] optics,[224-228] and thermal physics,[59, 166] such as tracking projectile motion, measuring focal lengths, or modeling cooling curves. Students use video or image analysis tools such as *Tracker* (14% usage), *ImageJ* (6%), and *FLIR* to analyze motion, optics, or thermal processes frame by frame. Though more time-consuming, camera-based methods promote deeper conceptual reasoning, error analysis, and multi-step modeling.

***Shared data processing techniques***. Both sensor- and camera-based SmartIPLs draw on a rich set of data analysis techniques, including:

- *Graph plotting:* To visualize trends in displacement, velocity, light intensity, or temperature. For example, acceleration vs. time graphs help verify Newton's laws in motion experiments, [39-43, 90] while light intensity vs. distance graphs confirm the inverse-square law in optics [217].
- *Frequency and Fourier analysis:* Used in acoustic and electronics labs to study resonance and signal composition. Applications like *Audio Kit* and *Audacity* allow students to measure sound frequency shifts due to Doppler effects [44, 45, 82, 135, 140] or analyze complex harmonic signals. [137, 148]
- *Video analysis:* Especially powerful in mechanics and modern physics, video-based tools like *Tracker* support precise position tracking and motion modeling, such as pendulum motion[120] or fluid oscillations[47]. These analyses promote visual learning and connect theoretical predictions with observed behavior.
- *Image analysis:* Used extensively in optics and fluid dynamics to study diffraction, interference, and surface tension. Tools like ImageJ allow pixel-level quantification, enabling experiments such as measuring droplet diameters[191] or analyzing light transmission through filters[61, 230].
- *Mathematical modeling and statistical fitting:* Increasingly prevalent in advanced topics such as electromagnetism and thermodynamics, these methods enable students to extract parameters like spring constants, damping coefficients, or energy transfer rates.[54, 67, 131] They bridge theory and experiment but require mathematical maturity and computational tools.
- *Real-time visualization and mapping:* GPS tracking, magnetic field mapping, and augmented reality interfaces offer immersive experiences. Examples include measuring 3D magnetic field vectors with Magna AR[65] or tracking orbital motion of the ISS via video capture.[247, 248]

As summarized in **Table 15**, each method serves distinct instructional purposes. Sensor-based labs emphasize immediacy, technical fluency, and live data interaction. Camera-based labs promote conceptual modeling, interpretation, and post-hoc reasoning. Together, they offer complementary pathways for engaging students in experimental design, data literacy, and physics inquiry, and adaptable for both in-person and remote environments.

**Table 15** Comparison between sensor- *versus* camera-based SmartIPLs.

| Aspect | Sensor-Based Method | Camera-Based Method |
|---|---|---|
| **I. Tools and Setup** | | |
| ***Primary Tools*** | Built-in sensors (accelerometer, gyroscope, microphone, etc.) | Smartphone camera, video/image capture tools |
| ***Common Apps*** | Phyphox, SPARKvue, Physics Toolbox | Tracker, ImageJ, VidAnalysis |
| ***Equipment Dependence*** | Requires sensor access and app compatibility | Requires stable mounting or lighting; less sensitive to phone model |
| **II. Experimental Design & Data Acquisition** | | |
| ***Data Collection Mode*** | Real-time and automated logging | Manual capture; analyzed post-experiment |
| ***Time Requirement*** | Short – fast setup and instant results | Longer – requires video setup, capture, and frame-by-frame analysis |
| ***Accuracy Profile*** | Good temporal accuracy; may vary by sensor quality | High spatial precision; depends on camera quality and calibration |
| ***Limitations*** | Sensor variability, possible over-automation | Time-intensive, steeper learning curve with analysis software |
| **III. Learning & Engagement** | | |
| ***Typical Applications*** | Harmonic motion, acceleration, sound, rotation, magnetic fields | Projectile motion, optics (diffraction), thermal dynamics, pendulum motion |
| ***Educational Focus*** | Emphasizes real-time measurement, signal trends, sensor awareness | Emphasizes modeling, uncertainty, spatial-temporal reasoning |
| ***Conceptual Depth*** | Moderate – often focused on data acquisition | High – fosters interpretation, system design, critical thinking |
| ***Interactivity*** | High – immediate feedback supports active exploration | Moderate – feedback is delayed but allows rich interpretation |
| ***Student Engagement*** | High – intuitive, app-driven experiences | Moderate to high – varies with student role in analysis |

### 4.3.2 Lab types: verification, inquiry, and project-based designs

SmartIPLs are compatible with a spectrum of laboratory designs:

***Verification labs***: Traditional physics experiments such as pendulum motion, free fall, and harmonic oscillations are easily replicated using built-in smartphone sensors. Kaps et al. demonstrated how these labs can be scaffolded through structured video guides, enabling both in-person and remote replication with conceptual reflection.[275] Zhao et al developed a curated set of eight SmartIPL labs, completed with video tutorials, lab manuals, and report templates, available at https://www.zhao-nano-lab.com/phys1211-1212.

***Inquiry-based labs***: Many SmartIPL frameworks encourage students to design and iterate their own investigations using apps like *Phyphox*, *Tracker*, or high-frame-rate video tools. For instance, Koenig and Bake redesigned a large introductory physics lab to support asynchronous, inquiry-based modules where students made predictions, tested models, and reflected on discrepancies.[28]

***Project-based labs***: Multi-week design tasks often integrate Arduino microcontrollers, 3D-printed components, LEGO structures, or other accessories with smartphone sensors to create custom experiments, enable real-time data acquisition, and investigate complex physical phenomena through hands-on engineering and modeling. Cicuta et al. illustrated how students used Arduino-smartphone systems to design original sensors, calibrate them using known physical models, and validate results through app-based data analysis.[285] Similarly, Zhao & Phang introduced project-based optics labs leveraging smartphones, where students investigated optical principles and instruments through hands-on experimentation.[23]

### 4.3.3 Instructional formats: in-person, remote, and hybrid

SmartIPLs support various instructional formats, offering flexibility across teaching contexts:

***In-person instruction and live demonstrations***: SmartIPLs could replace or augment traditional demonstrations and hands-on labs. Instructors can use smartphone apps to visualize physical phenomena in real-time, from harmonic motion to light intensity. [286] Studies by Colt et al. [287] and Kittiravechote & Sujarittham[288] demonstrated how real-time data sharing and hands-on replication could deepen conceptual understanding. Broader reviews highlighted the widespread success of in-person SmartIPLs. [276, 282] Research by González, Kuhn, Vogt, and others highlighted how in-person SmartIPLs help students physically experience forces and phenomena.[24, 45, 62, 82, 160, 261] Together, these studies affirm that well-supported in-person SmartIPLs offer a powerful blend of conceptual modeling and experiential learning.

***Home-based and remote labs***: Smartphones and everyday household materials have enabled students to conduct rigorous scientific experiments entirely from home. González et al.[24, 289] and Chevrier et al.[290] provided compelling evidence that students could investigate acceleration, optics, magnetism, and sound waves with conceptual accuracy and procedural autonomy. During the COVID-19 pandemic, Zhao successfully implemented smartphone-based optics labs using low-cost components and household items,[23] providing an effective and accessible model for remote learning. These approaches proved especially critical during periods of remote instruction necessitated by the pandemic.

***Hybrid and e-teaching modalities***: O'Brien proposed a three-layered e-teaching strategy that combined synchronous instruction, smartphone-enabled experimentation, and asynchronous feedback loops.[280] Students engaged in weekly experimental challenges with video walkthroughs and customized support. Lv and Peng implemented SmartIPLs using both synchronous sessions (for collective model discussion) and asynchronous modules (for self-paced experimentation).[291]

Their results showed that students maintained strong engagement with data modeling and interpretation even without live supervision, highlighting SmartIPLs' scalability across temporal formats.

### 4.3.3 Pedagogical modalities: individual, collaborative, instructor-led, and teacher training

SmartIPLs are adaptable to diverse teaching and learning models:

***Individual exploration***: Flipped classrooms or self-guided modules allow students to conduct experiments independently, often supported by interactive apps and instructional videos.[275]

***Collaborative and team-based labs***: Students work in defined roles, such as experimenter, data analyst, and presenter, while engaging in peer feedback throughout the lab process. This structure is particularly effective in project-based settings, where distributed responsibilities foster ownership and sustained engagement.[23, 285] Koenig & Bake further highlighted the social learning benefits of this approach, especially in facilitating collaborative hypothesis generation and data interpretation.[28]

***Instructor-led demonstrations***: In large-class environments, smartphones can be used by instructors to deliver dynamic, sensor-based physics demonstrations that promote active learning. Sans et al. demonstrated that smartphones can effectively substitute traditional lab equipment for topics such as harmonic motion, light intensity, and magnetic fields, allowing instructors to guide observation while maintaining engagement.[286] Expanding on this approach, Colt et al. integrated real-time smartphone-based experiments, like pendulum motion and free fall, into structured lectures, using live data displays to enhance conceptual clarity.[287] Kittiravechote and Sujarittham employed acoustic timing via the Phyphox app to measure gravitational acceleration, combining instructor-led demonstrations with student replication to promote model-based reasoning through multi-method data analysis.[288]

***Preservice teacher training***: Chevrier et al.[290] and Kramer[276] showed that embedding SmartIPLs into teacher preparation not only improves science content knowledge but also empowers educators to design their own mobile-based instructional strategies.

## 4.4. From Tools to Pedagogical Transformation

SmartIPLs are not merely technological innovations or cost-effective lab alternatives, they represent a pedagogical shift in how physics is taught, learned, and conceptualized. Their real significance lies not in the devices themselves, but in how they reshape the epistemology, practice, and accessibility of physics education. This shift reflects a broader movement in science education, as articulated by Holmes et al.[272], toward valuing modeling, reasoning, and reflection over rote procedure and content delivery. Within this framework, SmartIPLs function not merely as instruments for data collection, but as platforms that support metacognitive engagement, collaborative inquiry, and authentic scientific practice. As emphasized by Cicuta and Organtini [285], Zhao and Phang[23], and others, SmartIPLs can scaffold experimentation in ways that are inquiry-driven, adaptable, and embedded in students' lived experiences. For example, the Smartphysicslab initiative encourages preservice teachers to develop their own experimental setups using Arduino, Phyphox, and low-cost materials. [285, 292] These learners are not simply following procedures, they are engaging in design, reflection, and iterative refinement. Similarly, Liu et al.[269] and Hochberg et al. [261] found that SmartIPLs significantly enhanced students' scientific reasoning, autonomy, and enjoyment, particularly for those previously under-engaged in traditional lab environments.

This re-centering of learner agency and cognitive engagement underpins the transformation from *tool-centric* to *learning-centric* laboratory design. Tools serve pedagogy, not the other way around.[24, 270, 290] The smartphone becomes a dual-purpose object: a scientific sensor and an epistemic platform for sense-making, modeling, and feedback.[282] In this way, SmartIPLs can embody a culture of inquiry where students not only perform physics but begin to *think like physicists*.

This section unpacks the pedagogical implications of SmartIPLs, tracing how they could transform the goals, roles, and norms of physics laboratory instruction, from the position of the instructor to the nature of experimentation, the scope of accessibility, and the institutional pathways for scaling innovation.

#### 4.4.1 Shifting the instructor's role: from demonstrator to facilitator

SmartIPLs inherently promote active, student-centered learning. By enabling students to design, execute, and interpret experiments using tools they already own, SmartIPLs decentralize authority in the lab. The instructor's role evolves from provider of content and apparatus to guide, coach, and co-investigator. This mirrors trends in inquiry-based science education, where open-ended investigations, collaborative design, and formative feedback loops are prioritized over rigid procedure. [23, 24, 28] The portability and immediacy of smartphone tools support just-in-time learning, allowing instructors to respond to students' evolving hypotheses or unexpected data by modifying lab protocols in real time. Studies show that such instructional flexibility improves students' conceptual understanding and fosters productive failure and iterative learning.[261, 269, 280, 290]

### 4.4.2 Reconceptualizing what counts as "Doing physics"

SmartIPLs challenge traditional norms about what constitutes a legitimate physics experiment. Rather than relying on precision-grade instruments in fixed settings, students use sensors and cameras embedded in everyday devices to measure acceleration, frequency, magnetic field strength, or light intensity. This shift reinforces the everyday relevance of physics and legitimizes diverse forms of data collection and modeling. By aligning physics inquiry with students' lived experiences, SmartIPLs lower epistemic barriers and empower learners to pose their own questions, create their own apparatus, and interpret messy real-world data. This aligns with calls from science education researchers to promote authentic science practices, where students engage in modeling, argumentation, and abstraction.[293-295] In this context, SmartIPLs help bridge formal and informal learning environments. When learners collect and analyze their own data, they are no longer passive recipients of knowledge but active constructors of understanding.[296, 297]

### 4.4.3 Democratizing access and broadening participation

Perhaps the most transformative aspect of SmartIPLs lies in their potential to democratize access to high-quality science education. Because smartphones are already ubiquitous among students, they offer a low-barrier entry point to experimental physics, especially in under-resourced schools, remote settings, or during disruptions such as the COVID-19 pandemic.[23, 24, 280] SmartIPLs thus can serve as a counterweight to inequities in laboratory infrastructure, enabling greater participation from historically underrepresented groups in STEM. Several programs have successfully used SmartIPLs to expand physics access in high schools, teacher training institutes, and international contexts where traditional equipment is unavailable.[263, 265, 268, 276, 285, 290] Moreover, their compatibility with remote and hybrid learning makes them resilient to institutional disruption and flexible for lifelong learners or community colleges that rely on asynchronous formats.[28]

### 4.4.4 Scaling across curricula and institutions

SmartIPLs enable both horizontal and vertical scaling across physics curricula. Horizontally, they can be integrated into topics such as mechanics, waves, optics, thermodynamics, electromagnetism, and modern physics, thereby promoting coherence in data analysis methods and scientific reasoning strategies across content domains. Vertically, SmartIPLs support progression from high school to undergraduate education, and from general education courses to advanced project-based labs and capstone experiences.

Faculty communities have also begun to integrate SmartIPLs into interdisciplinary contexts, including chemistry[238, 298-300], biology[136, 137, 301-303], or engineering design[96, 99, 254]. These efforts align with a broader shift toward transdisciplinary STEM instruction. At the institutional level, some universities have formally adopted SmartIPLs in core lab sequences, summer bridge programs, and teacher preparation curricula, signaling a move toward systemic pedagogical transformation. [23, 28]

#### 4.4.5 Toward a culture of reflective and adaptive experimentation

Ultimately, SmartIPLs foster a culture of experimentation grounded in reflection, adaptability, and epistemic agency. Rather than emphasizing correctness or procedural rigidity, SmartIPLs invite students to ask: *What can I measure? What assumptions am I making? How should I design the setup? How valid is this result?* This mindset shifts the focus from following instructions to developing a scientific approach based on questioning, sense-making, and revision.

As students engage in model construction, prediction, testing, and refinement, they acquire more than just technical skills, they begin to see themselves as participants in the scientific enterprise. Research by Stroupe [304] and Kang & Windschitl[305] suggests that this sense of ownership over the scientific process is vital for sustained engagement and epistemic growth. In this light, SmartIPLs do not simply modernize lab instruction, they reimagine the physics lab as a space for cognitive empowerment, creativity, and equity.

These pedagogical transformations, from instructor roles to epistemic agency, have not only reshaped how physics is taught but have also led to measurable educational impacts. The next section summarizes recent assessments that validate the effectiveness of SmartIPLs across a range of educational contexts.

### 4.5. Evidence of Impact

As SmartIPLs gained traction, large-scale assessments began validating their impact across diverse educational settings. **Table 16** summarizes key studies from 2022–2024 that used validated instruments to assess science process skills, motivation, engagement, and conceptual understanding.

**Table 16**: Assessment results of SmartIPLs across educational settings.

| Study | Sample Size | Assessment Tool | Outcome | Ref. |
|---|---|---|---|---|
| **Zhetysu (2022)** | 50 | $\chi^2$ comparison | Smartphone labs > virtual labs for research skill development | 306 |
| **Göttingen (2023)** | 110 | Surveys | 86% improved understanding, strong student approval | 307 |
| **Charles Univ. (2023)** | 2,024 | Intrinsic Motivation Inventory | Value & effort > competence; smartphones reduce anxiety, esp. for girls | 308 |
| **Bayesian Analytics (2023)** | 72 | Bayesian model, Pearson $r$ | $r = 0.582$ between SPS and learning gains | 309 |
| **Sapienza & Trento (2023)** | ~ 88 | E-CLASS, survey | Equal/better learning, high engagement | 310 |
| **Wajo HS (2024)** | 24 | SPS test, $t$-test | Mean SPS = 72.08, $p < 0.05$ | 311 |
| **Walailak University (2024)** | 254 | Pre/post quizzes, Zoom feedback | Higher average grades, effective self-labs | 312 |

One of the earliest comparative studies was conducted by Zhanatbekova et al. at Zhetysu University in Kazakhstan who evaluated the effectiveness of smartphone-based labs (SBLs) versus virtual simulations (VLs) in fostering research skills among high school students.[306] Using a pre/post experimental design and chi-square statistical analysis, the study found that students who conducted real-world experiments using smartphones outperformed those using virtual labs in areas such as hypothesis formulation, experimental design, and data interpretation. The authors concluded that SBLs significantly enhanced the development of scientific research competencies. In a parallel effort, Tufino and Organtini introduced Arduino and smartphones into their undergraduate mechanics laboratory.[313] The redesigned course emphasized student-led experimental design, data acquisition, and Python-based analysis. The E-CLASS survey revealed that student conceptual understanding remained stable or improved relative to traditional labs, while engagement and ownership increased. Students reported a stronger grasp of measurement uncertainty, model fitting, and sensor limitations. The study marked a major milestone in validating that smartphone-based labs can match or exceed the educational value of conventional instruction.

Building on these early validations, Lahme et al. evaluated a series of smartphone-based digital experiments in an undergraduate physics course.[307] Students used sensors and video analysis tools like Tracker to analyze real-world motion. Students who engaged with these tasks reported that the experiments improved their understanding of physics concepts; 86% rated their experience positively. Tasks that involved real-world, student-designed setups—such as comparing sensor accuracy or analyzing the motion of a parachute—were particularly effective. At Charles University in Prague, Kácovský et al. conducted a large-scale investigation into the motivational factors that influence engagement in physics labs.[308] Surveying over 2,000 high school students with the Intrinsic Motivation Inventory (IMI), the study found that perceived value and personal effort were stronger drivers of engagement than perceived competence. The research emphasized the importance of BYOD (Bring Your Own Device) strategies, such as smartphone integration, which appeared to reduce anxiety—especially among female students—and promote confidence during experimentation. In the realm of middle school education, Jiang and colleagues applied Bayesian network modeling to explore how students develop science process skills during smartphone-based thermal experiments.[309] Conducted in the United States, the study recorded students' hands-on actions via an app and linked this behavioral data to conceptual learning outcomes. The results revealed a statistically significant correlation ($r$ = 0.582) between lab skill proficiency and learning gains. This work demonstrated that smartphone-based labs could not only engage students in hands-on science but also provide automated, real-time assessments of their learning. Meanwhile, Tufino et al., through the COSID-20 project at the University of Trento, implemented a variety of distance lab formats—including home-based smartphone experiments and simulation-based virtual labs—and collected extensive feedback from both high school and university students.[310] Among the various modalities, hands-on smartphone-based activities were rated the most effective. Students appreciated the opportunity to interact with real-world materials, troubleshoot experimental setups, and take ownership of their data collection and interpretation.

A complementary perspective came from Yurchenko et al. at Sumy State Pedagogical University, whose survey of physics teachers and pre-service educators revealed a striking contrast between students' and instructors' attitudes toward digital technologies.[314] While teachers expressed hesitancy, students reported that smartphones and other digital tools improved their understanding, made abstract physics concepts more tangible, and increased their motivation to engage with lab activities.

The momentum continued as Evains et al. investigated science process skill development among high school students in Indonesia using the Phyphox app for harmonic motion analysis.[311]

A post-test evaluation showed significant learning gains, with a mean score of 72.08 and statistical significance ($p < 0.05$). The study highlighted the feasibility of implementing SmartIPLs even in resource-constrained environments, where conventional lab setups might be impractical. Finally, in Thailand, Dam-O et al. at Walailak University implemented a fully online physics lab course for 254 first-year engineering students, incorporating Phyphox, Tracker, and common smartphone apps for light and sound measurement.[312] Students worked in small groups, conducted experiments at home, and presented their results via Zoom. Analysis of grades over multiple years showed improved student performance during the smartphone-integrated online semesters compared to pre-pandemic traditional lab cohorts. Students responded positively to the flexibility, interactivity, and accessibility of the new lab format.

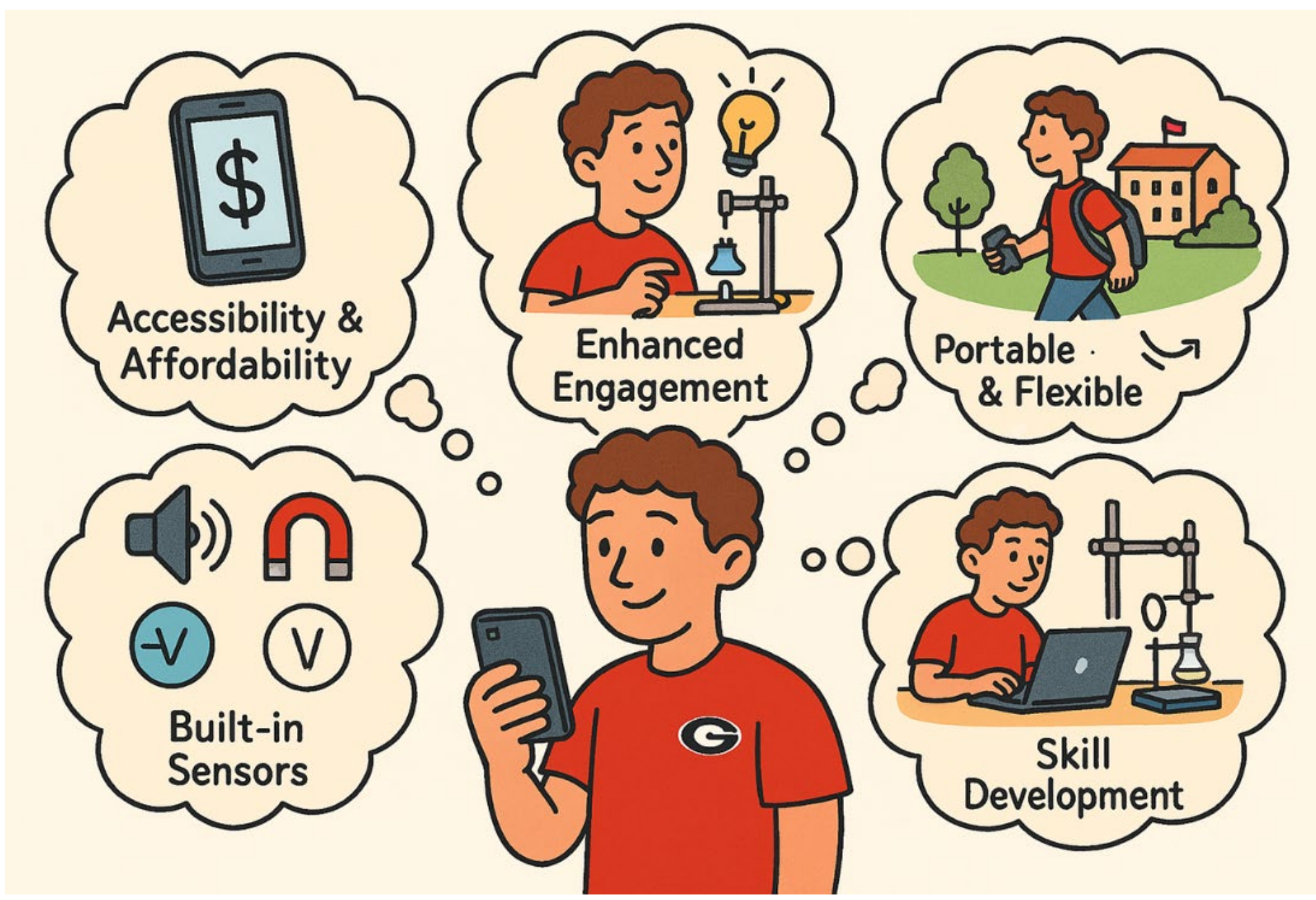

**Figure 5** Key educational benefits of SmartIPLs.

Based on those assessments, the shift from traditional to hybrid and smartphone-integrated labs can be characterized by clear advantages and a few persistent challenges. **Table 17** compares key features of smartphone-based and traditional lab formats, while **Figure 5** highlights the core educational benefits of SmartIPLs. These include greater accessibility and affordability, enhanced student engagement, portability and flexibility for learning in diverse settings, hands-on skill development, and the use of built-in smartphone sensors for real-time data collection. SmartIPLs also support remote learning and promote student independence through open-ended experimentation. Meanwhile, traditional labs maintain strengths in standardized equipment and structured instructional support, but often face limitations in terms of scheduling, location, and resource availability.

**Table 17.** Comparative features of SmartIPL *versus* traditional physics labs.

| Feature | SmartIPLs | Traditional Physics Labs |
|---|---|---|
| Accessibility | Can be done anywhere (home/classroom) using widely available devices (e.g., smartphones) [31, 311, 315] | Require physical attendance in well-equipped labs [31, 312] |
| Cost Efficiency | Uses built-in sensors and apps, reducing reliance on costly lab instruments [310, 311, 315] | Need specialized instruments (e.g., motion sensors, oscilloscopes) [312, 315] |
| Flexibility | Support asynchronous and remote participation [312, 315] | Fixed lab schedules; less adaptable to student needs[312] |
| Engagement | Encourages self-directed, real-world problem solving; boosts motivation[308, 311] | Often more rigid; limited student autonomy[31, 310] |
| Data Collection | Real-time data from accelerometers, gyroscopes, sound/light sensors [307, 311] | Data collected via dedicated hardware; often manual[315] |
| Experimental Skills | Enhances troubleshooting and hands-on learning; promotes creativity [307, 310] | Structured with limited student-designed elements[310] |
| Learning Outcomes | Comparable or better learning gains in SPS and conceptual understanding [310, 311] | Well-established effectiveness but limited flexibility[310] |
| Equity & Inclusivity | Familiarity with phones bridges gender gaps and supports diverse learners [31, 308] | Physical tools can intimidate students less confident in physics[308] |
| Instructor Workload | May reduce need for lab supervision; setup effort transferred to students [312, 315] | Requires TA/instructor guidance and equipment maintenance [312] |
| Assessment & Feedback | Apps like Phyphox, Tracker allow automatic data logging and instant feedback [309, 311] | Manual grading/report evaluation; less real-time interaction [309] |

In addition to logistical and structural benefits, empirical evidence shows that SmartIPLs lead to measurable educational gains. **Table 18** synthesizes core findings from recent assessments of SmartIPLs. Studies consistently report significant improvements in science process skills, comparable or better conceptual understanding, higher motivation and engagement, and improved gender equity through familiar, low-barrier technology. Autonomy and creativity were especially enhanced when students were tasked with designing or adapting experimental setups. However, many challenges remain: inconsistent sensor calibration, app-device compatibility, and the need for more structured guidance, particularly in early-stage deployments, are common concerns raised in the literature.

Together, these studies underscore the pedagogical value of SmartIPLs across diverse educational contexts. By combining accessibility with authentic inquiry and measurable learning outcomes, SmartIPLs not only modernize lab instruction but also promote equity, engagement, and scientific reasoning. Their success highlights the promise of learner-driven, mobile-first platforms to shape the future of STEM education.

**Table 18**. Summary of pedagogical outcomes from SmartIPL studies.

| Aspect | Finding | Data Support |
|---|---|---|
| **Science Process Skills (SPS)** | Improved significantly with smartphone-based experiments | SPS scores increased to a mean of 72.08 ($p < 0.05$) using Phyphox[311] |
| **Conceptual Understanding** | Comparable to traditional labs | No significant difference in E-CLASS scores and learning remained consistent[309, 310] |
| **Motivation & Engagement** | Significantly higher with smartphones | Perceived value strongly predicted motivation;[308] showed high student interest[310] |
| **Gender Equity** | Narrowed gender gaps | Girls felt more confident using familiar smartphone tools than traditional lab equipment.[308] |
| **Autonomy & Creativity** | Greater experimental freedom | Students performed better when designing and troubleshooting their own setups[307, 310] |
| **Challenges** | Inconsistent setups, need for guidance | Students often needed support in data analysis and experimental design[307, 312] |

## 5. Challenges and Mitigation Strategies

Despite their many advantages, SmartIPLs face a range of implementation challenges that must be addressed to ensure equitable and effective learning across diverse educational environments. These challenges can be broadly categorized into five areas: (1) measurement accuracy and calibration, (2) device heterogeneity and app compatibility, (3) equity and accessibility, (4) integration into large classrooms, and (5) digital well-being and cognitive support. Each area presents specific risks, but also offers pathways for mitigation through thoughtful design, scaffolding, and technology-supported interventions.

### 5.1. Measurement Accuracy and Calibration

One of the most commonly cited limitations in SmartIPL experiments, especially those involving video or camera-based tracking, is measurement error. Factors such as lighting, resolution, frame rate, and camera angle can significantly affect video-based motion analyses, leading to parallax distortion or timing inconsistencies, particularly in fast-moving systems like pendulums or projectiles (see **Table 9**).

Mitigation strategies include the use of tripods or phone stands for stabilization, proper alignment of the camera, the inclusion of scale references, and averaging results across multiple trials. Standardized best-practice tutorials, such as those provided in the UGA SmartPhone Intro Physics Lab YouTube Channel (https://www.youtube.com/@ugasmartphoneintrophysicsl1041) help reinforce these approaches.

Sensor-based experiments also require careful calibration. Misplaced phones in rotational setups or uncalibrated microphones can lead to significant errors. Instructors should emphasize calibration against known standards (e.g., using gravity for accelerometers) and provide 3D setup diagrams, printed templates, or augmented reality (AR) tools to help standardize alignment.

### 5.2 Device Heterogeneity and App Reliability

Smartphones differ significantly in their sensor performance, sampling rates, system integration, and compatibility across operating systems. This device heterogeneity could pose a significant challenge in sensor-based labs, where accelerometer drift, gyroscope instability, or inconsistent sampling frequencies can lead to unreliable or inconsistent data. For example, two students measuring acceleration on the same inclined plane may record different values depending on their device model or App behavior. Instructors should also be mindful of differences in app functionality across platforms. Some features available on Android versions of Apps like *Phyphox* or *SPARKvue* may be limited or missing on iOS, leading to disparities in lab execution and data interpretation.

To address these inconsistencies, educators can:

- Recommend well-tested, cross-platform apps (e.g., *Phyphox*, *Physics Toolbox*).
- Provide device-specific guidance and calibration routines.
- Encourage collaborative work for cross-validation among student groups.
- Implement cloud-based data submissions that allow instructors to verify results and identify anomalies.

These practices support consistent outcomes and reduce confusion stemming from hardware and software discrepancies.

### 5.3 Equity, Accessibility, and Digital Inclusion

While SmartIPLs are often praised for democratizing access to physics labs, this potential must be carefully balanced against real-world disparities in device access, internet connectivity, and app availability. Despite high smartphone ownership in some regions, access to compatible devices, reliable internet, and educational apps remains uneven, both globally and within affluent societies.[316, 317] Students from low-income households or rural areas may lack the smartphone quality, storage capacity, or connectivity needed to fully participate in SmartIPL activities.[318] In these contexts, reliance on advanced mobile technology may inadvertently deepen rather than bridge educational inequities.

To mitigate these disparities:

- Offering classroom loaner programs or peer device-sharing.
- Selecting free, offline-compatible apps that minimize data usage.
- Prioritizing camera-based experiments that require only basic video recording and analysis.
- Providing flexible options for participation, including extended deadlines and alternative submission formats (e.g., written reflections, screenshots, or photos of setup).
- Avoiding software that requires institutional licenses or desktop environments, such as MATLAB or GeoGebra, unless access is guaranteed for all students.

### 5.4 Cognitive Support and Digital Well-Being

Effective SmartIPL instruction must also consider students' cognitive load and psychological health associated with prolonged or unstructured screen use, especially among younger learners.[34, 319, 320] Excessive digital exposure has been linked to distraction, fatigue, and diminished psychological well-being. For early learners in particular, the overuse of mobile and interactive media can interfere with healthy sensory and cognitive development if not properly guided.[321] While SmartIPLs hold promise, they should be implemented thoughtfully with short, structured tasks that balance screen-based interaction with physical experimentation, reflection, and discussion.

Instructors should also consider the type and timing of guidance offered in SmartIPL-based learning, especially in virtual and remote contexts. Research has shown that effective guidance is critical for supporting student inquiry and avoiding cognitive overload in digital science labs.[322] In the context of SmartIPLs, this means embedding conceptual scaffolding, meaningful feedback, and structured opportunities for sense-making, which are especially important for developing critical thinking and scientific reasoning skills.[272]

Ultimately, equity in SmartIPLs requires more than technical design, it demands pedagogical intentionality. Educators must navigate complex terrain involving hardware access, learning contexts, attention spans, and digital wellbeing. When thoughtfully designed and implemented, SmartIPLs can provide meaningful, inclusive, and inquiry-based learning experiences but only if digital enthusiasm is matched by equity-driven planning and caution.

## 5.5 Integration into Large-Classroom Environments

The flexibility of SmartIPLs, allowing students to experiment independently, at home or in class, is both a strength and a challenge. In large, heterogeneous classrooms, this flexibility can lead to a wide range of experimental conditions.[23] For example, students conducting a pendulum lab may use strings of different lengths or materials, perform the experiment under varied lighting, or place the smartphone at inconsistent angles. Such diversity introduces variance in data quality, making instructor support and assessment more difficult.

To support scaling, instructors can

- use pre-lab setup templates to guide students while preserving creativity;
- tier support systems involving TAs, peer mentors, and discussion forums can facilitate troubleshooting;[276]
- assign group roles such as "data collector" or "analyst" encourages collaboration;
- grade with a flexible rubrics that emphasize process over precision accommodate experimental variability.

Tools like Google Sheets or collaborative platforms can be used for class-wide data aggregation and pattern recognition. Structured reporting formats and semi-automated feedback tools help instructors manage grading at scale.

Looking forward, artificial intelligence (AI) and open-source ecosystems hold promise for further supporting SmartIPL scalability. AI assistants could offer personalized feedback, identify anomalies in real time, and assist in setup validation. Open platforms may host reusable simulations, lab templates, and community-developed instructional guides, helping educators customize labs while maintaining instructional rigor (see **Section 6.3**).

## 6. Future Directions

The future of SmartIPLs holds immense promise beyond their initial surge during the COVID-19 pandemic. These tools have demonstrated lasting value by deepening conceptual understanding, boosting student engagement, and supporting flexible, accessible experimentation. As illustrated in **Figure 6**, the continued evolution of SmartIPLs will be shaped by six interconnected directions: curriculum integration, technical advancements, interdisciplinary applications, AI and data analysis, open-source platform development, and expanding access and inclusivity. Together, these directions reflect both the expanding capabilities of modern smartphones and the growing global demand for equitable, hands-on STEM education.

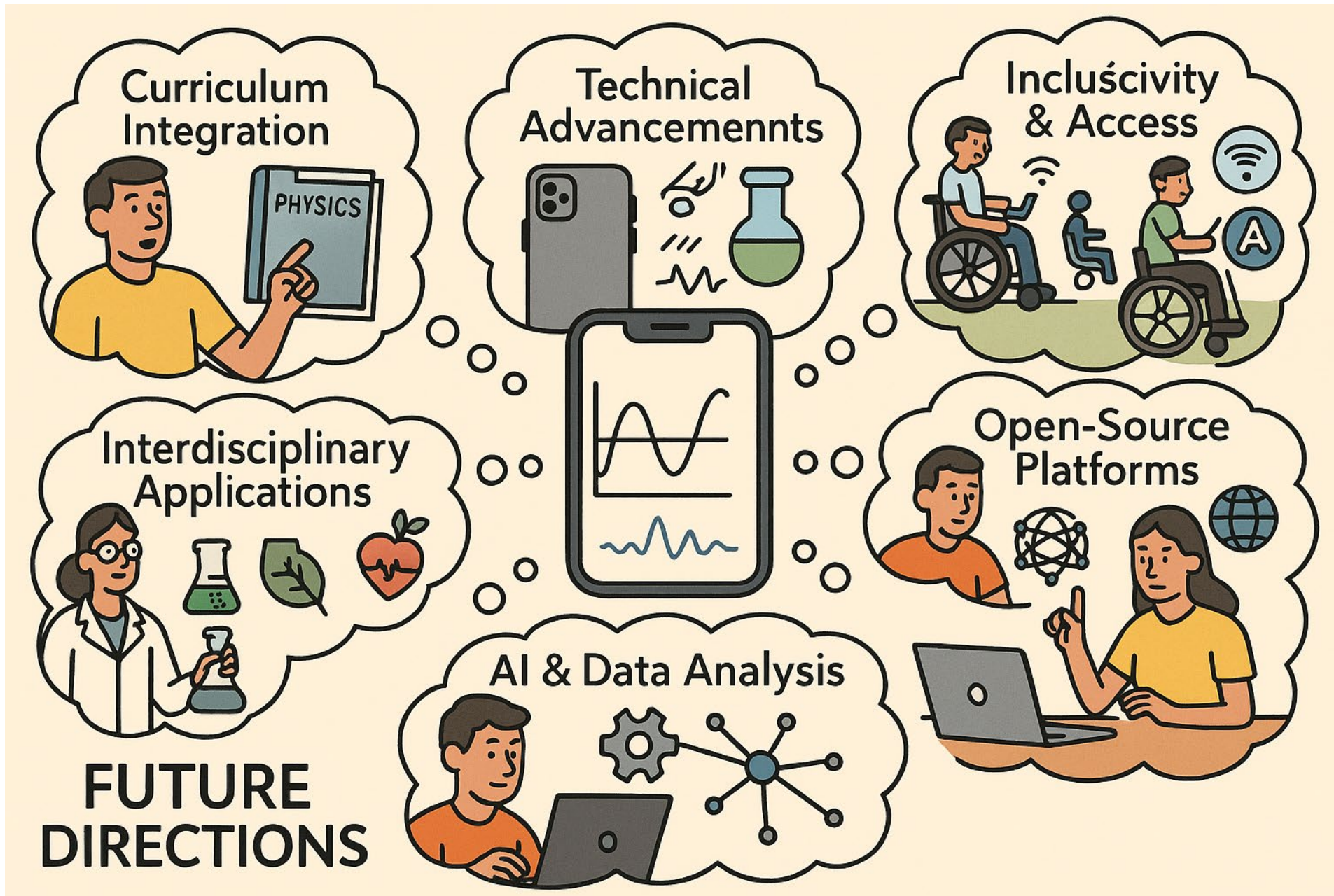

**Figure 6** Future directions for SmartIPLs: (1) Curriculum integration to expand coverage across underrepresented physics topics; (2) Technical advancements in sensors, imaging, and real-time processing; (3) Interdisciplinary applications bridging physics with chemistry, biology, engineering, and environmental science; (4) AI & data analysis to support personalized feedback and intelligent assessment; (5) Open-source platforms for resource sharing and collaboration; and (3) Inclusivity & access through low-spec compatibility and adaptive features.

### 6.1 Curriculum-Centered SmartIPL Development

The first major direction involves the systematic and comprehensive integration of SmartIPLs into physics curricula, ensuring that these labs align with instructional goals and fill content-specific gaps. As documented in **Table S1-S9**, while there is a wealth of SmartIPL experiments in mechanics, acoustics, and optics, other areas of the curriculum, such as fluid dynamics, electromagnetism, and modern physics, remain severely underrepresented.

For example, electrostatics, a foundational topic in high school and introductory college physics, has very few effective SmartIPL implementations (**Table 10**). In fluid dynamics, while surface tension and viscosity are sometimes addressed, experimental diversity and alignment with conceptual teaching remain limited (**Table 9**). The same is true for modern physics, where a few smartphone-based investigations touch on radiation or optics, but deeper explorations into quantum phenomena, atomic structure, or relativity are virtually nonexistent (**Table 13**). To address these content gaps, a coordinated research and development effort is needed to design SmartIPL modules that directly support course syllabi and learning outcomes. These modules should include clearly articulated learning objectives, experimental protocols, data analysis workflows, scaffolding tools, and aligned assessment rubrics. For successful classroom implementation, the materials must be adaptable across different learning environments—from high school to undergraduate physics courses—and include educator guides, safety notes, and multimedia supports. SmartIPLs should not remain as isolated enrichment tools but must become embedded into the instructional sequence, reinforcing key concepts and skill development within a coherent pedagogical framework.

Equally important is the alignment of SmartIPLs with national and international physics education standards (e.g., NGSS, AP Physics, or university-level learning outcomes), which can support their adoption at scale. Through partnerships among educators, researchers, and developers, it is possible to expand SmartIPL coverage to underrepresented domains, ensuring that students experience hands-on inquiry across the full spectrum of physics topics—not just those that happen to align with existing sensor capabilities.

### 6.2 Technical Advancements and Sensor Expansion

The next phase of SmartIPL development will be driven by ongoing advancements in smartphone hardware and software ecosystems. Today's smartphones already feature high-resolution cameras, multi-axis gyroscopes, sensitive microphones, barometers, and magnetometers. These components, in many cases, rival or even outperform traditional lab instruments in accessibility and convenience. Looking ahead, future devices are expected to deliver enhanced sensor precision, enabling more accurate measurements of motion, sound, light, pressure, and electromagnetic fields. Advancements in thermal and optical imaging will further support applications such as heat mapping, refractive index detection, and diffraction pattern analysis. In parallel, growing computational capabilities will allow for real-time processing of large data streams, integration of machine learning for pattern recognition, and high-resolution video analysis—all handled natively on the smartphone. Crucially, these hardware improvements will be complemented by a new generation of specialized physics apps. These tools will be capable of synchronizing multiple sensor inputs, conducting on-device data processing, visualizing trends in real time, and exporting results to cloud-based platforms for collaborative analysis and instruction. The integration of intuitive, app-based interfaces will empower students and instructors alike to design, execute, and interpret sophisticated experiments with minimal technical overhead.

These technological upgrades will significantly expand the experimental possibilities within physics education. In classical mechanics and acoustics, students will be able to explore complex systems such as damped or driven harmonic oscillators, interference patterns in sound waves, and motion dynamics with fine-grained temporal and spatial resolution. In thermal physics and fluid dynamics, enhanced imaging and computation will make it feasible to quantify temperature gradients, visualize fluid flow, and capture bubble dynamics using high-speed video and image analysis. For electromagnetism and electronics, improved sensors and external probe compatibility will enable detailed investigations of magnetic field distributions, induced EMF, and voltage-current characteristics in circuits. In optics, better camera resolution and image processing will support accurate studies of polarization, interference fringes, and diffraction phenomena. Meanwhile, modern physics and astronomy stand to benefit from the integration of smartphone-based spectroscopy, low-cost radiation detection, and real-time tracking of celestial events through image analysis. These capabilities, once confined to specialized labs, are rapidly becoming feasible in portable, user-friendly formats—broadening access to authentic experimental experiences and empowering learners to engage with complex scientific concepts through their everyday devices.

### 6.3 Future Integration of AI in SmartIPLs

As SmartIPLs generate increasingly rich datasets, the integration of AI, machine learning, and data science offers a promising direction for enhancing experimental physics education. Rather than a speculative add-on, AI is emerging as a viable tool for augmenting instruction, provided its applications are thoughtfully designed, critically evaluated, and aligned with pedagogical goals. In physics education, AI technologies have been used to support adaptive learning, real-time feedback, intelligent tutoring, and automated assessment.[323-325] Applied to SmartIPLs, these tools can help students detect inconsistencies in measurements, improve data interpretation, and receive timely, personalized guidance; all while supporting, not replacing, the instructional role of educators.[326, 327] AI systems can also enhance assessment by analyzing lab reports, sensor data, and modeling outputs to evaluate procedural accuracy, detect conceptual misunderstandings, and assess alignment between data and conclusions. This could generate comparable performance metrics across large student cohorts, enabling instructors to identify trends, support at-risk learners, and target instruction more effectively.[327, 328]

Despite these advances, empirical research specific to AI use in undergraduate physics labs remains limited, and concerns about algorithmic bias, data privacy, and transparency must be addressed.[329, 330] Thus, AI in SmartIPLs should currently be viewed as an evolving research direction rather than an established best practice. Nonetheless, introducing students to algorithmic thinking, computational modeling, and responsible data analysis within experimental settings aligns with broader STEM education goals. As SmartIPLs continue to evolve, the integration of intelligent systems may offer a path toward more authentic, inquiry-based, and data-informed science learning.

### 6.4 Open-Source Platforms and Curriculum Sharing

To sustain the momentum of SmartIPLs and expand their global reach, a centralized, open-source platform is needed to support the development, dissemination, and refinement of resources. Currently, materials are fragmented across journal articles, institutional sites, YouTube channels, and private collections, creating barriers for educators who wish to discover, adopt, or contribute high-quality instructional content. For both students and educators, the value of SmartIPLs lies not only in portability or affordability but in transforming passive labs into active, inquiry-driven experiences. Achieving this at scale requires a collaborative infrastructure where educators and learners can adapt, contribute, and build on shared practices. An open-source SmartIPL platform could provide this infrastructure by serving as a living repository of experiments, pedagogical tools, and community expertise. Rather than simply storing static content, it would amplify proven models such as "Dorm Room Labs"[25] which use smartphones and low-cost materials for hands-on exploration, and large-enrollment online labs[28], which maintained inquiry and collaboration during pandemic-era teaching.

Embedding Open Educational Resources (OER) principles into such a platform would further enhance access and innovation. OER have been shown to reduce costs, support localization, and broaden equity.[331, 332] More importantly, Open Educational Practices (OEP) encourage learners and educators to co-create, remix, and innovate through shared communities.[333, 334] A SmartIPL ecosystem that embraces OER and OEP could foster both scalability and pedagogical depth, particularly in resource-limited or remote contexts.

The platform itself could host searchable and editable lab guides, scaffolded inquiry prompts, video demonstrations, and analysis templates. It would support flexible levels of guidance, from step-by-step protocols to open-ended frameworks, while enabling learners to upload adaptations, datasets, and reflections. Educators could remix materials, share implementation tips, and exchange feedback in forums, fostering a culture of participatory teaching and learning. Additional features might include version-controlled repositories, reflection prompts, peer-review tools, and open-data experiments that facilitate cross-institutional collaboration.

These efforts would build on existing models while leaving room for future growth. For instance, AI-powered tools trained on community-generated content could provide optional support for experiment design, troubleshooting, or data interpretation, always as complements, not replacements, for human inquiry. Ultimately, a robust open SmartIPL platform would empower educators and learners alike, strengthening the collective capacity to deliver hands-on, inquiry-based science education. By lowering barriers and encouraging co-creation, it would transform smartphones into gateways for exploratory, transparent, and meaningful laboratory experiences worldwide.

### 6.5 Expanding Access, Inclusivity, and Public Engagement

The rapid growth of SmartIPLs has highlighted their potential to transform not only formal science education, but also informal learning and public engagement with physics. As SmartIPLs continue to evolve, the primary challenge is no longer technical feasibility, but ensuring they are equitable, inclusive, and inspiring. To maximize their impact, SmartIPLs must be accessible across socioeconomic and geographic contexts, adaptable to a range of educational levels, and engaging for both students and the general public.

Although smartphones are globally prevalent, disparities remain in terms of device capabilities, internet access, and user familiarity. To ensure equitable access, SmartIPLs must be designed with low-spec device compatibility and offline functionality to support learners with limited hardware or connectivity. Flexible experimental setups that rely on household or low-cost materials can enable meaningful experimentation without compromising scientific rigor. Additionally, SmartIPLs should incorporate multilingual interfaces and accessibility features to support learners with diverse language needs or physical challenges, such as visual or motor impairments.

Equity in SmartIPLs also involves tailoring content to different educational stages. For college and high school students, SmartIPLs can offer rich learning opportunities through theoretical modeling, quantitative analysis, and open-ended inquiry. For middle school learners, smartphone tools can be integrated into structured activities that help students recognize physical patterns through visual, auditory, and motion-based feedback, supporting the development of conceptual understanding through experimentation. For elementary students, kindergarteners, and the general public, SmartIPLs can provide accessible entry points into physics through observation-driven exploration and sensory-rich engagement, connecting movement, sound, and imagery to physical principles in ways aligned with early developmental needs. These uses are not intended to replace embodied or tactile experimentation, but rather to serve as inclusive, adaptable extensions of hands-on science, especially in settings where traditional lab infrastructure may be limited.

Beyond the classroom, SmartIPLs can hold promise for integration into competitive, project-based STEM initiatives. Their affordability and adaptability will make them ideal for challenges such as those found in the FIRST LEGO League, where students can design and test experiments, analyze sensor data, and present findings collaboratively. These formats will promote creativity, teamwork, and inclusivity, allowing SmartIPLs to serve as a gateway to deeper STEM exploration.

SmartIPLs could also be a powerful tool for public engagement, particularly when applied to real-world scenarios such as sports and everyday activities. With thoughtfully designed apps and guided activities, families and individuals could explore physics through hands-on, relatable experiences. For example, users could track the arc of a soccer ball or the hang time of a basketball dunk using video analysis, measure sprint reaction times or rotational motion in gymnastics with smartphone accelerometers, or analyze sound levels in a stadium using built-in microphones. These applications can turn familiar experiences into opportunities for scientific inquiry, sparking curiosity across age groups.

By blending physical intuition with sensor technology, SmartIPLs have the potential to make physics not just more accessible, but more exciting and relevant to daily life. Whether in classrooms, homes, or on sports fields, they can empower users to see the world through a scientific lens—fueling curiosity, fostering learning, and broadening participation in the scientific enterprise.

## 7. Conclusions

Over the past decade, SmartIPLs have transitioned from innovative teaching experiments into legitimate, scalable, and pedagogically powerful tools that support inquiry-based science education across a broad spectrum of learning environments. Drawing on built-in sensors, high-resolution cameras, and accessible apps, smartphones have empowered students and educators alike to conduct meaningful, hands-on experiments using ubiquitous, everyday technology. This review has systematically documented nearly 200 SmartIPL experiments spanning mechanics, optics, acoustics, thermodynamics, electromagnetism, and modern physics, while highlighting both their educational benefits and implementation challenges.

SmartIPLs offer a compelling blend of accessibility, affordability, flexibility, and interactivity, positioning them as a transformative solution for experimental instruction—particularly in resource-limited or remote learning contexts. Sensor-based experiments can provide real-time data acquisition for high-frequency, dynamic phenomena, while camera-based methods will foster deep conceptual modeling through video and image analysis. Together, these approaches support skill development in data collection, uncertainty analysis, graphical modeling, and critical reasoning.

Evidence from assessment studies demonstrates that SmartIPLs yield comparable or superior outcomes to traditional labs in terms of conceptual understanding, science process skills, student motivation, and gender equity. Yet the full potential of these labs remains untapped, especially in underrepresented content areas such as electrostatics, fluid dynamics, and modern physics. Addressing these gaps will require systematic curriculum design, expansion into upper-level and interdisciplinary domains, and the creation of age-appropriate lab pathways that serve students from elementary to undergraduate levels.

While this review focuses on applications of SmartIPLs in physics education, their adoption is also expanding in related STEM disciplines such as chemistry,[238, 298-300, 335] biology, [136, 137, 301-303] engineering, [96, 99, 254] and environmental & health science.[132, 139, 186, 188] These interdisciplinary extensions, often grounded in physical principles and smartphone sensor technologies, offer promising directions for future research and cross-curricular integration.

The integration of AI and data science into SmartIPLs could represent a critical next step. AI-powered tools can offer real-time feedback, automate analysis, and evolve by learning from student behaviors, acting as intelligent teaching assistants that personalize support and improve assessment. Furthermore, interdisciplinary applications in chemistry, biophysics, engineering, environmental science, and medical physics show that SmartIPLs are not limited to physics alone, but can foster cross-disciplinary thinking and STEM workforce readiness.

To scale adoption and innovation, SmartIPLs must be supported by open-source platforms and educator communities. Resources like the UGA SmartPhone Intro Physics Lab YouTube channel, UGA Modern Optics: Smartphone Projects, and the open-access SPIE book *Use of Smartphones in Optical Experimentation* provide foundational tools for global collaboration. These initiatives should be expanded into centralized repositories that facilitate content sharing, adaptation, and continuous improvement.

As educational institutions continue to explore blended and remote learning models, SmartIPLs could provide a scalable and inclusive platform for broadening access to quality experimental learning. Future work should continue to evaluate their pedagogical impact through controlled studies, address equity and digital wellness considerations, and explore their synergy with emerging technologies such as AI-driven feedback systems and adaptive learning environments.

Finally, SmartIPLs hold great promise not only for formal education but also for public engagement with science. Apps and activities that capture the physics of sports, music, and daily life can inspire curiosity in learners of all ages and backgrounds. By designing inclusive, culturally relevant, and competition-friendly lab experiences, educators can bridge the gap between science and society.

In summary, SmartIPLs are no longer a novelty, they represent a paradigm shift in physics (or even STEM) education, making experimentation more engaging, equitable, and responsive to the needs of 21st-century learners. As smartphone technology advances and educator networks grow, SmartIPLs are poised to become the foundation of a more interactive, intelligent, and inclusive future for experimental science learning.

**Acknowledgments**: The author was supported by the Learning Technology Grant of University of Georgia.

**The authors have no conflicts to disclose.**

## References

1. E. A. Dare, K. Keratithamkul, B. M. Hiwatig and F. Li, "Beyond content: The role of STEM disciplines, real-world problems, 21st century skills, and STEM careers within science teachers' conceptions of integrated STEM education," Education Sciences **11** (11), 737 (2021).
2. American Association of Physics Teachers, "Goals of the Introductory Physics Laboratory," American Journal of Physics **66** (6), 483-485 (1998).
3. N. Holmes and C. Wieman, "Introductory physics labs: We can do better," Physics Today **71** (1), 38 (2018).
4. J. Piaget and M. Cook, *The origins of intelligence in children*. (International universities press New York, 1952).
5. L. S. Vygotsky, *Mind in society: The development of higher psychological processes*. (Harvard university press, 1978).
6. A. Collins, J. S. Brown and A. Holum, "Cognitive Apprenticeship: Making thinking visible. ," American Educator **15**, 6-11, 38-46 (1991).
7. D. H. Jonassen and S. M. Land, *Theoretical foundations of learning environments*. (Routledge, New York, 2012).
8. National Science Teachers Association, 2019.
9. C. Wieman and H. Perkins, "Transforming physics education," Physics Today **58**, 36 (2005).
10. C. Wieman, in *Forum for the Future of Higher Education* (2008), pp. 61.
11. J. M. Wilson, "The CUPLE Physics Studio," Physics Teacher **32**, 518-523 (1994).
12. P. W. Laws, "Millikan Lecture 1996: Promoting active learning based on physics education research in introductory physics courses. ," Am J Phys **65**, 14-21 (1997).
13. J. W. Belcher, 2001.
14. H. Roy, "Studio vs Interactive Lecture Demonstration – Effects on Student Learning," Bioscene **29**, 3-6 (2003).
15. E. F. Redish, *Teaching Physics With the Physics Suite*. (John Willey & Sons, 2003).
16. R. R. Hake, in *Handbook of Design Research Methods in Mathematics, Science, and Technology Education*, edited by A. E. Kelly, R. A. Lesh and J. Y. Baek (Routledge, Erlbaum, 2008).
17. P. J. Enderle, S. A. Southerland and J. A. Grooms, "Exploring the context of change: Understanding the kinetics of a studio physics implementation effort," PHYSICAL REVIEW SPECIAL TOPICS - PHYSICS EDUCATION RESEARCH **9**, 010114 (2013).
18. R. J. Beichner and J. M. Saul, in *Successful Pedagogies* (Project Kaleidoscope), pp. 61-66.
19. P. W. Laws, "Calculus-based physics without lectures," Physics today **44** (12), 24-31 (1991).
20. K. Cummings, J. Marx, R. Thornton and D. Kuhl, "Evaluating innovation in studio physics," American journal of physics **67** (S1), S38-S44 (1999).
21. R. Beichner, L. Bernold, E. Burniston, P. Dail, R. Felder, J. Gastineau, M. Gjertsen and J. Risley, "Case study of the physics component of an integrated curriculum," American Journal of Physics **67** (S1), S16-S24 (1999).
22. L. Darmendrail, Université de Genève, 2023.
23. Y. Zhao and Y. S. Phang, *Use of smartphones in optical experimentation*. (SPIE Press, 2022).
24. M. Á. González, M. Á. González, M. E. Martín, C. Llamas, Ó. Martínez, J. Vegas, M. Herguedas and C. Hernández, "Teaching and learning physics with smartphones," Journal of cases on Information Technology (JCIT) **17** (1), 31-50 (2015).

25. M. B. Moldwin, "Dorm room labs for introductory large-lecture science classes for nonscience majors," Journal of College Science Teaching **47** (5), 36-41 (2018).
26. S. Wye, "Teaching remote laboratories using smart phone sensors: determining the density of air," Physics Education **58** (1), 015002 (2022).
27. C. Stampfer, H. Heinke and S. Staacks, "A lab in the pocket," Nature Reviews Materials **5** (3), 169-170 (2020).
28. K. Koenig and S. Bake, "Maintaining essential course features when moving introductory physics labs online," The Physics Teacher **59** (6), 394-396 (2021).
29. M. F. J. Fox, J. R. Hoehn, A. Werth and H. J. Lewandowski, "Lab instruction during the COVID-19 pandemic: Effects on student views about experimental physics in comparison with previous years," Physical Review Physics Education Research **17** (1) (2021).
30. P. Klein, L. Ivanjek, M. N. Dahlkemper, K. Jeličić, M. A. Geyer, S. Küchemann and A. Susac, "Studying physics during the COVID-19 pandemic: Student assessments of learning achievement, perceived effectiveness of online recitations, and online laboratories," Physical Review Physics Education Research **17** (1) (2021).
31. S. Z. Lahme, P. Klein, A. Lehtinen, A. Müller, P. Pirinen, L. Rončević and A. Sušac, "Physics lab courses under digital transformation: A trinational survey among university lab instructors about the role of new digital technologies and learning objectives," Physical Review Physics Education Research **19** (2), 1-27 (2023).
32. E. Tufino and G. Organtini, "Evaluation of the effectiveness of an introductory mechanics Lab with Arduino and smartphone," arXiv preprint arXiv:2311.00606 (2023).
33. S. Gill, (https://prioridata.com/data/smartphone-stats/#:~:text=As%20of%202024%2C%20there%20are,second%20with%20659%20million%20users., 2025).
34. J. M. Twenge and W. K. Campbell, "Associations between screen time and lower psychological well-being among children and adolescents: Evidence from a population-based study," Preventive medicine reports **12**, 271-283 (2018).
35. A. K. Przybylski and N. Weinstein, "Digital screen time limits and young children's psychological well-being: Evidence from a population-based study," Child development **90** (1), e56-e65 (2019).
36. E. C. Hall, *Journey to the moon: the history of the Apollo guidance computer*. (Aiaa, 1996).
37. M. L. Simmons and H. J. Wasserman, "Performance comparison of the CRAY-2 and CRAY X-MP/416 supercomputers," The Journal of Supercomputing **4**, 153-167 (1990).
38. K. Yao, (https://www.hardwarezone.com.sg/feature-apple-iphone16-a18-pro-performance-review, 2024), Vol. 2025.
39. P. Vogt and J. Kuhn, "Analyzing free fall with a smartphone acceleration sensor," Phys. Teach. **50**, 182–183 (2012).
40. J. Kuhn and P. Vogt, "Smartphones as experimental tools: Different methods to determine the gravitational acceleration in classroom physics by using everyday devices," European Journal of Physics Education **4** (1), 16-27 (2013).
41. A. Abdulayeva, presented at the 2021 IEEE International Conference on Smart Information Systems and Technologies (SIST), 2021 (unpublished).
42. A. Mazzella and I. Testa, "An investigation into the effectiveness of smartphone experiments on students’ conceptual knowledge about acceleration," Physics education **51** (5), 055010 (2016).

43. P. Vogt and J. Kuhn, "Acceleration sensors of smartphones," Frontiers in Sensors **2**, 1-9 (2014).
44. J. S. Ardid-Ramírez, S. Marquez and M. Ardid Ramírez, "Use of sound recordings and analysis for physics lab practices," INTED2021 Proceedings, 7687-7693 (2021).
45. P. Vogt, J. Kuhn and S. Müller, "Experiments Using Cell Phones in Physics Classroom Education: The Computer-Aided g Determination," The Physics Teacher **49** (6), 383-384 (2011).
46. R. Hurtado-Velasco, Y. Villota-Narvaez, D. Florez and H. Carrillo, "Video analysis-based estimation of bearing friction factors," European Journal of Physics **39** (6) (2018).
47. R. P. Smith and E. H. Matlis, "Gravity-driven fluid oscillations in a drinking straw," American Journal of Physics **87** (6), 433-435 (2019).
48. H. Crompton, in *Handbook of mobile learning*, edited by Z. L. Berge and L. Y. Muilenburg (Routledge, Florence, KY, 2013), pp. 3–14.
49. M. L. Crescente and D. Lee, "Critical issues of m-learning: design models, adoption processes, and future trends," Journal of the Chinese Institute of Industrial Engineers **28**, 111–123 (2011).
50. R. Robinson and J. Reinhart, *Digital Thinking and Mobile Teaching: Communicating, Collaborating, and Constructing in an Access Age*. (Bookboon, Denmark, 2014).
51. T. G. and R. M., *Using Network and Mobile Technology to Bridge Formal and Informal Learning*. (Woodhead/Chandos Publishing Limited, Cambridge, UK, 2013).
52. R. Oller, in *EDUCAUSE Center for Analysis and Research (Research Bulletin)* (2012).
53. I. E. Allen and J. Seaman, 2013.
54. J. Kuhn and P. Vogt, "Analyzing spring pendulum phenomena with a smart-phone acceleration sensor," The Physics Teacher **50** (8), 504-505 (2012).
55. P. Vogt and J. Kuhn, "Analyzing radial acceleration with a smartphone acceleration sensor," Physics Teacher **51** (3), 182-183 (2013).
56. J. A. Sans, F. J. Manjón, A. L. J. Pereira, J. A. Gomez-Tejedor and J. A. Monsoriu, "Oscillations studied with the smartphone ambient light sensor," European Journal of Physics **34** (6), 1349-1354 (2013).
57. E. Arribas, I. Escobar, C. P. Suarez, A. Najera and A. Beléndez, "Measurement of the magnetic field of small magnets with a smartphone: a very economical laboratory practice for introductory physics courses," European Journal of Physics **36** (6) (2015).
58. M. Monteiro and A. C Martí, "Using smartphone pressure sensors to measure vertical velocities of elevators, stairways, and drones," Physics Education **52** (1), 015010 (2017).
59. M. R. Silva, P. Martín-Ramos and P. P. da Silva, "Studying cooling curves with a smartphone," The Physics Teacher **56** (1), 53-55 (2018).
60. P. Onorato, T. Rosi, E. Tufino, C. Caprara and M. Malgieri, "Quantitative experiments in a distance lab: studying blackbody radiation with a smartphone," European Journal of Physics **42** (4) (2021).
61. P. Onorato, T. Rosi, E. Tufino, S. Toffaletti and M. Malgieri, "Selective Light Transmittance in a Glue Stick During a Distance Lab," The Physics Teacher **62** (3), 219-222 (2024).
62. J. Kuhn and P. Vogt, "Smartphones as mobile minilabs in physics," Smartphones as Mobile Minilabs in Physics (2022).
63. P. Onorato and L. M. Gratton, "Measuring the Raman spectrum of water with a smartphone, laser diodes and diffraction grating," European Journal of Physics **41** (2) (2020).

64. S. Gröber, A. Molz and J. Kuhn, "Using smartphones and tablet PCs for β−-spectroscopy in an educational experimental setup," European Journal of Physics **35** (6), 065001 (2014).
65. R. E. Vieyra, C. Megowan-Romanowicz, D. J. O'Brien, C. Vieyra and M. C. Johnson-Glenberg, in *Theoretical and Practical Teaching Strategies for K-12 Science Education in the Digital Age* (2023), pp. 131-152.
66. P. Pirinen, P. Klein, S. Z. Lahme, A. Lehtinen, L. Rončević and A. Susac, "Exploring digital signal processing using an interactive Jupyter notebook and smartphone accelerometer data," European Journal of Physics **45** (1) (2023).
67. F. Bouquet, C. Dauphin, F. Bernard and J. Bobroff, "Low-cost experiments with everyday objects for homework assignments," Physics Education **54** (2), 025001 (2019).
68. C. E. Wieman, W. K. Adams and K. K. Perkins, "PhET: Simulations that enhance learning," Science **322** (5902), 682-683 (2008).
69. B. Thaller, *Visual quantum mechanics: selected topics with computer-generated animations of quantum-mechanical phenomena*. (Springer Science & Business Media, 2007).
70. P. Falstad, "Math and physics applets," URL: http://www. falstad. com/mathphysics. html (2006).
71. L. A. Testoni and G. Brockington, "The use of smartphones to teach kinematics: An inexpensive activity," Physics Education **51**, 063008 (2016).
72. A. Y. Nuryantini, A. Sawitri and B. W. Nuryadin, "Constant speed motion analysis using a smartphone magnetometer," Physics Education **53** (6) (2018).
73. C. Karakotsou and I. Zafiriadis, "Teaching uniform linear motion using an Arduino sensor and a smartphone device," Physics Education **58** (2) (2023).
74. P. Martín Ramos, M. Ramos Silva and P. S. P. d. Silva, "Smartphones in the teaching of Physics Laws: Projectile motion," RIED. Revista Iberoamericana de Educación a Distancia (2017).
75. J. Kuhn, P. Vogt and F. Theilmann, "Going nuts: Measuring free-fall acceleration by analyzing the sound of falling metal pieces," The Physics Teacher **54** (3), 182-183 (2016).
76. J. Kim, L. Bouman, F. Cayruth, C. Elliott, B. Francis, E. Gogo, C. Hyman, A. Marshall, J. Masters and W. Olano, "A measurement of gravitational acceleration using a metal ball, a ruler, and a smartphone," The Physics Teacher **58** (3), 192-194 (2020).
77. J. Zhang, J. Zhang, Q. Chen, X. Deng and W. Zhuang, "Measurement of g Using a Steel Ball and a Smartphone Acoustic Stopwatch," The Physics Teacher **61** (1), 74-75 (2023).
78. K. Forinash and R. F. Wisman, "Photogate timing with a smartphone," The Physics Teacher **53** (4), 234-235 (2015).
79. S. Kapucu, "Finding the acceleration and speed of a light-emitting object on an inclined plane with a smartphone light sensor," Physics Education **52** (5) (2017).
80. P. Pathak and Y. Patel, "Analyzing a Free-Falling Magnet to Measure Gravitational Acceleration Using a Smartphone's Magnetometer," The Physics Teacher **60** (6), 441-443 (2022).
81. M. Santamaría Lezcano, E. S. Cruz de Gracia and T. J. A. Mori, "Gravitational Acceleration—A Smartphone Approach with the Magnetic Ruler," The Physics Teacher **62** (3), 191-193 (2024).
82. J. Kuhn and P. Vogt, "Applications and examples of experiments with mobile phones and smartphones in physics lessons," Frontiers in Sensors **1** (4), 67-73 (2013).
83. S. L. Bevill and K. K. Bevill, 2015 (unpublished).

84. P. Klein, J. Kuhn, A. Müller and S. Gröber, in *Multidisciplinary research on teaching and learning* (Springer, 2015), pp. 270-288.
85. D. Howard and M. Meier, "Meeting laboratory course learning goals remotely via custom home experiment kits," The Physics Teacher **59** (6), 404-409 (2021).
86. O. Schwarz, P. Vogt and J. Kuhn, "Acoustic measurements of bouncing balls and the determination of gravitational acceleration," The Physics Teacher **51** (5), 312-313 (2013).
87. C. Baldock and R. Johnson, "Investigation of kinetic friction using an iPhone," Physics Education **51** (6), 065005 (2016).
88. S. Tsoukos, P. Lazos, P. Tzamalis, A. Kateris and A. Velentzas, "How Effectively Can Students' Personal Smartphones be Used as Tools in Physics Labs?," International Journal of Interactive Mobile Technologies **15** (14) (2021).
89. A. Çoban and M. Erol, "Teaching and determination of kinetic friction coefficient using smartphones," Physics Education **54**, 025019 (2019).
90. C. Fahsl and P. Vogt, "Determination of the drag resistance coefficients of different vehicles," The Physics Teacher **56** (5), 324-325 (2018).
91. P. Pathak and Y. Patel, "Determination of the friction coefficient of an inclined plane using the Doppler effect and smartphones," Physics Education **55** (6) (2020).
92. S. Kapucu, "A simple experiment to measure the maximum coefficient of static friction with a smartphone," Physics Education **53** (5), 053006 (2018).
93. M. Kousloglou, A. Molohidis, K. Nikolopoulou and E. Hatzikraniotis, "Mobile inquiry-based science learning (m-IBSL): Employment of the Phyphox application for an experimental study of friction," Teaching Science **68** (2), 14-18 (2022).
94. K. Hochberg, S. Gröber, J. Kuhn and A. Müller, "The spinning disc: studying radial acceleration and its damping process with smartphone acceleration sensors," Physics Education **49** (2), 137 (2014).
95. M. Loth, C. Gibbons, S. Belaiter and J. B. Clarage, "Stable and Unstable Rotational Dynamics of a Smartphone," Physics Teacher **55** (7), 431-434 (2017).
96. A. Shakur and J. Kraft, "Measurement of Coriolis Acceleration with a Smartphone," The Physics Teacher **54** (5), 288-290 (2016).
97. M. Monteiro, C. Cabeza and A. C. Martí, "Exploring phase space using smartphone acceleration and rotation sensors simultaneously," European Journal of Physics **35** (4) (2014).
98. M. Monteiro, C. Cabeza, C. Stari and A. C. Marti, "Smartphone sensors and video analysis: two allies in the physics laboratory battle field," Journal of Physics: Conference Series **1929** (1) (2021).
99. S. Z. Lahme, P. Klein, A. Lehtinen, A. Müller, P. Pirinen, A. Susac and B. Tomrlin, presented at the PhyDid B: Didaktik der Physik: Beiträge zur DPG-Frühjahrstagung, 2022 (unpublished).
100. M. Monteiro, C. Cabeza, A. C. Marti, P. Vogt and J. Kuhn, in *Smartphones as Mobile Minilabs in Physics* (2022), pp. 107-111.
101. M. Patrinopoulos and C. Kefalis, "Angular velocity direct measurement and moment of inertia calculation of a rigid body using a smartphone," Physics Teacher **53** (9), 564-565 (2015).
102. R. Pörn and M. Braskén, "Interactive modeling activities in the classroom—rotational motion and smartphone gyroscopes," Physics Education **51** (6) (2016).
103. M. S. Wheatland, T. Murphy, D. Naoumenko, D. v. Schijndel and G. Katsifis, "The mobile phone as a free-rotation laboratory," American Journal of Physics **89** (4), 342-348 (2021).

104. F. Tornaría, M. Monteiro and A. C. Marti, "Understanding coffee spills using a smartphone," Physics Teacher **52** (8), 502-503 (2014).
105. U. Pili and R. Violanda, "Measuring average angular velocity with a smartphone magnetic field sensor," The Physics Teacher **56** (2), 114-115 (2018).
106. A. Barrera-Garrido, "A smartphone inertial balance," Physics Teacher **55** (4), 248-249 (2017).
107. P. Klein, A. Müller, S. Gröber, A. Molz and J. Kuhn, "Rotational and frictional dynamics of the slamming of a door," American Journal of Physics **85** (1), 30-37 (2017).
108. D. Lopez, I. Caprile, F. Corvacho and O. Reyes, "Study of a variable mass Atwood's machine using a smartphone," The Physics Teacher **56** (3), 182-183 (2018).
109. I. Salinas, M. H. Gimenez, J. A. Monsoriu and J. A. Sans, "Demonstration of the parallel axis theorem through a smartphone," Physics Teacher **57** (5), 340-341 (2019).
110. C. Puttharugsa, S. Khemmani, P. Utayarat and W. Luangtip, "Investigation of the rolling motion of a hollow cylinder using a smartphone," European Journal of Physics **37** (5) (2016).
111. P. Wattanayotin, C. Puttharugsa and S. Khemmani, "Investigation of the rolling motion of a hollow cylinder using a smartphone's digital compass," Physics Education **52** (4), 045009 (2017).
112. U. Dilek and S. K. Şengören, "A new position sensor to analyze rolling motion using an iPhone," Physics Education **54** (4) (2019).
113. J. Carlos Castro-Palacio, L. Velázquez-Abad, M. H. Giménez and J. A. Monsoriu, "Using a mobile phone acceleration sensor in physics experiments on free and damped harmonic oscillations," American Journal of Physics **81** (6), 472-475 (2013).
114. U. Pili, "A dynamic-based measurement of a spring constant with a smartphone light sensor," Physics Education **53** (3) (2018).
115. M. H. Giménez, I. Salinas, J. A. Monsoriu and J. C. Castro-Palacio, "Direct Visualization of Mechanical Beats by Means of an Oscillating Smartphone," Physics Teacher **55** (7), 424-425 (2017).
116. D. Nichols, "Measuring Mass with a Rubber Band and a Smartphone," The Physics Teacher **60** (7), 608-609 (2022).
117. J. Kuhn, "Relevant Information About Using a Mobile Phone Acceleration Sensor in Physics Experiments," American Journal of Physics **82** (2), 94-94 (2014).
118. W.-K. Wong, T.-K. Chao, P.-R. Chen, Y.-W. Lien and C.-J. Wu, "Pendulum experiments with three modern electronic devices and a modeling tool," Journal of Computers in Education **2** (1), 77-92 (2015).
119. E. Momox and C. Ortega De Maio, "Computer-based learning in an undergraduate physics course: Interfacing a mobile phone and matlab to study oscillatory motion," American Journal of Physics **88** (7), 535-541 (2020).
120. J. C. Sanders, "The effects of projectile mass on ballistic pendulum displacement," American Journal of Physics **88** (5), 360-364 (2020).
121. S. Reinhold and M. Ziese, "Frequency doubling in a pendulum," European Journal of Physics **42** (2) (2021).
122. X. Deng, J. Zhang, Q. Chen, J. Zhang and W. Zhuang, "Measurement of g using a pendulum and a smartphone proximity sensor," The Physics Teacher **59** (7), 584-585 (2021).
123. A. Kaps, T. Splith and F. Stallmach, "Shear Modulus Determination Using the Smartphone in a Torsion Pendulum," Physics Teacher **59** (4), 268-271 (2021).

124. T. Splith, A. Kaps and F. Stallmach, "Phase plot of a gravity pendulum acquired via the MEMS gyroscope and magnetic field sensors of a smartphone," American Journal of Physics **90** (4), 314-316 (2022).
125. R. Mathevet, N. Lamrani, L. Martin, P. Ferrand, J. P. Castro, P. Marchou and C. M. Fabre, "Quantitative analysis of a smartphone pendulum beyond linear approximation: A lockdown practical homework," American Journal of Physics **90** (5), 344-350 (2022).
126. A.-M. Pendrill, "Serious physics on a playground swing—with toys, your own body, and a smartphone," The Physics Teacher **61** (5), 355-359 (2023).
127. J. Kuhn, P. Vogt and A. Müller, "Analyzing elevator oscillation with the smartphone acceleration sensors," Physics Teacher **52** (1), 55-56 (2014).
128. L. Tuset-Sanchis, J. C. Castro-Palacio, J. A. Gómez-Tejedor, F. J. Manjón and J. A. Monsoriu, "The study of two-dimensional oscillations using a smartphone acceleration sensor: example of Lissajous curves," Physics Education **50** (5), 580-586 (2015).
129. J. C. Castro-Palacio, L. Velázquez-Abad, F. Giménez and J. A. Monsoriu, "A quantitative analysis of coupled oscillations using mobile accelerometer sensors," European Journal of Physics **34** (3), 737-744 (2013).
130. Y. T. Wang, X. T. Duan, M. Z. Shao, C. L. Wang and H. Zhang, "An asymmetrical double torsion pendulum for studying coupled harmonic motion," American Journal of Physics **88** (9), 760-768 (2020).
131. M. Monteiro, C. Cabeza and A. C. Marti, "Rotational energy in a physical pendulum," The Physics Teacher **52** (3), 180-181 (2014).
132. M. Kubsch, J. Nordine and D. Hadinek, "Using smartphone thermal cameras to engage students' misconceptions about energy," The Physics Teacher **55** (8), 504-505 (2017).
133. T. Pierratos and H. M. Polatoglou, "Study of the conservation of mechanical energy in the motion of a pendulum using a smartphone," Physics Education **53** (1) (2018).
134. W. Namchanthra, S. Khemmani, S. Wicharn, S. Plaipichit, C. Pipatpanukul and C. Puttharugsa, "Analyzing a torsion pendulum using a smartphone's sensors: mechanical energy conservation approach," Physics Education **54** (6) (2019).
135. J. Kuhn and P. Vogt, "Analyzing acoustic phenomena with a smartphone microphone," Physics Teacher **51** (2), 118-119 (2013).
136. A. Müller, P. Vogt, J. Kuhn and M. Müller, "Cracking knuckles - A smartphone inquiry on bioacoustics," Physics Teacher **53** (5), 307-308 (2015).
137. M. A. González and M. A. González, "Smartphones as experimental tools to measure acoustical and mechanical properties of vibrating rods," European Journal of Physics **37** (4) (2016).
138. C. Florea, "Brief Analysis of Sounds Using a Smartphone," Physics Teacher **57** (4), 214-215 (2019).
139. E. Macho-Stadler and M. J. Elejalde-Garcia, "Measuring the Acoustic Response of Classrooms with a Smartphone," The Physics Teacher **58** (8), 585-588 (2020).
140. K. Ludwig-Petsch and J. Kuhn, "Shepard scale produced and analyzed with mobile devices," Physics Teacher **59** (5), 378-379 (2021).
141. O. Robin, "Teaching acoustics using smartphones," Canadian Acoustics **50** (3), 28-29 (2022).
142. A. Y. Nuryantini, R. Zakwandi and M. A. Ariayuda, in *THE 4TH INTERNATIONAL CONFERENCE ON LIFE SCIENCE AND TECHNOLOGY (ICoLiST)* (2023).
143. J. Allen, A. Boucher, D. Meggison, K. Hruby and J. Vesenka, "Inexpensive Audio Activities: Earbud-based Sound Experiments," The Physics Teacher **54** (8), 500-502 (2016).

144. S. H. Hawley and R. E. McClain, "Visualizing Sound Directivity via Smartphone Sensors," Physics Teacher **56** (2), 72-74 (2018).
145. M. Osorio, C. Pereyra, D. Gau and A. Laguarda, "Measuring and characterizing beat phenomena with a smartphone," European Journal of Physics **39** (2), 025708 (2018).
146. A. A. Gallitto, O. R. Battaglia, G. Cavallaro, G. Lazzara, L. Lisuzzo and C. Fazio, "Exploring historical scientific instruments by using mobile media devices," The Physics Teacher **60** (3), 202-206 (2022).
147. S. O. Parolin and G. Pezzi, "Kundt's tube experiment using smartphones," Physics Education **50** (4), 443-447 (2015).
148. M. Hirth, S. Gröber, J. Kuhn and A. Müller, "Harmonic Resonances in Metal Rods – Easy Experimentation with a Smartphone and Tablet PC," The Physics Teacher **54** (3), 163-167 (2016).
149. N. Z. Abidin and S. Tho, "The development of an innovative resonance experiment using smartphones with free mobile software applications for tertiary education," International Journal of Education and Development using ICT **14** (1) (2018).
150. A. A. Soares, R. F. Cantão, J. B. Pinheiro Jr and F. G. Castro, "Sound waves in a tube: measuring sound pressure profiles using smartphones," Physics Education **57** (5) (2022).
151. M. Monteiro, A. C. Marti, P. Vogt, L. Kasper and D. Quarthal, "Measuring the acoustic response of Helmholtz resonators," The Physics Teacher **53** (4), 247-249 (2015).
152. L. Darmendrail and A. Müller, "Helmholtz in the kitchen: a frying pan as a volume resonator," European Journal of Physics **41** (3) (2020).
153. M. Hirth, A. Müller and J. Kuhn, "Side window buffeting: a smartphone investigation on a car trip," European Journal of Physics **42** (6) (2021).
154. M. Monteiro, C. Stari and A. C. Martí, "A home-lab experiment: resonance and sound speed using telescopic vacuum cleaner pipes," Physics Education **58** (1) (2022).
155. M. Monteiro, C. Stari, C. Cabeza and A. C. Marti, "A bottle of tea as a universal Helmholtz resonator," The Physics Teacher **56** (9), 644-645 (2018).
156. A. Yavuz, "Measuring the speed of sound in air using smartphone applications," Physics Education **50** (3), 281-284 (2015).
157. S. Staacks, S. Hütz, H. Heinke and C. Stampfer, "Simple Time-of-Flight Measurement of the Speed of Sound Using Smartphones," The Physics Teacher **57** (2), 112-113 (2019).
158. S. Hellesund, "Measuring the speed of sound in air using a smartphone and a cardboard tube," Physics Education **54** (3) (2019).
159. T. Thongsuk and A. Intanin, "Smartphones as detector the speed of sound: A classroom explanation and demonstration," Journal of Physics: Conference Series **2145** (1) (2021).
160. P. Klein, M. Hirth, S. Gröber, J. Kuhn and A. Müller, "Classical experiments revisited: smartphones and tablet PCs as experimental tools in acoustics and optics," Physics Education **49**, 412 - 418 (2014).
161. J. A. Gómez-Tejedor, J. C. Castro-Palacio and J. A. Monsoriu, "The acoustic Doppler effect applied to the study of linear motions," European Journal of Physics **35** (2) (2014).
162. H. Zhao, X. Duan, S. Xie, L. Cheng, F. Wang, W. Xu, Y. Zhang, W. Zhang and C. Zhao, "Sensitive and selective evaluation of the 3-hydroxy-2-butanone biomarker via gas sensors based on WO3 nanomaterials with multiple morphologies," CYTA - Journal of Food **23** (1) (2025).

163. J. Haglund, F. Jeppsson, D. Hedberg and K. J. Schönborn, "Students' framing of laboratory exercises using infrared cameras," Physical Review Special Topics—Physics Education Research **11** (2), 020127 (2015).
164. J. Jiménez-Muñoz, J. Sobrino, G. Sòria, J. Delegido and S. Bañauls, "The role of emissivity during the cooling of a body: an experimental design for a laboratory classroom," European Journal of Physics **38** (1), 015102 (2016).
165. M. Vollmer and K.-P. Möllmann, presented at the Education and training in optics and photonics, 2017 (unpublished).
166. P. Martín-Ramos, M. Susano, P. S. P. d. Silva and M. R. Silva, in *Proceedings of the 5th International Conference on Technological Ecosystems for Enhancing Multiculturality* (2017), pp. 1-5.
167. A. Bonanno, G. Bozzo and P. Sapia, "Physics meets fine arts: a project-based learning path on infrared imaging," European Journal of Physics **39** (2), 025805 (2018).
168. M. Vollmer and K.-P. Möllmann, "Infrared cameras as accessories to smartphones: facts you need to know," Physics Education **53** (6), 065019 (2018).
169. C. P. Wong and R. Subramaniam, "Seeing in a different light—using an infrared camera to teach heat transfer and optical phenomena," Physics Education **53** (3), 035007 (2018).
170. S. Kang, H. Choi, S. Park, C. Park, J. Lee, U. Lee and S.-J. Lee, presented at the The 25th Annual International Conference on Mobile Computing and Networking, 2019 (unpublished).
171. X. Xu, M. Wu and X. Wang, "Smartphone visualization of thermal phenomena with thermal imaging accessories," Journal of Chemical Education **96** (11), 2545-2552 (2019).
172. S. Oss, "Infrared visualization of lumped and non-lumped thermal transient processes in an introductory laboratory," European Journal of Physics **42** (1) (2021).
173. S. Oss, "Vintage thermology and modern-day infrared imaging," Physics Education **56** (2), 025025 (2021).
174. F. Jiao and L. Jiao, presented at the Journal of Physics: Conference Series, 2022 (unpublished).
175. R. Jiang, X. Ding, S. H. Sung, D. Bulseco, C. Xie and C. Li, "A New Type of Interactive Video for Physics Education," The Physics Teacher **60** (8), 656-659 (2022).
176. A. Zhdanov and A. Pyayt, "Using a Mobile Phone to Demonstrate Thermal Properties of Materials," The Physics Teacher **60** (8), 660-662 (2022).
177. S. J. Hong, B. Im, J. Park, H. Kim and H.-G. Yoon, "Experimental investigation of the educational uses of thermal cameras," Physics Education **58** (1), 015016 (2022).
178. M. Dutra, M. Monteiro and A. C. Marti, "Cooling of an object by forced convection," European Journal of Physics **45** (4), 045101 (2024).
179. R. Guido, M. Dutra, M. Monteiro and A. C. Marti, "Seeing the invisible: convection cells revealed with thermal imaging," European Journal of Physics **44** (5) (2023).
180. A.-M. Pendrill, "With an infrared camera in an amusement park: heating and cooling of magnetic brakes," Physics Education **59** (1), 015033 (2024).
181. O. Robin, M. Courchesne, D. Derome, F. Turcotte and A. Bourget, "Mobile learning in engineering using smartphones: Two examples in acoustics and thermal courses," Proceedings of the Canadian Engineering Education Association (CEEA) (2024).
182. A. S. Thakur, S. Dubey, A. S. Thakur, V. Srivastava and R. Vaish, "Pencil-drawn tunable electrical resistance and Joule heating demonstration using a smartphone-based thermal imaging camera," Physics Education **59** (5), 055011 (2024).

183. C. de Cabo-Rodríguez, C. Torremocha-Serra, A. n. Ferradanes-Martínez, M. a. Gelpi, P. Dafonte-Rodríguez, L. Alonso-Marañón, S. Castro-García and J. M. Bermúdez-García, "Thermal Imaging for Teaching Materials Chemistry: Phase Change Materials toward Energy Transition," Journal of Chemical Education (2025).
184. J. Vandermarlière, "On the inflation of a rubber balloon," The Physics Teacher **54** (9), 566-567 (2016).
185. A. Müller, M. Hirth and J. Kuhn, "Tunnel pressure waves–A smartphone inquiry on rail travel," The Physics Teacher **54** (2), 118-119 (2016).
186. R. E. Vieyra, C. Vieyra and S. Macchia, "Kitchen Physics: Lessons in Fluid Pressure and Error Analysis," The Physics Teacher **55** (2), 87-90 (2017).
187. S. Macchia, "Analyzing Stevin's law with the smartphone barometer," The Physics Teacher **54** (6), 373-373 (2016).
188. M. Monteiro, P. Vogt, C. Stari, C. Cabeza and A. C. Marti, "Exploring the atmosphere using smartphones," The Physics Teacher **54** (5), 308-309 (2016).
189. M. Monteiro, F. Tornaría and A. C. Martí, "Experimental analysis of the free surface of a liquid in a rotating frame," European Journal of Physics **41** (3) (2020).
190. P. Dam-O, T. Eadkong and P. Channuie, "Lab at Home of fluid draining with Tracker," arXiv preprint arXiv:2203.07894 (2022).
191. N. A. Goy, Z. Denis, M. Lavaud, A. Grolleau, N. Dufour, A. Deblais and U. Delabre, "Surface tension measurements with a smartphone," Physics Teacher **55** (8), 498-499 (2017).
192. M. X. Wei, S. Huang, J. Wang, H. H. Li, H. J. Yang and S. H. Wang, "The study of liquid surface waves with a smartphone camera and an image recognition algorithm," European Journal of Physics **36** (6) (2015).
193. A. Delvert, P. Panizza and L. Courbin, "Measuring the viscosity of air with soapy water, a smartphone, a funnel, and a hose: An experiment for undergraduate physics students," American Journal of Physics **90** (1), 64-70 (2022).
194. J. S. Bobowski, "Hydraulic Analogues Illustrating the Charging of a Capacitor and Ohm's Law: Labs for Online Learning Environments," The Physics Teacher **59** (7), 560-565 (2021).
195. R. Hurtado-Gutiérrez and Á. Tejero, "Measuring capacitor charge and discharge using an LED and a smartphone," European Journal of Physics **44** (6) (2023).
196. J. R. Groff, "Estimating RC Time Constants Using Sound," The Physics Teacher **57** (6), 393-396 (2019).
197. J. Lincoln, "Biot–Savart law with a smartphone: Phyphox app," The Physics Teacher **62** (1), 72-73 (2024).
198. S. Arabasi and H. Al-Taani, "Measuring the Earth's magnetic field dip angle using a smartphone-aided setup: a simple experiment for introductory physics laboratories," European Journal of Physics **38** (2) (2017).
199. Y. Ogawara, S. Bhari and S. Mahrley, "Observation of the magnetic field using a smartphone," Physics Teacher **55** (3), 184-U194 (2017).
200. I. Escobar, R. Ramirez-Vazquez, J. Gonzalez-Rubio, A. Belendez and E. Arribas, "Magnetic Field of a Linear Quadrupole Using the Magnetic Sensors Inside the Smartphones," (2018).
201. I. Escobar, R. Ramirez-Vazquez, J. Gonzalez-Rubio, A. Belendez and E. Arribas, "Smartphones Magnetic Sensors within Physics Lab," (2018).
202. K. D. Sullivan, A. Sen and M. C. Sullivan, "Investigating the magnetic field outside small accelerator magnet analogs via experiment, simulation, and theory," American Journal of Physics **91** (6), 432-439 (2023).

203. M. Monteiro, C. Stari, C. Cabeza and A. C. Marti, "Magnetic field 'flyby' measurement using a smartphone's magnetometer and accelerometer simultaneously," Physics Teacher **55** (9), 580-581 (2017).
204. M. Pirbhai, "Smartphones and Tracker in the e/m experiment," Physics Education **55** (1), 015001 (2019).
205. A. Ouariach, M. El Hadi, A. El Moussaouy, A. Hachmi, H. Magrez and D. Bria, "Development of an educational low-cost teslameter by using Arduino and Smartphone application," Physics Education **55** (3) (2020).
206. A. A. Soares and T. O. Reis, "Studying Faraday's law of induction with a smartphone and personal computer," Physics Education **54** (5), 055006 (2019).
207. F. G. Tomasel and M. C. Marconi, "Rolling magnets down a conductive hill: Revisiting a classic demonstration of the effects of eddy currents," American Journal of Physics **80** (9), 800-803 (2012).
208. G. A. Sobral, "Development of a metal detector for smartphones and its use in the teaching laboratory," Physics Education **53** (4), 045006 (2018).
209. S. R. Pathare, M. K. Raghavendra and S. Huli, "Low-Cost Alternative for Signal Generators in the Physics Laboratory," The Physics Teacher **55** (5), 301-305 (2017).
210. K. Forinash and R. F. Wisman, "Smartphones as portable oscilloscopes for physics labs," The Physics Teacher **50** (4), 242-243 (2012).
211. R. Ramos and C. Devers, "The iPad as a virtual oscilloscope for measuring time constants in RC and LR circuits," Physics Education **55** (2), 023003 (2020).
212. I. Torriente-García, A. C. Martí, M. Monteiro, C. Stari, J. C. Castro-Palacio and J. A. Monsoriu, "RLC series circuit made simple and portable with smartphones," Physics Education **59** (1) (2023).
213. K. Forinash and R. F. Wisman, "Smartphones—Experiments with an External Thermistor Circuit," The Physics Teacher **50** (9), 566-567 (2012).
214. S. Dyer, presented at the 2018 ASEE Zone IV Conference, 2018 (unpublished).
215. C. Avilés-Cruz and J. Villegas-Cortez, "A smartphone-based augmented reality system for university students for learning digital electronics," Computer Applications in Engineering Education **27** (3), 615-630 (2019).
216. T. Nagy, G. Vadai and Z. Gingl, "Digital phonocardiographic experiments and signal processing in multidisciplinary fields of university education," European Journal of Physics **38** (5) (2017).
217. B. Cobb, presented at the Astronomical Society of the Pacific Conference Series, 2022 (unpublished).
218. J. A. Sans, J. Gea-Pinal, M. H. Gimenez, A. R. Esteve, J. Solbes and J. A. Monsoriu, "Determining the efficiency of optical sources using a smartphone's ambient light sensor," European Journal of Physics **38** (2) (2017).
219. I. Salinas, M. H. Giménez, J. A. Monsoriu and J. C. Castro-Palacio, "Characterization of linear light sources with the smartphone's ambient light sensor," Physics Teacher **56** (8), 562-563 (2018).
220. M. Monteiro, C. Stari, C. Cabeza and A. C. Martí, "The polarization of light and Malus' law using smartphones," The Physics Teacher **55** (5), 264-266 (2017).
221. T. Rosi and P. Onorato, "Video analysis-based experiments regarding Malus' law," Physics Education **55** (4) (2020).

222. L. J. Thoms, G. Colicchia and R. Girwidz, "Using the Naked Eye to Analyze Polarized Light From a Smartphone," Physics Teacher **59** (5), 337-339 (2021).
223. C.-M. Chiang and H.-Y. Cheng, "Use smartphones to measure Brewster's angle," The Physics Teacher **57** (2), 118-119 (2019).
224. A. Girot, N.-A. Goy, A. Vilquin and U. Delabre, "Studying Ray Optics with a Smartphone," The Physics Teacher **58** (2), 133-135 (2020).
225. M. C. Sullivan, "Using a smartphone camera to explore ray optics beyond the thin lens equation," American Journal of Physics **90** (8), 610-616 (2022).
226. J. Wang and W. Q. Sun, "Measuring the focal length of a camera lens in a smart-phone with a ruler," Physics Teacher **57** (1), 54-54 (2019).
227. J. Freeland, V. R. Krishnamurthi and Y. Wang, "Learning the lens equation using water and smartphones/tablets," The Physics Teacher **58** (5), 360-361 (2020).
228. Y. S. Phang and Y. Zhao, "Determining the Focal Length of Converging and Diverging Lenses Using a Smartphone," The Physics Teacher **60** (8), 703-705 (2022).
229. W. Baird, J. Secrest, C. Padgett, W. Johnson and C. Hagrelius, "Smartphones and Time Zones," The Physics Teacher **54** (6), 351-353 (2016).
230. M. Mitsushio, "Laboratory-on-a-Smartphone," Analytical Sciences **36** (2), 141-142 (2019).
231. K. Malisorn, S. Wicharn, S. Plaipichit, C. Pipatpanukul, N. Houngkamhang and C. Puttharugsa, "Demonstration of light absorption and light scattering using smartphones," Physics Education **55** (1) (2020).
232. A. Nuryantini, B. Nuryadin and U. Umairoh, presented at the IOP Conference Series: Materials Science and Engineering, 2021 (unpublished).
233. P. Onorato, T. Rosi, E. Tufino, C. Caprara and M. Malgieri, "Using smartphone cameras and ambient light sensors in distance learning: the attenuation law as experimental determination of gamma correction," Physics Education **56** (4) (2021).
234. K. V. Oskolok, O. V. Monogarova and A. V. Garmay, "Molecular Optical Analyzers Based on Smartphones for High School and Universities," Journal of Chemical Education **98** (6), 1937-1945 (2021).
235. K. Rabosky, C. Inglefield and K. Spirito, "Interference and Diffraction in Modern Technology: A New Approach for an Introductory Physics Laboratory Experiment," The Physics Teacher **58** (9), 646-648 (2020).
236. L. Feenstra, C. Julia and P. Logman, "A Lego® Mach–Zehnder interferometer with an Arduino detector," Physics Education **56** (2) (2021).
237. E. Tufino, L. Gratton and S. Oss, "Two simple experiments using FFT with digital devices in the introductory physics laboratory," The Physics Educator **5** (03), 2350013 (2023).
238. E. K. Grasse, M. H. Torcasio and A. W. Smith, "Teaching UV–Vis spectroscopy with a 3D-printable smartphone spectrophotometer," Journal of Chemical Education **93** (1), 146-151 (2016).
239. C. Balado Sánchez, R. P. Díaz Redondo, A. Fernández Vilas and A. M. Sánchez Bermúdez, "Spectrophotometers for labs: A cost-efficient solution based on smartphones," Computer Applications in Engineering Education **27** (2), 371-379 (2018).
240. Y.-G. Ju, "Fabrication of a low-cost and high-resolution papercraft smartphone spectrometer," Physics Education **55** (3), 035005 (2020).
241. A. R. R. Castellannos, H. Castellanos and C. Alvarez-Salazar, "Using homemade spectrometers to perform accurate measurements of discrete and continuous spectra," arXiv preprint arXiv:2201.07110 (2022).

242. V. V. Mayer and E. I. Varaksina, "Holographic grating and smartphone-based optical spectrograph for educational research," European Journal of Physics **41** (5) (2020).
243. L.-J. Thoms, G. Colicchia and R. Girwidz, "Color reproduction with a smartphone," The Physics Teacher **51** (7), 440-441 (2013).
244. G. Colicchia and H. Wiesner, "Looking into the Eye with a Smartphone," Physics Teacher **53** (2), 106-108 (2015).
245. T. Hergemöller and D. Laumann, "Smartphone Magnification Attachment: Microscope or Magnifying Glass," Physics Teacher **55** (6), 361-364 (2017).
246. P. Preechaburana and S. Amloy, "Smartphone-based surface plasmon resonance imaging for near-field concentration mapping," European Journal of Physics **42** (4) (2021).
247. M. Meissner and H. Haertig, "Smartphone astronomy," Physics Teacher **52** (7), 440-441 (2014).
248. L. Schellenberg, "Revisiting smartphone astronomy," Physics Teacher **53** (1), 4-5 (2015).
249. A. Barrera-Garrido, "Analyzing planetary transits with a smartphone," Phys. Teach. **53**, 179-181 (2015 ).
250. J. Durelle, J. Jones, S. Merriman and A. Balan, "A smartphone-based introductory astronomy experiment: Seasons investigation," Physics Teacher **55** (2), 122-123 (2017).
251. J. Tronicke and M. H. Trauth, "Classroom-sized geophysical experiments: magnetic surveying using modern smartphone devices," European Journal of Physics **39** (3) (2018).
252. H. Caerols, R. A. Carrasco and F. A. Asenjo, "Using smartphone photographs of the Moon to acquaint students with non-Euclidean geometry," American Journal of Physics **89** (12), 1079-1085 (2021).
253. S. J. Spicker, A. Küpper and A. Bresges, "astro-lab@ home–Bringing Science to the Sofa," Commun. Astron. Phys. J. **31**, 37-41 (2022).
254. M. Westerfield, *Building IPhone and IPad Electronic Projects: Real-world Arduino, Sensor, and Bluetooth Low Energy Apps in TechBASIC*. (" O'Reilly Media, Inc.", 2013).
255. J. M. Kinser, *Kinematic Labs with Mobile Devices*. (Morgan & Claypool Publishers, 2015).
256. M. E. Auer, A. K. Azad, A. Edwards and T. De Jong, *Cyber-physical laboratories in engineering and science education*. (Springer, 2018).
257. G. Organtini, *Physics experiments with Arduino and smartphones*. (Springer, 2021).
258. I. A. D. Astuti, D. Dasmo, N. Nurullaeli and I. B. Rangka, "The impact of pocket mobile learning to improve critical thinking skills in physics learning," Journal of Physics: Conference Series **1114** (1), 80-86 (2018).
259. F. Moosvi, S. A. Reinsberg and G. W. Rieger, "Can a hands-on physics project lab be delivered effectively as a distance lab?," International Review of Research in Open and Distance Learning **20** (1), 22-42 (2019).
260. C. L. Countryman, "The educational impact of smartphone implementation in introductory mechanics laboratories," 95-98 (2015).
261. K. Hochberg, J. Kuhn and A. Müller, "Using smartphones as experimental tools—effects on interest, curiosity, and learning in physics education," Journal of Science Education and Technology **27** (5), 385-403 (2018).
262. K. Nikolopoulou and M. Kousloglou, "Mobile learning in science: A study in secondary education in Greece," Creative Education **10** (06), 1271 (2019).
263. E. Luh Sukariasih, L. Sahara, L. Hariroh and S. Fayanto, "Studies the use of smartphone sensor for physics learning," Int. J. Sci. Technol. Res **8**, 862-870 (2019).

264. L. Darmendrail, A. Gasparini and A. Müller, "Mobile devices as experimental tools in physics education: some historical and educational background," arXiv preprint arXiv:2303.07294 (2023).
265. H. Rediansyah, "The Model of Physics Experiments Using Smartphone Sensors in Senior High School 1 Pangkalpinang," SEAQIS Journal of Science Education **2** (01), 34-38 (2022).
266. R. Alfonte Zapana and M. d. C. Córdova Martínez, presented at the Proceedings of the 2020 4th International Conference on Education and E-Learning, 2020 (unpublished).
267. M. Bülbül, N. Yiğit and B. Garip, presented at the Journal of Physics: Conference Series, 2016 (unpublished).
268. W.-K. Wong, T.-K. Chao, Y.-W. Lien and C.-J. Wu, presented at the International Conference on Computers in Education, 2012 (unpublished).
269. C.-Y. Liu, C.-J. Wu, W.-K. Wong, Y.-W. Lien and T.-K. Chao, "Scientific modeling with mobile devices in high school physics labs," Computers & Education **105**, 44-56 (2017).
270. C. Countryman and M. Paesler, "Enhancing Students' Understanding of Scientific Equipment: Smartphones in the Laboratory," arXiv preprint arXiv:1607.03736 (2016).
271. C. Orban and R. Teeling-Smith, "Computational thinking in introductory physics," The Physics Teacher **58** (4), 247-251 (2020).
272. N. G. Holmes, C. E. Wieman and D. A. Bonn, "Teaching critical thinking," Proceedings of the National Academy of Sciences **112** (36), 11199-11204 (2015).
273. L. F. Motiwalla, "Mobile learning: A framework and evaluation," Computers & education **49** (3), 581-596 (2007).
274. M. Milner-Bolotin, O. Aminov, W. Wasserman and V. Milner, "Pushing the boundaries of science demonstrations using modern technology," Canadian Journal of Physics **98** (6), 571-578 (2020).
275. A. Kaps, T. Splith and F. Stallmach, "Implementation of smartphone-based experimental exercises for physics courses at universities," Physics education **56** (3), 035004 (2021).
276. D. Kramer, "Teaching physics with phones: A game changer?," Physics Today **77** (1), 23-25 (2024).
277. L. Darmendrail and A. Müller, presented at the INTED2018 Proceedings, 2018 (unpublished).
278. C. I. Larnder, F. Nebia, M. Livingstone and S. Huang, "Physics, smartphones and 3D-print technology: A digital-transition case study in science education," Revue internationale des technologies en pédagogie universitaire **17** (1), 43-56 (2020).
279. S. Becker, P. Klein, A. Gößling and J. Kuhn, "Using mobile devices to enhance inquiry-based learning processes," Learning and Instruction **69**, 101350 (2020).
280. D. J. O'Brien, "A guide for incorporating e-teaching of physics in a post-COVID world," American journal of physics **89** (4), 403-412 (2021).
281. J. S. Bruner, "On learning mathematics1," Mathematics Teacher **53** (8), 610-619 (1960).
282. Y. Sirisathitkul and C. Sirisathitkul, "Smartphones as smart tools for science and engineering laboratory: A review," Iraqi Journal of Science, 2240-2249 (2023).
283. N. R. Council, B. o. S. Education and C. o. G. o. I. t. N. G. S. Standards, "Guide to implementing the next generation science standards," (2015).
284. N. R. Council, D. o. Behavioral, S. Sciences, B. o. S. Education, N. C. o. S. E. Standards and Assessment, *National science education standards*. (National Academies Press, 1995).
285. P. Cicuta and G. Organtini, presented at the Journal of Physics: Conference Series, 2022 (unpublished).

286. J. A. Sans, F. J. Manjón Herrera, V. P. Cuenca Gotor, M. H. Giménez Valentín, I. Salinas, J. A. Barreiro Diez, J. A. Monsoriu Serra and J. A. Gómez Tejedor, presented at the 1ST INTERNATIONAL CONFERENCE ON HIGHER EDUCATION ADVANCES (HEAD'15), 2015 (unpublished).
287. M. Colţ, C. Radu, O. Toma, C. Miron and V.-A. Antohe, "Integrating smartphone and hands-on activities to real experiments in physics," Rom. Reports Phys **72** (4), 1-12 (2020).
288. A. Kittiravechote and T. Sujarittham, presented at the Journal of Physics: Conference Series, 2021 (unpublished).
289. M. Á. González and M. Á. González, presented at the Selected papers from the 20th international conference on multimedia in physics teaching and learning, 2016 (unpublished).
290. J. Chevrier, L. Madani and A. Bsiesy, presented at the LearnxDesign Chicago 2015, 2015 (unpublished).
291. Z. Lv and Z.-y. Peng, presented at the Proceedings of the 2021 6th International Conference on Distance Education and Learning, 2021 (unpublished).
292. A. Kaps and F. Stallmach, "Development and didactic analysis of smartphone-based experimental exercises for the smart physics lab," Physics Education **57** (4), 045038 (2022).
293. M. Windschitl, J. Thompson, M. Braaten and D. Stroupe, "Proposing a core set of instructional practices and tools for teachers of science," Science education **96** (5), 878-903 (2012).
294. M. Windschitl, J. Thompson and M. Braaten, "How novice science teachers appropriate epistemic discourses around model-based inquiry for use in classrooms," Cognition and Instruction **26** (3), 310-378 (2008).
295. R. Duschl, "Science education in three-part harmony: Balancing conceptual, epistemic, and social learning goals," Review of research in education **32** (1), 268-291 (2008).
296. M. Milner-Bolotin and V. Milner, "Increasing student science, technology, engineering and mathematics engagement through phyphox activities: Three practical examples," Future in Educational Research **3** (1), 111-126 (2025).
297. M. Milner-Bolotin, in *STEM Teachers and Teaching in the Digital Era: Professional Expectations and Advancement in the 21 st Century Schools* (Springer, 2020), pp. 201-219.
298. B.-Y. Chang, "Smartphone-based Chemistry Instrumentation: Digitization of Colorimetric Measurements," Bulletin of the Korean Chemical Society **33** (2), 549-552 (2012).
299. M. Rezazadeh, S. Seidi, M. Lid, S. Pedersen-Bjergaard and Y. Yamini, "The modern role of smartphones in analytical chemistry," TrAC Trends in Analytical Chemistry **118**, 548-555 (2019).
300. M. Pohanka, "Photography by Cameras Integrated in Smartphones as a Tool for Analytical Chemistry Represented by an Butyrylcholinesterase Activity Assay," Sensors (Basel) **15** (6), 13752-13762 (2015).
301. A. Stopczynski, C. Stahlhut, M. K. Petersen, J. E. Larsen, C. F. Jensen, M. G. Ivanova, T. S. Andersen and L. K. Hansen, "Smartphones as pocketable labs: visions for mobile brain imaging and neurofeedback," Int J Psychophysiol **91** (1), 54-66 (2014).
302. V. Lang and A. Šorgo, in *INTED2024 Proceedings* (2024), pp. 465-468.
303. V. Lang, M. Melanšek and A. Šorgo, in *2024 47th MIPRO ICT and Electronics Convention (MIPRO)* (2024), pp. 420-424.
304. D. Stroupe, "Naming and disrupting epistemic injustice across curated sites of learning," Journal of the Learning Sciences **31** (2), 317-334 (2022).

305. H. Kang and M. Windschitl, "How does practice-based teacher preparation influence novices' first-year instruction?," Teachers College Record **120** (8), 1-44 (2018).
306. N. Zhanatbekova, A. Abdulayeva, Y. Andasbayev, Z. Zhiyembayev and M. Urazova, "Formation of research skills of students when performing laboratory work in physics: Virtual laboratory vs smartphone-based laboratory," Cypriot Journal of Educational Sciences **17** (12), 4303-4310 (2022).
307. S. Z. Lahme, P. Klein, A. Lehtinen, A. Müller, P. Pirinen, L. Rončević and A. Su, "Evaluating digital experimental tasks for physics laboratory courses Didaktik der Physik Frühjahrstagung – Hannover 2023 Evaluating digital experimental tasks for physics laboratory courses," (June) (2023).
308. P. Kácovský, M. Snětinová, M. Chvál, J. Houfková and Z. Koupilová, "Predictors of students' intrinsic motivation during practical work in physics," International Journal of Science Education **45** (10), 806-826 (2023).
309. S. Jiang, X. Huang, S. H. Sung and C. Xie, "Learning Analytics for Assessing Hands-on Laboratory Skills in Science Classrooms Using Bayesian Network Analysis," Research in Science Education **53** (2), 425-444 (2023).
310. E. Tufino and G. Organtini, "Evaluation of the effectiveness of an introductory mechanics Lab with Arduino and smartphone," (2023).
311. A. C. Evains, S. Anggereni and M. S. L, "Enhancing Science Process Skills in Physics Education : The Impact of the Phyphox Smartphone Application in High School Laboratories A . Introduction Integrating technology in education , particularly in physics , represents a crucial Previous research ha," **3** (1), 9-18 (2024).
312. P. Dam-O, Y. Sirisathitkul, T. Eadkhong, S. Srivaro, C. Sirisathitkul and S. Danworaphong, "Online physics laboratory course: United Kingdom Professional Standards Framework perspective from Walailak University, Thailand," Distance Education **45** (1), 122-140 (2024).
313. G. Organtini and E. Tufino, "Effectiveness of a Laboratory Course with Arduino and Smartphones," Education Sciences **12** (12) (2022).
314. A. Yurchenko, Y. Khvorostina, V. Shamonia, M. Soroka and O. Semenikhina, "Digital Technologies in Teaching Physics: An Analysis of Existing Practices," 2023 46th ICT and Electronics Convention, MIPRO 2023 - Proceedings, 666-671 (2023).
315. P. Onorato, T. Rosi, E. Tufino, V. Ambrosini, S. Toffaletti, C. Caprara, M. Di Mauro, L. M. Gratton and S. Oss, "Bringing the Physics Laboratory at home: Tools and methodologies for distance learning at the University of Trento," Nuovo Cimento della Societa Italiana di Fisica C **46** (6), 1-16 (2023).
316. M. Antoninis, B. Alcott, S. Al Hadheri, D. April, B. Fouad Barakat, M. Barrios Rivera, Y. Baskakova, M. Barry, Y. Bekkouche and D. Caro Vasquez, "Global Education Monitoring Report 2023: Technology in education: A tool on whose terms?," (2023).
317. J. Van Dijk, *The digital divide*. (John Wiley & Sons, 2020).
318. V. S. Katz, C. Gonzalez and K. Clark, "Digital inequality and developmental trajectories of low-income, immigrant, and minority children," Pediatrics **140** (Supplement_2), S132-S136 (2017).
319. C. L. Odgers and M. R. Jensen, "Annual research review: Adolescent mental health in the digital age: Facts, fears, and future directions," Journal of Child Psychology and Psychiatry **61** (3), 336-348 (2020).
320. V. Rideout, A. Peebles, S. Mann and M. Robb.

321. J. S. Radesky, J. Schumacher and B. Zuckerman, "Mobile and interactive media use by young children: the good, the bad, and the unknown," Pediatrics **135** (1), 1-3 (2015).
322. Z. C. Zacharia, C. Manoli, N. Xenofontos, T. De Jong, M. Pedaste, S. A. van Riesen, E. T. Kamp, M. Mäeots, L. Siiman and E. Tsourlidaki, "Identifying potential types of guidance for supporting student inquiry when using virtual and remote labs in science: A literature review," Educational technology research and development **63** (2), 257-302 (2015).
323. I. Roll and R. Wylie, "Evolution and revolution in artificial intelligence in education," International journal of artificial intelligence in education **26** (2), 582-599 (2016).
324. X. Zhai, X. Chu, C. S. Chai, M. S. Y. Jong, A. Istenic, M. Spector, J.-B. Liu, J. Yuan and Y. Li, "A Review of Artificial Intelligence (AI) in Education from 2010 to 2020," Complexity **2021** (1), 8812542 (2021).
325. F. Mahligawati, E. Allanas, M. Butarbutar and N. Nordin, presented at the Journal of Physics: Conference Series, 2023 (unpublished).
326. W. Holmes, M. Bialik and C. Fadel, *Artificial intelligence in education promises and implications for teaching and learning*. (Center for Curriculum Redesign, 2019).
327. S. Z. Salas-Pilco, K. Xiao and X. Hu, "Artificial intelligence and learning analytics in teacher education: A systematic review," Education sciences **12** (8), 569 (2022).
328. C. Wieman and N. G. Holmes, "Measuring the impact of an instructional laboratory on the learning of introductory physics," American Journal of Physics **83** (11), 972-978 (2015).
329. M. Chassignol, A. Khoroshavin, A. Klimova and A. Bilyatdinova, "Artificial Intelligence trends in education: a narrative overview," Procedia Computer Science **136**, 16-24 (2018).
330. C. A. Eden, O. N. Chisom and I. S. Adeniyi, "Integrating AI in education: Opportunities, challenges, and ethical considerations," Magna Scientia Advanced Research and Reviews **10** (2), 006-013 (2024).
331. T. Caswell, S. Henson, M. Jensen and D. Wiley, "Open educational resources: Enabling universal education," International Review of Research in Open and Distributed Learning **9** (1), 1-11 (2008).
332. S. Mishra, "Open educational resources: Removing barriers from within," Distance education **38** (3), 369-380 (2017).
333. U.-D. Ehlers, "Extending the territory: From open educational resources to open educational practices," Journal of open, flexible and distance learning **15** (2), 1-10 (2011).
334. D. Wiley, T. Bliss and M. McEwen, "Open educational resources: A review of the literature," Handbook of research on educational communications and technology, 781-789 (2013).
335. A. J. Williams and H. E. Pence, "Smart Phones, a Powerful Tool in the Chemistry Classroom," Journal of Chemical Education **88** (6), 683-686 (2011).